\documentclass[showpacs,preprintnumbers,amsmath,amssymb,floatfix]{revtex4}
\usepackage[german, english]{babel}
\usepackage{amsmath}
\usepackage{graphicx}
\usepackage{epsf}
\usepackage[hang,nooneline]{subfigure}
\newcounter{fig}

\begin{document}

\title{ Geometry of Spinning Ellis Wormholes}

\author{
  Xiao Yan Chew, Burkhard Kleihaus, 
  and  Jutta Kunz}
\affiliation{
  Institut f\"ur  Physik, Universit\"at Oldenburg, 
  D-26111 Oldenburg, Germany
}

\date{\today}
\pacs{04.20.Jb, 04.40.-b, 04.40.Nr}

\vspace{1.5truecm}

\begin{abstract}
We give a detailed account of the
properties of spinning Ellis wormholes,
supported by a phantom field.
The general set of solutions depends on three parameters,
associated with the size of the throat, the rotation 
and the symmetry of the solutions.
For symmetric wormholes the global charges
possess the same values in both asymptotic regions,
while this is no longer the case for non-symmetric wormholes.
We present mass formulae for these wormholes, study
their quadrupole moments, and discuss the geometry of their throat
and their ergoregion. We demonstrate, that these wormholes
possess limiting configurations corresponding
to an extremal Kerr black hole. 
Moreover, we analyze the geodesics of these wormholes,
and show that they possess bound orbits.
\end{abstract}

\maketitle

\section{Introduction}

General Relativity allows for 
Lorentzian wormholes \cite{Visser:1995cc},
which can be either non-traversable or traversable.
Non-traversable wormholes were first discussed 
in 1935 by Einstein and Rosen \cite{Einstein:1935tc}.
Here the famous `Einstein-Rosen bridge'
represents an interesting feature of the Schwarzschild spacetime,
connecting two asymptotically flat universes by a throat.
In the 1950's Wheeler discussed the intriguing possibility 
to connect two distant regions within a single universe 
\cite{Wheeler:1957mu,Wheeler:1962}.  Unfortunately,
this kind of wormholes does not allow for their passage
\cite{Kruskal:1959vx,Fuller:1962zza,Redmount:1985,Eardley:1974zz,Wald:1980nk}.

Transversable wormholes in General Relativity were first obtained 
by Ellis \cite{Ellis:1973yv,Ellis:1979bh} and Bronnikov \cite{Bronnikov:1973fh}
(see also \cite{Kodama:1978dw}).
However, these need the presence of a new kind of matter field
coupled to gravity. Such a phantom field carries the wrong sign
in front of its kinetic term, and consequently its
energy-momentum tensor violates the null energy condition.
The conditions for the existence of traversable wormholes 
as well as the properties of such static wormholes
were discussed later in detail by Morris and Thorne
\cite{Morris:1988cz},
who addressed also the possibility of time-travel
\cite{Morris:1988tu}.

While phantom fields have been employed frequently in cosmology
in recent years, since they give rise to an
accelerated expansion of the Universe \cite{Lobo:2005us},
it was also pointed out 
\cite{Hochberg:1990is,Fukutaka:1989zb,Ghoroku:1992tz,Furey:2004rq,Bronnikov:2009az, Kanti:2011jz,Kanti:2011yv},
that the presence of a phantom field can be evaded 
in the construction of wormholes,
when gravity theories with higher curvature terms are considered.

Wormholes have also been considered from an astrophysical point of view,
since they represent hypothetical objects that can be searched for
observationally
\cite{Abe:2010ap,Toki:2011zu,Takahashi:2013jqa}.
Cramer et al. \cite{Cramer:1994qj} and Perlick \cite{Perlick:2003vg},
for instance, have investigated 
wormholes as gravitational lenses, 
Tsukamoto et al. have addressed the Einstein rings of wormholes
\cite{Tsukamoto:2012xs},
Bambi \cite{Bambi:2013nla}
and Nedkova et al.  \cite{Nedkova:2013msa} 
have studied the shadow of wormholes,
while Zhou et al. \cite{Zhou:2016koy}
determined the iron line profile in the X-ray reflected spectrum 
of a thin accretion disk around wormholes.
The astrophysical signatures of mixed
neutron star--wormhole systems 
have been addressed by Dzhunushaliev et al.
\cite{Dzhunushaliev:2011xx,Dzhunushaliev:2012ke,Dzhunushaliev:2013lna,Dzhunushaliev:2014mza,Aringazin:2014rva,Dzhunushaliev:2016ylj}.

Here we consider rotating wormholes supported by a phantom field,
which represent solutions of the coupled Einstein-phantom field equations.
Slowly rotating perturbative solutions of these equations were
obtained by Kashargin and Sushkov 
\cite{Kashargin:2007mm,Kashargin:2008pk}. This was followed by
a brief discussion of rapidly rotating non-perturbative wormholes
given by two of us \cite{Kleihaus:2014dla}.
In the present paper we give a detailed account of the properties
of these rotating wormholes. Moreover, we discuss rotating
non-symmetric wormholes for the first time.
Let us note, that the rotating wormhole metric
presented by Teo \cite{Teo:1998dp} 
was not obtained as a solution of a coupled system of 
Einstein-matter equations.

The paper is structured as follows. We begin with the theoretical
setting in section II, 
which includes the presentation of the action and the Ansatz, 
the discussion of the equations of motion and the boundary conditions, 
the derivation of the global charges and mass formulae
as well as the derivation of the quadrupole moment. 
We then give
a discussion of the geometric properties of the wormholes, 
in particular, we define the position of the throat
as a minimal surface,
and we demonstrate the violation of the energy conditions.

Subsequently in section III
we present our numerical results for symmetric wormholes,
i.e., those wormholes whose global charges are the same on both
sides of the wormhole. These wormholes then possess one less parameter.
In particular, we consider families of wormhole configurations
for fixed equatorial throat radius, and demonstrate that 
these solutions approach an extremal Kerr black hole, as the
angular momentum tends towards its maximal value.
We then exhibit embeddings of wormhole spacetimes and 
demonstrate, that the Gaussian curvature of the throat
turns negative at the poles, when the rotation is sufficiently fast,
analogously to the curvature of the horizon of Kerr black holes.
Finally, we examine the effective potential and 
the geodesics of rotating wormholes, which in contrast
to static Ellis wormholes allow for bound orbits.

Non-symmetric rotating wormholes are presented in section IV.
They depend on three parameters, and have distinct properties
when the asymmetry parameter $\gamma$ is negative or positive.
We analyze the domain of existence of these non-symmetric
wormholes, and demonstrate that the families of non-symmetric
wormholes also approach an extremal Kerr black hole. 
We then address the somewhat surprising fact, that the
rotational velocity of the throat can exceed the velocity of light
for positive $\gamma$.  
We show how the location of the throat changes as the asymmetry
of the solutions is varied. Finally, we consider 
embeddings of non-symmetric wormholes and geodesics in these
spacetimes.
We end with our conclusions in section V, and present the
metric functions for large positive asymmetry parameter in the
Appendix.

\section{Theoretical Setting}

\subsection{Action and Ansatz}

We now turn to wormhole solutions in General Relativity,
obtained with a phantom field $\psi$.
The corresponding action reads
\begin{eqnarray}
S= {1\over 16\pi G_{ }}\int d^4x \sqrt{-g} \left(R +
2g^{\mu\nu}\partial_{\mu}\psi \partial_{\nu}\psi 
  \right) \ ,
\end{eqnarray}
leading to the Einstein equations
\begin{equation}
R_{\mu\nu} = - 2 \partial_\mu \psi \partial_\nu \psi 
	  \ ,
\label{einsteinequ}	 
\end{equation}
and the phantom field equation
\begin{eqnarray}
\partial_\mu\left(\sqrt{-g}g^{\mu\nu}\partial_\nu \psi\right) 
& = & 0 \ . 
\label{equpsi}
\end{eqnarray}

To construct stationary rotating spacetimes we employ the line element
\begin{equation}
ds^2 = -e^{f} dt^2 +p^2 e^{-f} 
\left( e^{\nu} \left[d\eta^2 +h d\theta^2\right]
                    + h \sin^2\theta \left(d\phi -\omega dt\right)^2\right) \ ,
\label{lineel}
\end{equation}
where  $f$, $p$, $\nu$ and $\omega$ are functions only of
$\eta$ and $\theta$, and $h=\eta^2 + \eta_0^2$ 
is an auxiliary function. 
We note that the coordinate $\eta$ takes positive and negative
values, i.e.~$-\infty< \eta < \infty$. 
The limits $\eta\to \pm\infty$
correspond to two distinct asymptotically flat regions.
The phantom field $\psi$ also depends only on the coordinates
$\eta$ and $\theta$.

\subsection{Equations of Motion and Boundary Conditions}

When we substitute the above Ansatz into the general set of equations of motion
we obtain a system of non-linear partial differential equations (PDEs).
As observed in \cite{Kleihaus:2014dla}
the PDE for the function $p$ decouples,
and has the simple form
\begin{equation}
p_{,\eta,\eta} +\frac{3\eta}{h}p_{,\eta} + 
\frac{2 \cos\theta}{h\sin\theta}p_{,\theta}+
\frac{1}{h}p_{,\theta,\theta} = 0 \ .
\label{pde_p}
\end{equation}
A solution which satisfies the boundary conditions 
$p(\eta \to \infty) =p(\eta \to -\infty) = 1$ and 
$\partial_\theta p(\theta =0) =\partial_\theta p(\theta =\pi) = 0$ 
is given by 
\begin{equation}
p=1 \ .
\label{p=1}
\end{equation}

Consequently, we obtain the phantom field equation
\begin{equation}
\partial_\eta\left(h \sin\theta \partial_\eta \psi\right) 
+\partial_\theta\left(\sin\theta \partial_\theta \psi\right) 
=0 \ .
\label{pde_psi}
\end{equation}
A first integral can be obtained under the assumption 
$\partial_\theta \psi=0$, 
\begin{equation}
\partial_\eta \psi = \frac{D}{h}\ ,
\label{sol_psie}
\end{equation}
where $D$ denotes the phantom scalar charge.
This leads to the closed form solution for $\psi$
\begin{equation}
\psi = D \left(\arctan(\eta/\eta_0)-\pi/2\right) \ , 
\label{exsol}
\end{equation}
where the integration constant was chosen to satisfy
$\psi(\infty) = 0$.

We now turn to the Einstein equations.
With $p=1$ and $\partial_\theta \psi=0$ 
the equations $R_{\phi\phi}=0$, $R_{\theta\theta}=0$, 
$R_{t\phi}=0$, $R_{tt}=0$ and $R_{\eta\theta}=0$
are independent of the scalar field.
They yield three second order PDEs for the functions 
$f$, $\omega$ and $\nu$, and a constraint,
\begin{eqnarray}
0 & = & 
 \partial_\eta\left( h \sin\theta \partial_\eta f\right)
+\partial_\theta\left( \sin\theta \partial_\theta f\right)
-h \sin^3\theta e^{-2 f}\left(h (\partial_\eta\omega)^2 
                              + (\partial_\theta\omega)^2 \right) \ ,
\label{pdef}\\
0 & = & 
 \partial_\eta\left(h^2\sin^3\theta e^{-2 f} \partial_\eta\omega\right)		      
+\partial_\theta\left(h\sin^3\theta e^{-2 f} \partial_\theta\omega\right) \ ,		      
\label{pdeo}\\
0 & = & 
 \partial_\eta\left( h \sin\theta \partial_\eta \nu\right)
 +\sin\theta \partial_{\theta\theta} \nu -\cos\theta \partial_\theta\nu
 -h \sin^3\theta e^{-2 f}\left(h (\partial_\eta\omega)^2 
                              -2 (\partial_\theta\omega)^2 \right) \ ,
\label{pdenu}\\
0 & = & 
-h \sin\theta \partial_\eta f \partial_\theta f
+ h \cos\theta \partial_\eta \nu + \eta\sin\theta \partial_\theta \nu
+ h^2\sin^3\theta e^{-2 f}\partial_\eta \omega \partial_\theta \omega \ .
\label{constraint}
\end{eqnarray}

In order to obtain regular solutions we have to impose boundary conditions
in the asymptotic regions $\eta \to \pm \infty$ and on the axis 
$\theta = 0, \pi$.
For $\eta \to +\infty$ we require the metric to approach Minkowski spacetime, i.e.,
\begin{equation}
\left. f\right|_{\eta \to \infty} = 0 \ , \ 
\left. \omega\right|_{\eta \to \infty} = 0 \ , \ 
\left. \nu\right|_{\eta \to \infty} = 0 \ .
\label{bcinfty}
\end{equation}
In the limit $\eta \to -\infty$ we allow for finite values of the
functions $f$ and $\omega$,
\begin{equation}
\left. f\right|_{\eta \to -\infty} = \gamma \ , \ 
\left. \omega\right|_{\eta \to -\infty} = \omega_{-\infty} \ , \ 
\left. \nu\right|_{\eta \to -\infty} = 0 \ .
\label{bcmininfty}
\end{equation}
We refer to the solutions as symmetric, when $\gamma=0$,
and non-symmetric, when $\gamma \ne 0$.
Note, that in the limit $\eta \to -\infty$ the spacetime  becomes 
Minkowskian after a suitable coordinate transformation is performed.
Regularity on the symmetry axis requires
\begin{equation}
\left.\partial_\theta f\right|_{\theta = 0} = 0 \ , \ 
\left.\partial_\theta \omega \right|_{\theta = 0} = 0 \ , \ 
\left. \nu\right|_{\theta = 0} = 0 \ ,
\label{bcaxis}
\end{equation}
together with the analogous conditions at $\theta = \pi$.

Static wormhole solutions correspond to $\omega_{-\infty}=0$. 
They are known in closed form,
\begin{equation}
f = \frac{\gamma}{2} \left(1-\frac{2}{\pi}\arctan\left(\frac{\eta}{\eta_0}\right) \right) \ , \
\omega = 0  \ , \ \nu = 0  \ . \
\label{statsol}
\end{equation}

Rotating wormhole solutions of the above equations 
have been studied in \cite{Kleihaus:2014dla} for 
the symmetric case only, while we here address the non-symmetric case, as well.
Substitution of the solutions for $f$, $\omega$ and $\nu$ in  $R_{\eta\eta}$ shows, that
\begin{equation}
R_{\eta\eta} = - 2\frac{D^2}{h^2} \ ,
\label{sol_psie2}
\end{equation}
where the scalar charge $D$ 
depends on the mass and the angular momentum of the spacetime.
 Explicitly we find
\begin{equation}
D^2  =  \frac{h}{4}\left[ h  (\partial_\eta f)^2 - (\partial_\theta f)^2 \right]
         -\frac{h}{2}\left(\eta \partial_\eta \nu-
	                  \frac{\cos\theta}{\sin\theta}\partial_\theta \nu\right)
         -\frac{h^2}{4}\sin^2\theta e^{-2 f}
	 \left[ h (\partial_\eta\omega)^2 
               - (\partial_\theta\omega)^2 \right] +\eta_0^2 \ .
\label{eqd2}
\end{equation}

\subsection{Mass, Angular Momentum and Mass Formulae}

The mass and angular momentum of the wormhole
solutions are engraved in the asymptotic
form of the metric tensor.
For an asymptotically flat metric, 
the mass and the angular momentum 
can be read off from the
components $g_{tt}$ and $g_{t\varphi}$,
\begin{equation}
g_{tt} \underset{\eta \to \pm \infty} 
\longrightarrow - \left(1\mp\frac{2\mu_\pm}{\eta}\right)  \ , \ \ \
g_{t\varphi} \underset{\eta \to \pm \infty} 
\longrightarrow -\frac{2 J_\pm  \sin^2\theta}{\eta} \ .
\label{mass_angmom}
\end{equation}

In the limit $\eta \to +\infty$ the wormhole metric becomes Minkowskian,
as seen from the asymptotic behavior of the metric functions.
In particular,
\begin{eqnarray}
& &   
f \to -\frac{2 \mu_+}{\eta} \ , \ \ \ 
\omega \to \frac{2J_+}{\eta^3} \ , \ \ \ 
{\rm as} \ \eta \to +\infty \ , 
\label{asymp1} 
\end{eqnarray}
so we can read off the mass $\mu_+$ and the angular momentum $J_+$ directly.

The analogous asymptotic expansion 
at the other side of the wormhole reads
\begin{eqnarray}
& &
f \to \gamma+\frac{2 \mu_-}{\eta}\ , \ \ \ 
\omega \to \omega_{-\infty}+ \frac{2J_-}{\eta^3}\ , \ \ \  
\label{asymp2}
{\rm as } \ \eta \to - \infty \ . 
\end{eqnarray}
However, in order 
to identify the mass $\bar \mu_-$ 
and the angular momentum $\bar J_-$
at $\eta \to -\infty$,
we first have to perform a coordinate transformation
to obtain an asymptotically flat spacetime in this limit.
After applying
\begin{eqnarray}
\label{trans}
\bar t = e^{\gamma/2} t \ , \ \ \
\bar \eta = e^{-\gamma/2} \eta \ , \ \ \
\bar \phi = \phi - \omega_{-\infty} t \ ,
\end{eqnarray}
we obtain
$\bar \mu_-$ and $\bar J_-$ in terms of the parameters
$\mu_-$ and $J_-$,
\begin{eqnarray}
\label{mom-}
\bar J_- = J_- e^{-2\gamma} \ , \ \ \
\bar \mu_- = \mu_- e^{-\gamma/2} \ .
\end{eqnarray}

Let us now derive some relations between the global charges.
We first note that Eq.~(\ref{pdeo}) is in the form of a conservation law,
and integrate Eq.~(\ref{pdeo}) over the full range of $\eta$ and $\theta$,
$-\infty < \eta  < \infty$, $0 < \theta < \pi $. 
This leads to
\begin{equation}
\int_0^{\pi}\left[h^2 e^{-2f} 
\sin^3\theta \omega_{,\eta}\right]_{\eta\to \infty}
d \theta = 
\int_0^{\pi}\left[h^2 e^{-2f} 
\sin^3\theta \omega_{,\eta}\right]_{\eta\to -\infty}
d \theta \ .
\label{omeqi2}
\end{equation}
Inserting the asymptotic expansions for $\omega$ and $f$, Eqs.~(\ref{asymp1})-(\ref{asymp2}),
we find
\begin{equation}
J_+ = e^{-2\gamma} J_-    = \bar J_- \ .
\label{j-m-rel1}
\end{equation}
Thus the angular momenta 
on both sides of the wormhole are equal.

To derive a second relation, we 
multiply  Eq.~(\ref{pdeo}) by $\omega$ and substract it from 
Eq.~(\ref{pdef}). This yields again an equation in the form of a conservation law,
\begin{equation}
\partial_\eta\left( h\sin\theta 
   \left[
   \partial_\eta f - e^{-2f} h \sin^2\theta \omega \partial_\eta\omega
   \right]\right)
+
\partial_\theta\left( \sin\theta 
   \left[
   \partial_\theta f - e^{-2f} h \sin^2\theta \omega \partial_\theta\omega
   \right]\right)
=0 \ .
\label{pdefo}
\end{equation}
Integrating Eq.~(\ref{pdefo}) and taking into account the 
asymptotic expansions Eqs.~(\ref{asymp1})-(\ref{asymp2}),
we find
\begin{equation}
\mu_+ + \mu_- = 2 \omega_{-\infty} e^{-2\gamma} J_- = 2 \omega_{-\infty} J_+ \ .
\label{j-m-rel}
\end{equation}
For the static wormhole solutions (\ref{statsol}),
this relation reduces to $\mu_+ = - \mu_-$, where
the mass $\mu_+$ is related to $\gamma$ by $\mu_+ = \gamma \eta_0 / 2 \pi$.

In the symmetric case, we regain a mass formula akin to the Smarr formula
for black holes. Here we integrate both formulae
from the throat ($\eta = 0$) to infinity, and take into account that
$\partial_\eta f$ vanishes at the throat.
Denoting $\omega(0) = \omega_0$,
we find \cite{Kleihaus:2014dla}
\begin{equation}
\label{smarr}
\mu_+ = \omega_{-\infty} J_+ = 2 \omega_0 J_+ ,
\end{equation}
where $\omega_{-\infty} = 2 \omega_0$ follows from the symmetry 
and the choice of boundary conditions.
Consequently, also the relation 
$\mu_- = 2 \omega_0 J_-$ holds in this case.
Thus on both sides of the wormhole the mass is the same.
Note, that the mass relation (\ref{smarr}) 
then agrees with the one for extremal Kerr black holes,
when $\omega_0$ is identified with the horizon angular velocity.

To find a relation of $D^2$ to the mass we evaluate Eq.~(\ref{eqd2})
in the limit $\eta \to \infty$. Here we must take into account 
the asymptotic behaviour of the function $\nu$, 
\begin{equation}
\nu \to - c_2 \frac{\sin^2\theta}{\eta^2} \ .
\label{asympnu}
\end{equation}
This yields
\begin{equation}
c_2  = \mu_+^2 +\eta_0^2 - D^2 \ ,
\label{eqd2m}
\end{equation}
which reduces to $0 = \mu_+^2 +\eta_0^2 -D^2$
in the static case.

\subsection{Quadrupole Moment}

In order to extract the quadrupole moment $Q$ of the rotating wormhole
solutions, we need to consider the expansion of the metric
in the asymptotic regions. For simplicity,
let us only consider $\eta \to + \infty$ here.
The asymptotic expansion of the metric functions is then given by
\begin{eqnarray}
f & = & -\frac{2 \mu_+}{\eta} +\frac{2}{3}\frac{ \mu_+\eta_0^2}{\eta^3}
+\frac{2}{3}\frac{2 \mu_+\eta_0^2 - 3 f_3}{\eta^3} P_2(\cos\theta)
+{\cal O}(\eta^{-4}) \ ,
\label{expan_f}
\\
\nu  & = & -c_2 \frac{\sin^2\theta}{\eta^2} +{\cal O}(\eta^{-3})
 \ ,
\label{expan_nu}
\\
\omega  & = & 2 \frac{J_+}{\eta^3} +{\cal O}(\eta^{-4})
 \ ,
\label{expan_om}
\end{eqnarray}
where $P_2(\cos\theta)$ is the second Legendre polynomial and 
$f_3$ and $c_2$ are constants.

The quadrupole moment can now be derived
employing the definitions of Geroch and Hansen
\cite{Geroch:1970cd,Hansen:1974zz},
and following later work
\cite{Hoenselaers:1992bm,Sotiriou:2004ud,Pappas:2012ns}
according to the steps 
outlined in \cite{Kleihaus:2016dui} 
we first transform the metric to
quasi-isotropic coordinates.
This yields
\begin{equation}
ds^2 = -e^f dt^2
+\left(1+\frac{r_0^2}{r^2}\right)^2 e^{-f}
\left[ e^\nu (dr^2 + r^2 d\theta^2) 
      +r^2 \sin^2\theta(d\phi - \omega dt)^2\right] \ ,
\label{ds_iso1}
\end{equation}
where $r$ is related to $\eta$ by 
$\eta/\eta_0 = (r/r_0-r_0/r)/2$, and $r_0=\eta_0/2$.
We compare this with 
\begin{equation}
ds^2 = -e^{2\nu_0} dt^2
+e^{2(\nu_1-\nu_0)}
\left[ e^{2\nu_2} (dr^2 + r^2 d\theta^2) 
      +r^2 \sin^2\theta(d\phi - \omega dt)^2\right]
\label{ds_iso2}
\end{equation}
and identify
\begin{equation}
\nu_0 = f/2\ , \ \ \ 
e^{\nu_1} = 1+\frac{r_0^2}{r^2} \ , \ \ \ 
\nu_2 = \nu/2\ .
\label{identmet}
\end{equation}
Using the expansion
$\eta^{-1} = r^{-1}(1+(r/r_0)^{-2} +{\cal O}((r/r_0)^{-4}))$
in the asymptotic region
we find by comparison with Eqs.~(A9)-(A11) in \cite{Kleihaus:2016dui}
\begin{eqnarray}
\nu_0 & = &
 -\frac{\mu_+}{r} +\frac{1}{12}\frac{\mu_+\eta_0^2 }{r^3}
+\frac{2 \mu_+\eta_0^2 - 3 f_3}{3 r^3} P_2(\cos\theta)
+{\cal O}(r^{-4}) 
\nonumber\\
& = &
 -\frac{\mu_+}{r} +\frac{1}{3}\frac{d_1 \mu_+}{r^3}
-\frac{M_2}{r^3} P_2(\cos\theta)
+{\cal O}(r^{-4}) 
\ ,
\label{expan_nu0}
\\
\nu_1 & = & \frac{\eta_0^2}{4r^2} +{\cal O}(r^{-3}) 
\nonumber\\
& = &
  \frac{d_1}{r^2} +{\cal O}(r^{-3}) 
\ ,
\label{expan_nu1}\\
\nu_2 & = & 
 -c_2 \frac{\sin^2\theta}{2 r^2} +{\cal O}(r^{-3})
\nonumber\\
& = &
 -\frac{4 \mu_+^2 +16 d_1 - q_2}{8 r^2} \sin^2\theta+{\cal O}(r^{-3})
\ ,
\label{expan_nu2}
\end{eqnarray}
where we have slightly changed the notation of \cite{Kleihaus:2016dui}:
$D_1 \to d_1$, $q^2 \to -q_2$.
Now $d_1$, $q_2$ and $M_2$ can be expressed in terms of 
$f_3$ and $c_2$,
\begin{equation}
d_1 = \frac{\eta_0^2}{4} \ , \ \ \ 
q_2 = 4(\mu_+^2 +\eta_0^2 - c_2) = 4 D^2 \ , \ \ \ 
M_2 = f_3- \frac{2\mu_+\eta_0^2}{3}\ .
\end{equation}
Thus we finally arrive at the quadrupole moment
\begin{equation}
Q = -f_3 + \mu_+\eta_0^2 +\frac{\mu_+(\mu_+^2 - D^2)}{3} \  
\label{quadmom}
\end{equation}
(see e.g.~\cite{Kleihaus:2016dui} for more details).

\subsection{Geometric Properties}

Let us next address the geometric properties of the wormhole solutions.
We first consider the equatorial (or circumferential) radius ${R}_e(\eta)$
as a function of the radial coordinate
\begin{equation}
\label{Re_jf}
{R}_e(\eta) = \sqrt{\eta^2+\eta_0^2}\left[e^{-f/2}\right]_{\theta=\pi/2} \ .
\end{equation}
Because of the rotation, the throat of the wormholes should deform and 
its circumference should be largest in the equatorial plane. 
Therefore a study of ${R}_e(\eta)$ should reveal the location of the throat
in the equatorial plane.

Consequently, we define the throat of the wormhole as the surface of minimal
area, which intersects the equatorial plane at the circle of
minimal circumferential radius $R_e$, i.e.,
\begin{equation}
R_e = \min_{-\infty \leq \eta \leq \infty}
\left\{ \sqrt{h} e^{-f/2} |_{\theta = \pi/2} \right\}
\ ,
\label{remin}
\end{equation}
for the line element Eq.~(\ref{lineel}).

We now parametrize the surface of the throat by 
$(\eta_t(\theta),\theta, \varphi)$. Then the line element on the surface
reads 
\begin{equation}
d\sigma^2 = e^{-f+\nu} 
\left[\eta_t'^2 + h\right] d\theta^2
+e^{-f} h \sin^2\theta d\varphi^2 \ ,
\label{dsig2}
\end{equation}
where we defined $\eta_t' =d \eta_t/d\theta$,
and the functions $f$, $\nu$ and $h$ are regarded as functions of $\eta_t$ and $\theta$.
The area of the surface is now given by the integral of the square root
of the determinant of the metric tensor
\begin{equation}
A_\sigma = \int{L_\sigma} d\theta d\varphi
\label{arsig}
\end{equation}
with
\begin{equation}
L_\sigma=\sqrt{\eta_t'^2 + h}
\sqrt{h}\sin\theta e^{-f+\nu/2} \ .
\label{Lsig}
\end{equation}

The function $\eta_t(\theta)$ is determined as solution
of the Euler-Lagrange equation
\begin{equation}
\frac{d}{d\theta}\frac{\partial L_\sigma}{\partial \eta_t'} 
-\frac{\partial L_\sigma}{\partial \eta_t} = 0 \ .
\label{ELsig}
\end{equation}
This yields the ordinary differential equation
\begin{equation}
\eta_t''
+h \partial_\eta s - 2\eta_t
-\left[\partial_\theta s-\frac{\cos\theta}{\sin\theta}\right] \eta_t'
+\left[ \partial_\eta s - \frac{3\eta_t}{h}\right]\eta_t'^2
-\left[\partial_\theta s-\frac{\cos\theta}{\sin\theta}\right]\frac{1}{h} \eta_t'^3
=0  \ ,
\label{eqetat}
\end{equation}
where we introduced $s=f-\nu/2$.
We solve this equation numerically for boundary conditions $\eta_t'(0)=0$, required by regularity,
and $\eta_t(\pi/2) = \eta_e$, where $\eta_e$ is the coordinate of the throat in the equatorial plane,
determined by the minimum condition for the circumferential radius.

We note that in the symmetric case the throat is located at the constant coordinate $\eta_t=0$, since
$\eta_e=0$ and $\partial_\eta s =0$ at $\eta =0$ for all $\theta$.
For the static wormholes $\nu=0$ and $\partial_\theta f=0$.
In this case the minimum condition Eq.~(\ref{remin}) yields
$h \partial_\eta f- 2 \eta_e = 0$, which implies $h \partial_\eta s- 2 \eta_e = 0$ in Eq.~(\ref{eqetat}). 
Consequently the solution of Eq.~(\ref{eqetat}) is in this case $\eta_t=\eta_e$.
Therefore the static non-symmetric wormholes also possess a throat with constant coordinate (as expected).
For the rotating non-symmetric wormholes, however, it is not obvious, 
that the throat should be located at a constant radial coordinate, as well.
Therefore the dependence of the coordinate $\eta_t$ on $\theta$ must be
examined numerically.
We further note, that we do not find rotating Ellis wormhole solutions with multiple throats,
separated by bellies (or equators).

To gain further information on the geometry of the rotating symmetric wormholes, 
we consider in addition the polar radius $R_p$ 
\begin{equation}
R_p  =\frac{\eta_0}{\pi} 
\int_0^\pi {\left. e^{(\nu-f)/2}\right|_{\eta=0}  d\theta} \ , 
\label{rpolar}
\end{equation}
and the areal radius $R_A$ 
\begin{equation}
R_A^2=\frac{\eta_0^2}{2}   
\int_0^\pi {\left. e^{\nu/2-f}\right|_{\eta=0} \sin \theta  d\theta} \ .
\label{rarea}
\end{equation}

Denoting the angular velocity of the throat by
$\Omega=\omega_0$, 
the rotational velocity
of the throat in the equatorial plane is given by
\begin{equation}
v_e= {R_e\Omega} \ . 
\label{ve}
\end{equation}
It was shown in \cite{Kleihaus:2014dla} that $v_e\le 1$ for the symmetric
wormholes.

\subsection{Violation of the Null Energy Condition}

Let us next demonstrate the violation of the Null Energy Condition (NEC)
and consider the quantity
\begin{equation}
\Xi = R_{\mu\nu} k^\mu k^\nu \ ,
\label{xidef}
\end{equation}
with null vector  \cite{Kashargin:2008pk}
\begin{equation}
k^\mu = \left(e^{-f/2}, e^{f/2-\nu/2}, 0, \omega e^{-f/2}\right) \ .
\label{kmudef}
\end{equation}
Taking into account the Einstein equations and the phantom field equation
we obtain
\begin{equation}
\Xi = - 2 D^2\, \frac{e^{f-\nu}}{h^2}  \ .
\label{xi}
\end{equation}
Since $\Xi$ is non-positive the NEC is violated everywhere.
In order to obtain a global scale invariant quantity as measure of the NEC violation we
define for later reference for the symmetric wormholes 
\begin{equation}
Y = \frac{1}{R_e}\int{\Xi \sqrt{-g} d\eta d\theta d\varphi} 
   =-8\pi \frac{D^2}{R_e} \int_0^\infty \frac{d\eta}{\eta^2+\eta_0^2}
   =-4\pi^2 \frac{D^2}{R_e \eta_0} \ .
\label{Yxi}
\end{equation}

\section{Symmetric Wormholes}

In the following we present the properties of symmetric rotating wormhole
solutions in General Relativity,
which are characterized by $\gamma=0$ \cite{Kleihaus:2014dla}.
We then continue with the non-symmetric case,
where for fixed $\eta_0$
the solutions depend on the parameters $\gamma$ and $\omega_{-\infty}$.

\subsection{Global Charges}

\begin{figure}[t!]
\begin{center}
\mbox{(a) \hspace*{0.48\textwidth}(b)}\\
\mbox{\hspace*{-1.0cm}
\includegraphics[height=.24\textheight, angle =0]{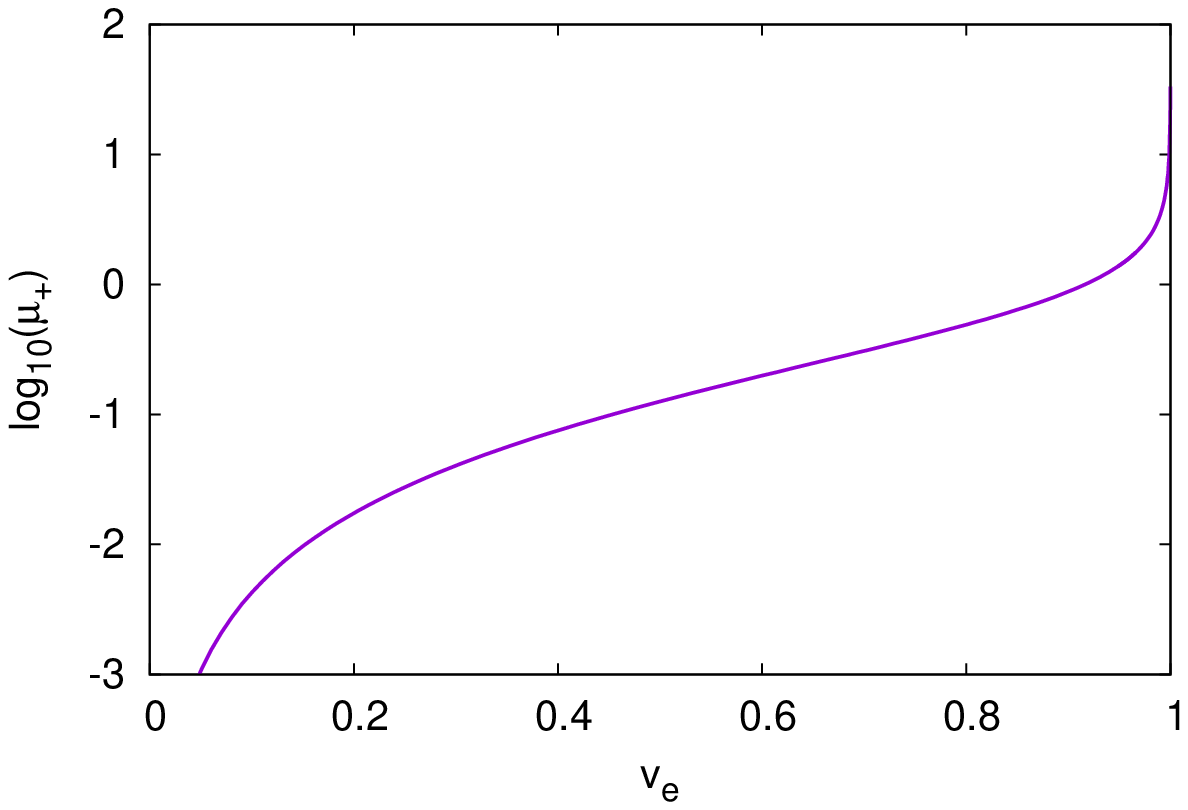}
\hspace*{-0.3cm}
\includegraphics[height=.24\textheight, angle =0]{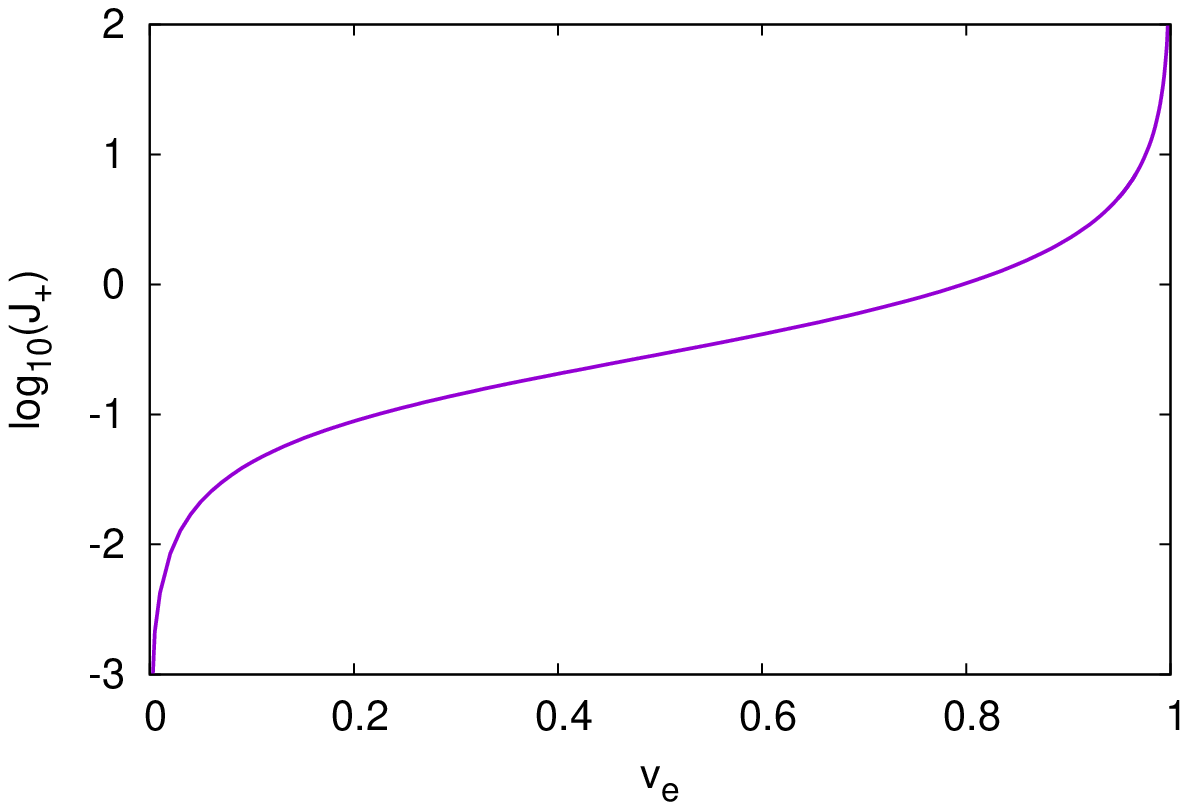}
}
\mbox{(c) \hspace*{0.48\textwidth}(d)}
\mbox{\hspace*{-1.0cm}
\includegraphics[height=.24\textheight, angle =0]{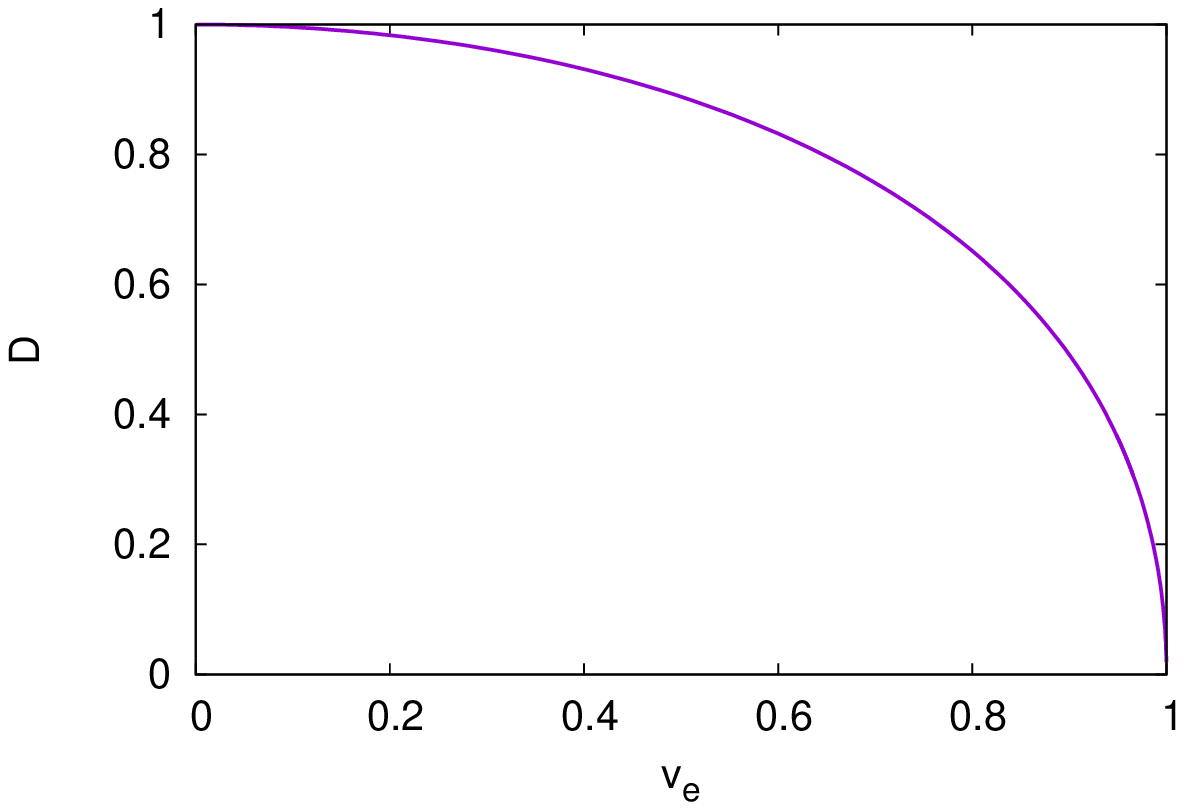}
\hspace*{-0.3cm}
\includegraphics[height=.24\textheight, angle =0]{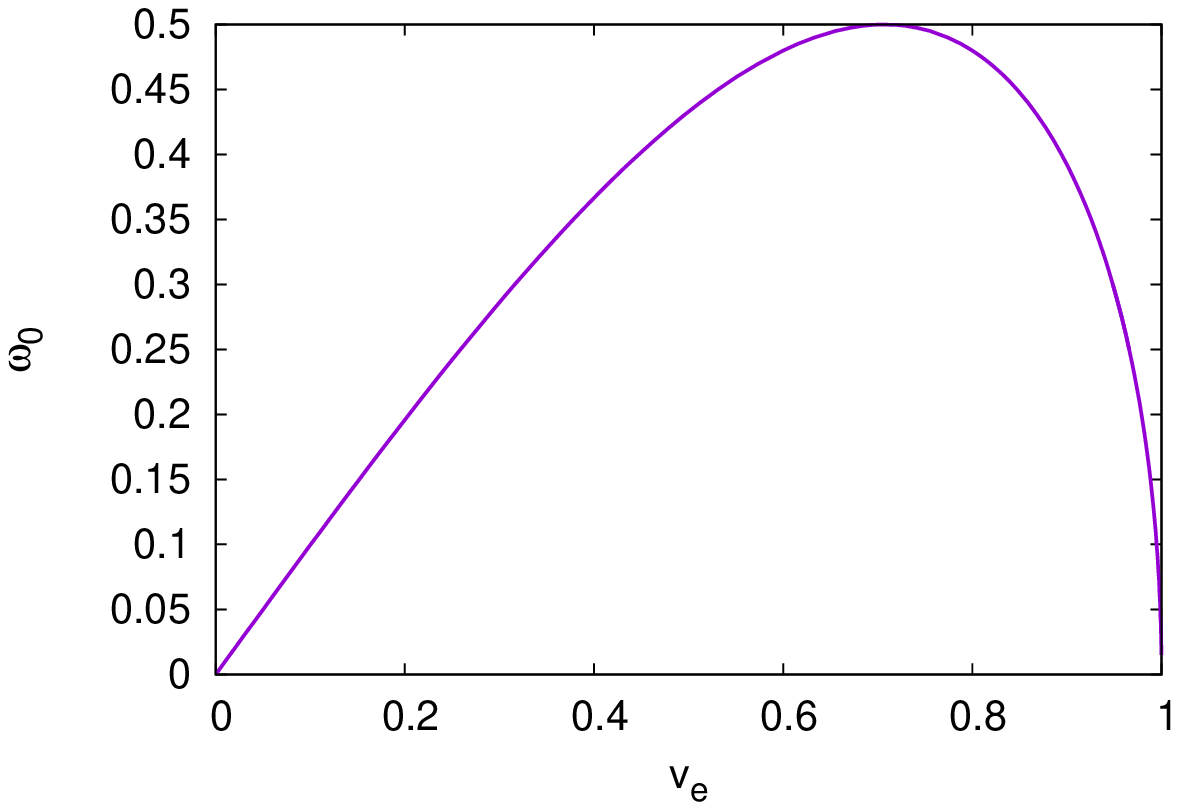}
}
\end{center}
\vspace*{-0.5cm}
\caption{
Properties of symmetric rotating wormholes at fixed throat parameter $\eta_0=1$:
the mass $\mu_+$ (a), the angular momentum $J_+$ (b),
the scalar charge $D$ (c), and the angular velocity of the throat $\omega_0$ (d)
versus the rotational velocity $v_e$ of the throat in the equatorial plane.}
\label{Fig1}
\end{figure}

Symmetric rotating Ellis wormholes depend on 
the throat parameter $\eta_0$ and the angular velocity of the throat $\omega_0$.
Let us first fix the throat parameter $\eta_0=1$
and vary $\omega_0$, starting from the static Ellis wormhole.
Then the mass $\mu_+$ and angular momentum $J_+$ increase from their Ellis value 
of zero with increasing angular velocity. At the same time, the 
rotational velocity $v_e$ of the throat in the equatorial plane increases
as it should, since it is proportional to $\omega_0$. The
centrifugal force causes in addition an increase of the equatorial circumferential
radius of the throat $R_e$ and thus of $v_e$.

We exhibit in Fig.~\ref{Fig1} the dependence of the
mass $\mu_+$ and the angular momentum $J_+$ on
the rotational velocity $v_e$ for these wormholes.
As the rotational velocity $v_e$ tends to its limiting value of one,
where the throat would be rotating with the velocity of light
the mass and the angular momentum both diverge.
Also shown in Fig.~\ref{Fig1} is the dependence 
of the scalar charge $D$ (we have chosen $D$ to be non-negative) on the rotational velocity $v_e$,
and the relation between $\omega_0$ and $v_e$.

The scalar charge is maximal for static wormholes
and decreases monotonically to zero, as the family of wormhole solutions
approaches its limiting configuration, where 
the rotational velocity reaches the speed of light.
The angular velocity of the throat, however,
does not increase monotonically with $v_e$.
Instead, it reaches a maximal value and then
decreases again towards zero.
The reason for the occurrence of two branches
with respect to $\omega_0$ is our use of `isotropic' coordinates.
The analogous pattern arises for rotating black holes,
when the isotropic horizon radius is held fixed, while
the angular momentum increases monotonically
(see e.g.~\cite{Kleihaus:2000kg}).

\begin{figure}[h!]
\begin{center}
\mbox{(a) \hspace*{0.48\textwidth}(b)}\\
\mbox{\hspace{-1.0cm}
\includegraphics[height=.24\textheight, angle =0]{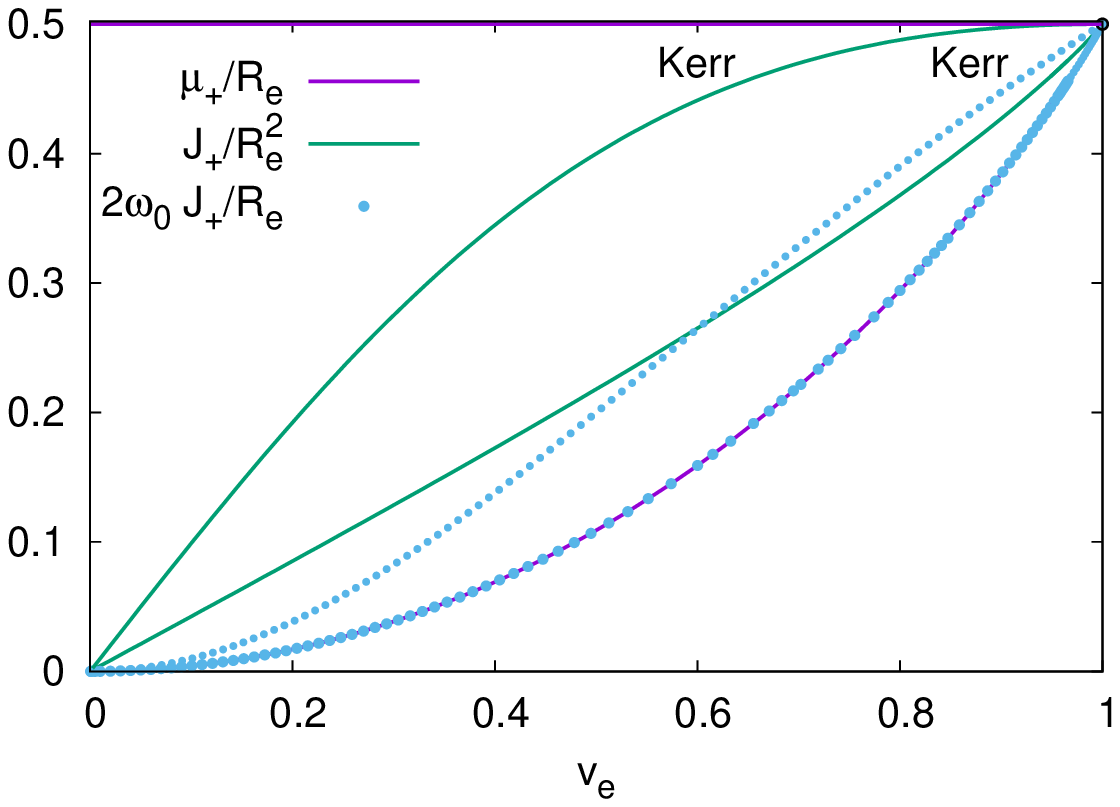}
\hspace*{-0.3cm}
\includegraphics[height=.24\textheight, angle =0]{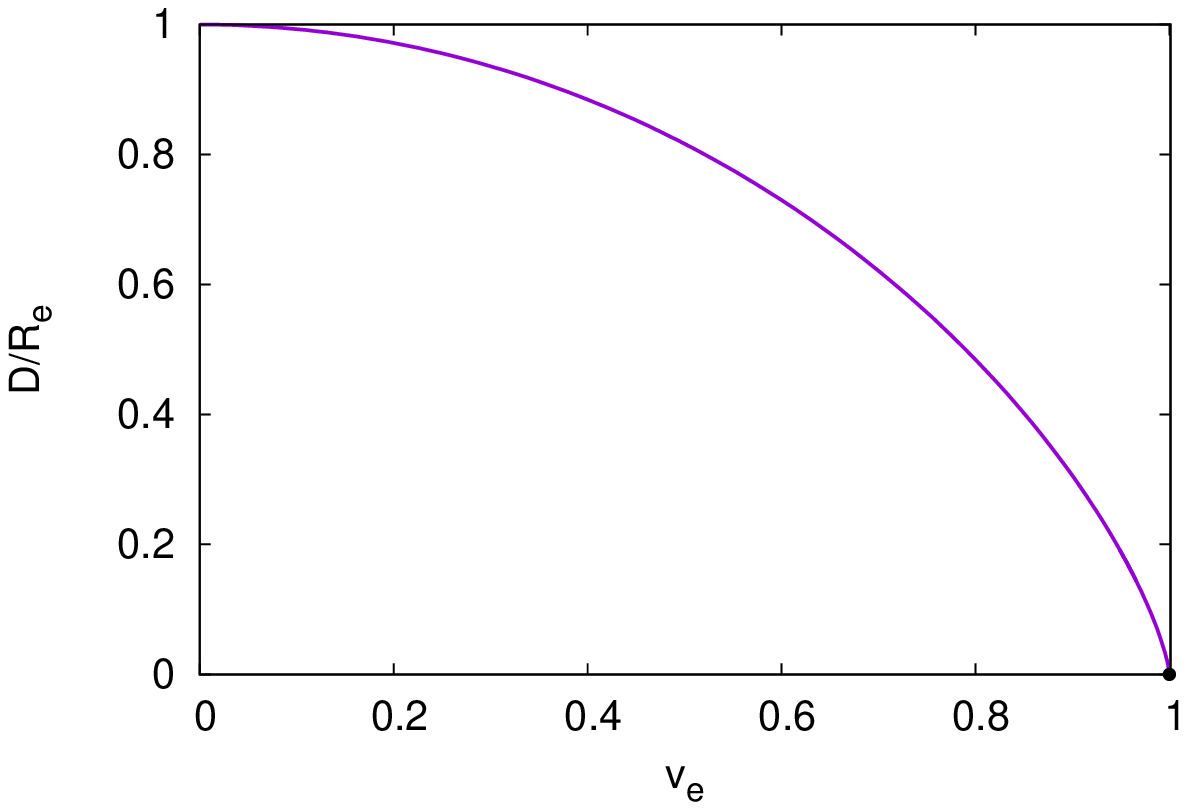}
}
\end{center}
\vspace{-0.5cm}
\caption{
Properties of symmetric rotating wormholes at fixed equatorial throat radius $R_e$:
the scaled mass $\mu_+/R_e$, the scaled angular momentum $J_+/R_e^2$
and the quantity $2 \omega_0 J_+/R_e$ (a),
the scaled scalar charge $D/R_e$  (b)
versus the rotational velocity $v_e$ of the throat in the equatorial plane.
Also shown are the respective curves for the Kerr black holes,
while the black dots mark the respective extremal Kerr values.}
\label{Fig2}
\end{figure}

To get a more physical picture of the families of symmetric rotating  wormholes
and better understand the limiting behavior for $v_e \to 1$,
let us now fix the equatorial throat radius $R_e$.
Then the family of symmetric rotating wormholes evolves from the static
Ellis wormhole with $\mu_+/R_e=J_+/R_e^2=0$ to a limiting configuration
with $\mu_+/R_e =J_+/R_e^2=1/2$.
This is seen in Fig.~\ref{Fig2}, where 
the scaled mass $\mu_+/R_e$, the scaled angular momentum $J_+/R_e^2$ and 
the scaled scalar charge $D/R_e$ 
are shown
versus the rotational velocity $v_e$ of the throat in the equatorial plane.
Again, the scalar charge decreases monotonically and tends to zero for $v_e \to 1$.
By exhibiting the r.h.s.~of Eq.~(\ref{smarr}) as well, we demonstrate, that
all solutions satisfy the Smarr-type relation (\ref{smarr}) with high accuracy.
Also shown in Fig.~\ref{Fig2} are the corresponding quantities for the 
Kerr black holes for fixed equatorial horizon radius, $\mu/R_e = 1/2$,
 $J/R_e^2 = v_e/(1+v_e^2)$ and $\omega_{\rm H} J /R_e =  v_e^2/(1+v_e^2)$.

\subsection{Limit $v_e \to 1$}

\begin{figure}[t!]
\begin{center}
\mbox{(a) \hspace*{0.48\textwidth}(b)}\\
\mbox{\hspace{-1.0cm}
\includegraphics[height=.24\textheight, angle =0]{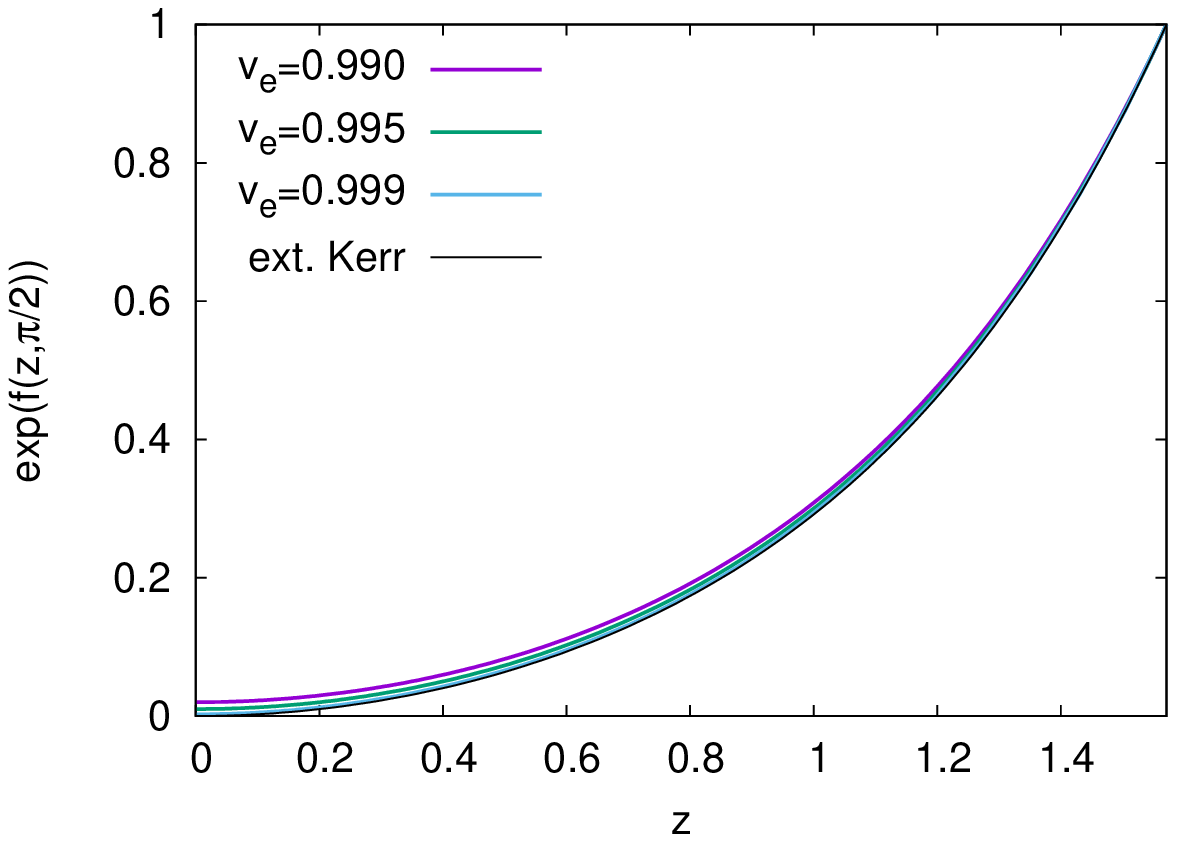}
\hspace*{-0.3cm}
\includegraphics[height=.24\textheight, angle =0]{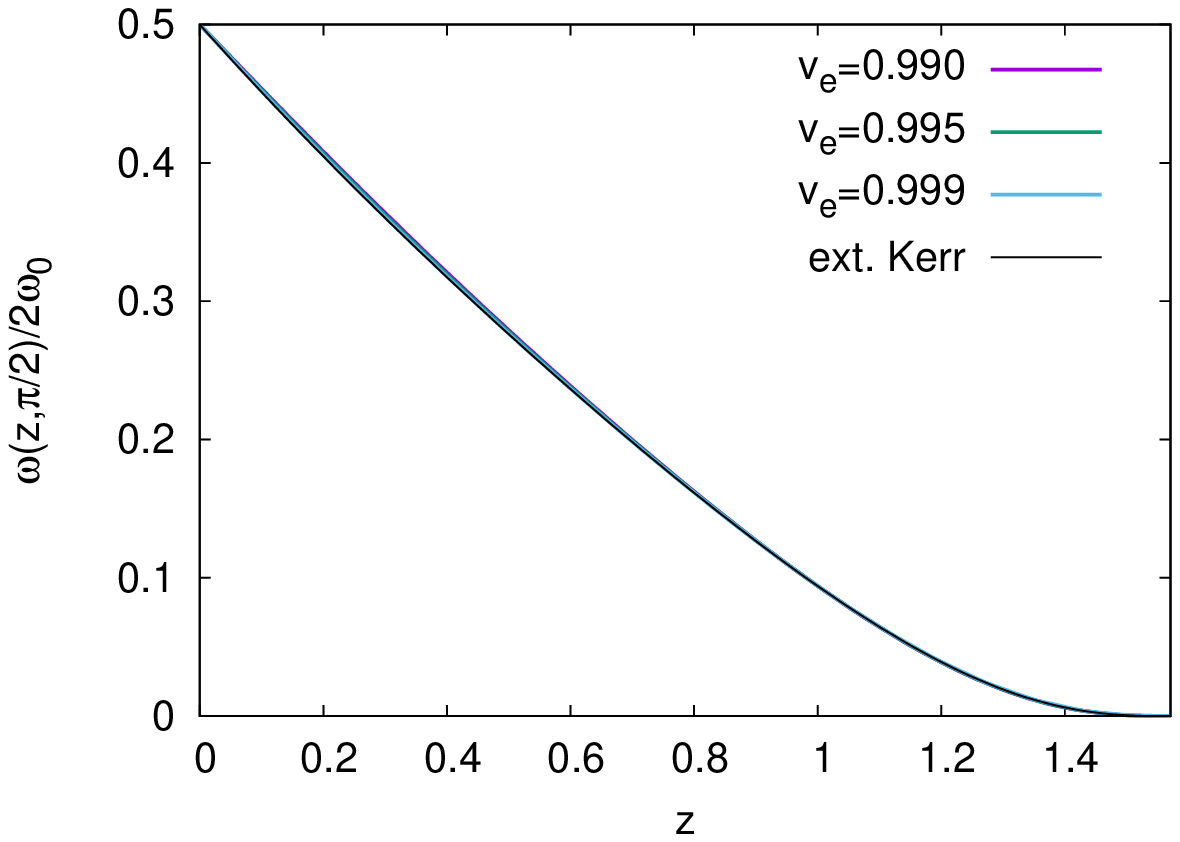}
}\\
\mbox{(c) }\\
\includegraphics[height=.24\textheight, angle =0]{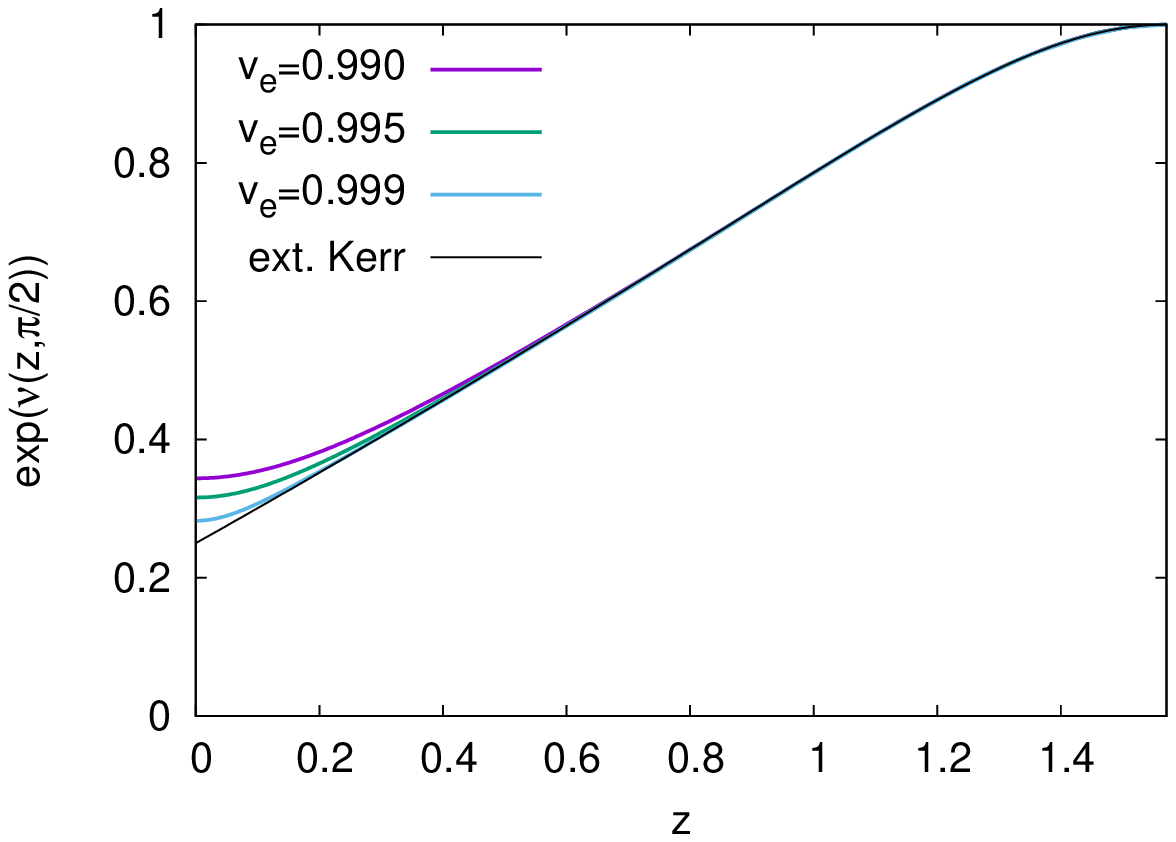}
\end{center}
\vspace{-0.5cm}
\caption{
Properties of symmetric rotating wormholes at fixed angular velocity
$\omega(0,\pi/2)=1/2$:
the limiting $v_e \to 1$ behavior of the metric functions $f$ (a), 
$\omega/(2\omega_0)$ (b) and $\nu$ (c)
versus the radial coordinate $z=\arctan(2\omega_0 \eta)$ 
in the equatorial plane $\theta=\pi/2$
for three values of  the rotational velocity of the throat in the equatorial plane,
$v_e=0.990, \, 0.995, \, 0.999$. The respective functions of the extremal
Kerr solution with horizon  angular velocity $\omega_{\rm H}=1/2$ are also shown.}
\label{Fig3}
\end{figure}

Let us now demonstrate, that in the limit $v_e \to 1$ an extremal Kerr black hole is
approached. We will consider $\eta \geq 0$ only; the discussion for $\eta \leq 0$
is analogous.
First of all, the gobal charges mass and angular momentum precisely assume
the respective Kerr values in the limit, while the scalar charge vanishes.
(Note, that a circumferential radius $R_e=2$ corresponds to a Boyer-Lindquist 
horizon radius $r_{\rm H}=1$, and an extremal black hole has $\mu=J/\mu = r_{\rm H}$.)
Thus asymptotically the extremal Kerr metric is approached.
However, in the limit $v_e \to 1$ the metric tends not only asymptotically
but everywhere to the extremal Kerr metric.
This is demonstrated in Fig.~\ref{Fig3}. 
Here we fix the angular velocity of the throat and the
horizon, $\omega_0=\omega_{\rm H} =1/2$. This requires a scaling
of the radial coordinate $\eta \to 2\omega_0 \eta/\eta_0$ and metric function
$\omega \to \omega/(2\omega_0)$, which leaves the product $\omega\eta$
and $v_e$ invariant.
Then 
the metric functions $f$, $\omega$ and $\nu$ are shown
versus the radial coordinate $z=\arctan(2\omega_0 \eta)$ for $\theta=\pi/2$
and $v_e=0.990,\, 0.995, \, 0.999$
together with the limiting extremal Kerr solution.

As the dimensionless rotational velocity $v_e$ reaches its maximal value,
$v_e=1$,
the hypersurface $\eta=0$, which corresponds to the throat
of the wormholes, must change its character.
Since the limiting solution represents an extremal black hole,
a degenerate horizon must form instead.
Moreover, the phantom field vanishes identically in this limit,
as it should for a Kerr black hole.
Consequently, the Smarr relation (\ref{smarr})
for extremal Kerr black holes is recovered,
with $\omega_0$ denoting the horizon angular velocity.

\subsection{NEC, Quadrupole Moment and Moment of Inertia}

\begin{figure}[t!]
\begin{center}
\mbox{(a) \hspace*{0.48\textwidth}(b)}\\
\mbox{\hspace{-1.0cm}
\includegraphics[height=.24\textheight, angle =0]{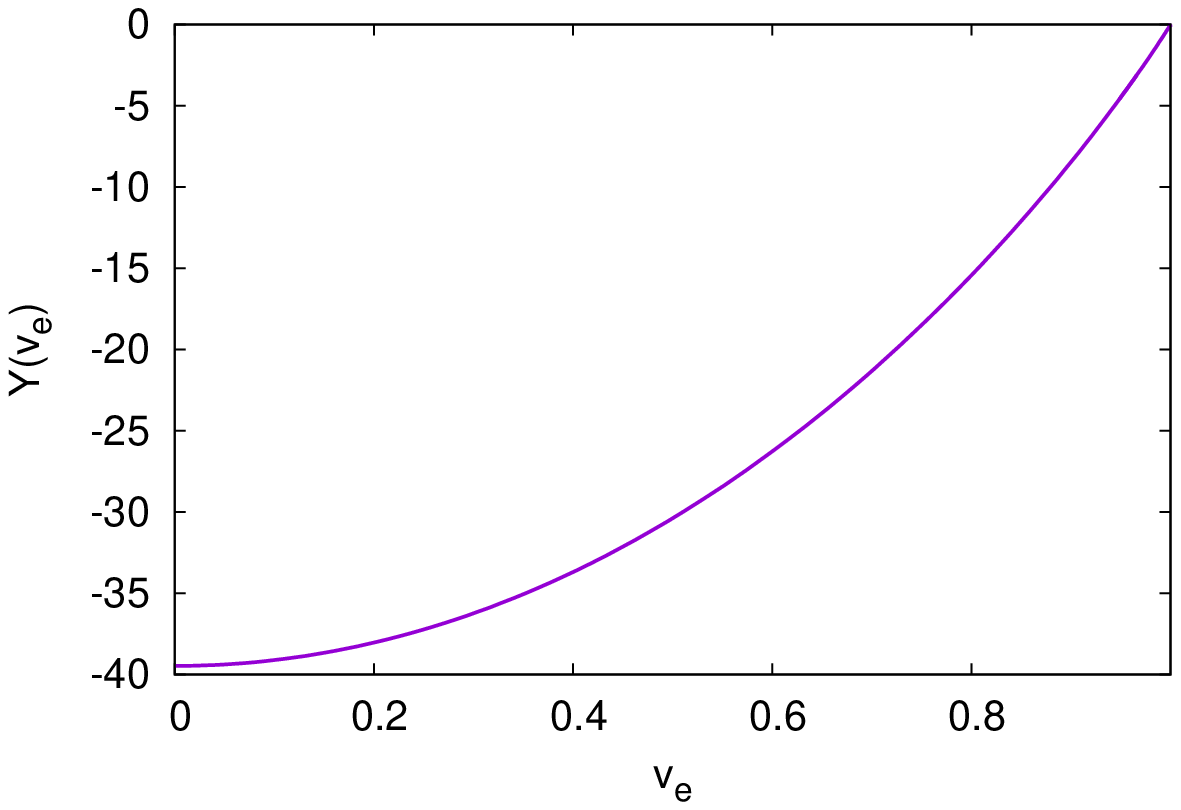}
\hspace{-0.3cm}
\includegraphics[height=.24\textheight, angle =0]{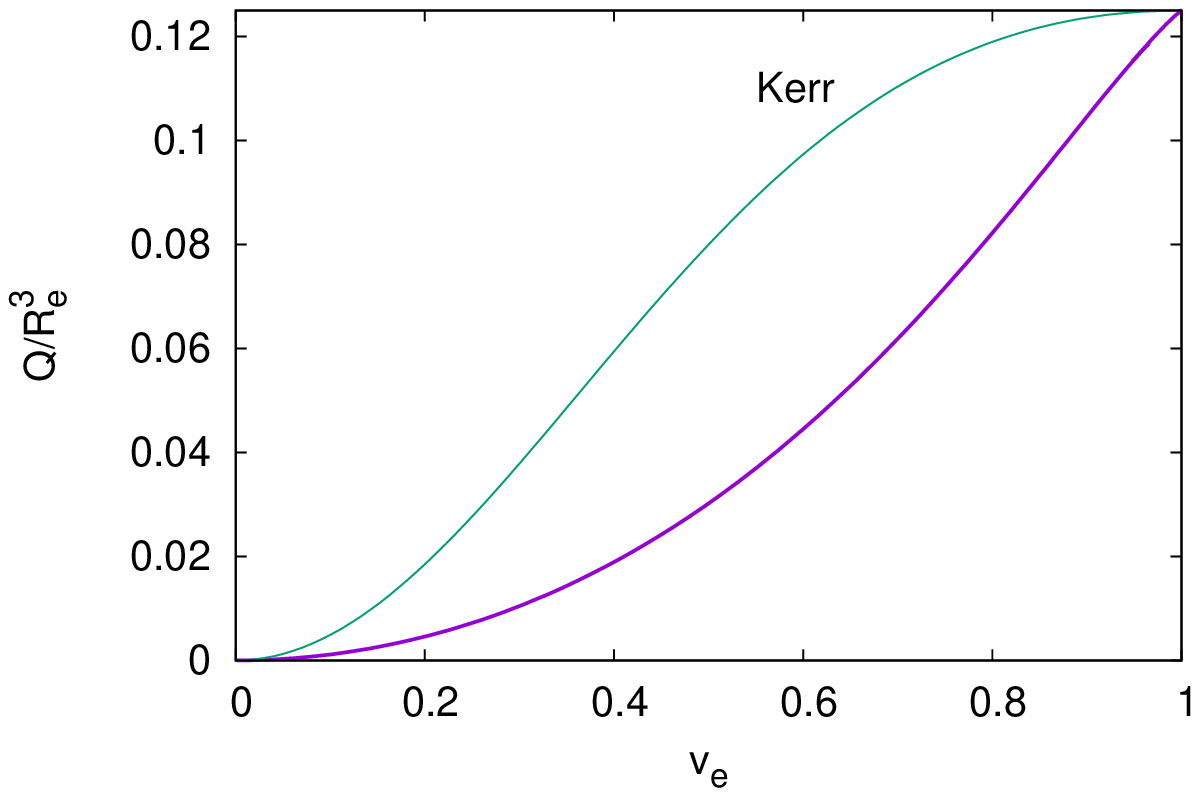}
}
\mbox{(c) \hspace*{0.48\textwidth}(d)}
\mbox{\hspace{-1.0cm}
\includegraphics[height=.24\textheight, angle =0]{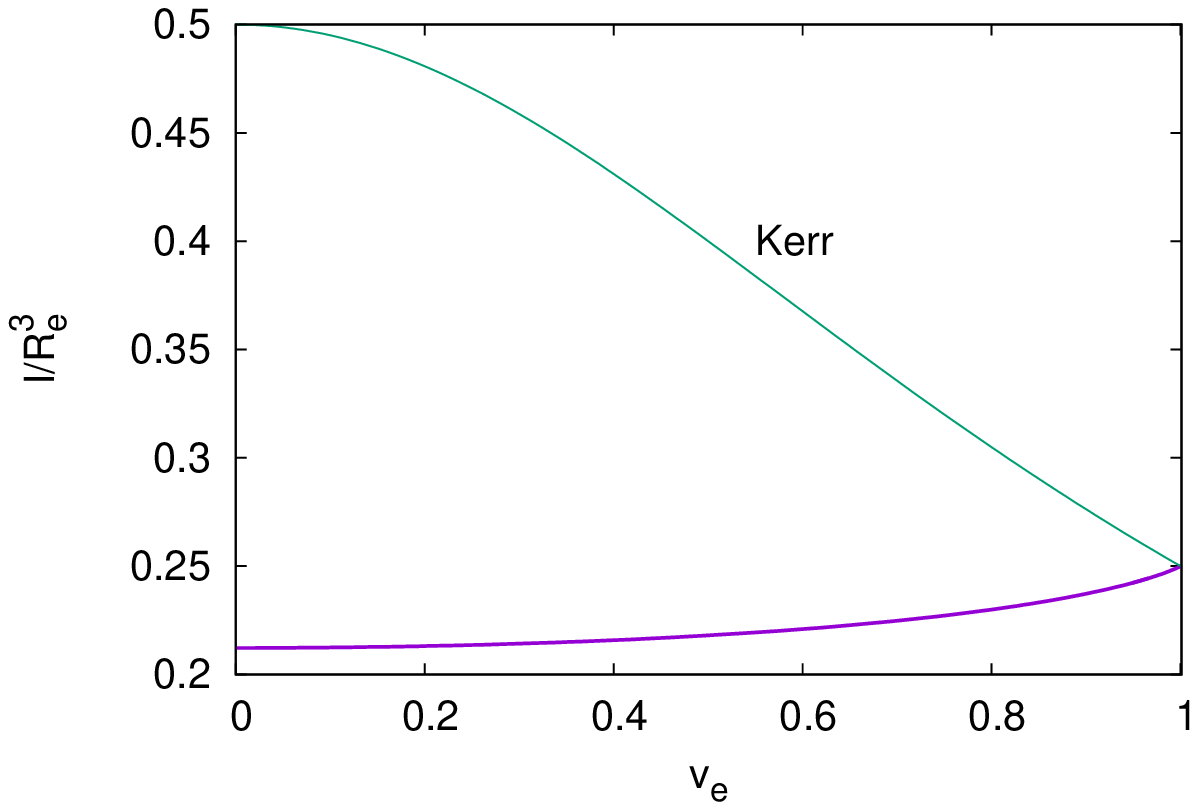}
\hspace{-0.3cm}
\includegraphics[height=.24\textheight, angle =0]{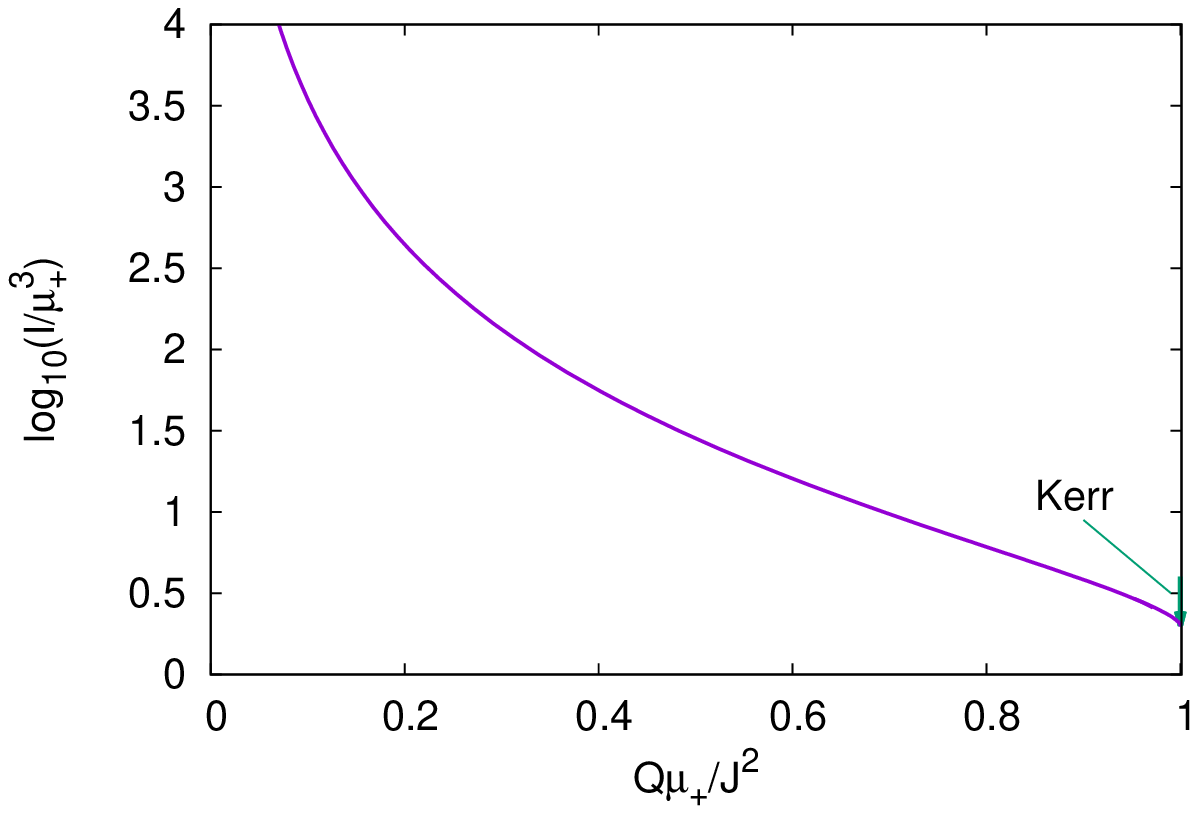}
}
\end{center}
\vspace{-0.5cm}
\caption{
Properties of symmetric rotating wormholes at fixed equatorial throat radius $R_e$:
The quantity $Y$, Eq.~(\ref{xi}),
giving a measure for the violation of the NEC (a),
the scaled quadrupole moment $Q/R_e^3$ (b), and 
the scaled moment of inertia $I/R_e^3$ (c)
versus the rotational velocity $v_e$ of the throat in the equatorial plane;
the dimensionless moment of inertia $I/\mu_+^3$
versus the dimensionless quadrupole moment $Q \mu_+/J^2$ (d).
Also shown are the respective curves for the Kerr black holes.
\label{Fig4}
}
\end{figure}

In Fig.~\ref{Fig4} we show the quantity $Y$, Eq.~(\ref{xi}),
which gives for the symmetric rotating wormhole solutions
a scale independent measure for the violation of the NEC,
as a function of the rotational velocity of the throat $v_e$. 
We observe that the violation of the NEC is strongest
in the static limit $v_e=0$
and disappears in the limit $v_e=1$,
where the extremal Kerr solution is approached.

The observation that with increasing rotation the violation of the NEC decreases and
the family of rotating wormholes ends in an extremal black hole,
appears to be a generic feature.
It was also observed for rotating Ellis wormholes in 5 dimensions,
which end in an extremal Myers-Perry black hole
\cite{Dzhunushaliev:2013jja}.
Moreover, the family of
electrically charged static Ellis wormholes in 4 dimensions
ends in an extremal Reissner-Nordstr\"om black hole
with increasing electric charge 
\cite{Hauser:2013jea}.

Let us now turn to the quadrupole moment $Q$, given in
Eq.~(\ref{quadmom}), as obtained by following the procedure
of Geroch and Hansen \cite{Geroch:1970cd,Hansen:1974zz}
by extracting it from the asymptotic expansion
in appropriate coordinates (see e.g.~\cite{Kleihaus:2014lba,Kleihaus:2016dui}).
We exhibit the quadrupole moment in Fig.~(\ref{Fig4})
for the family of
symmetric rotating wormholes at fixed equatorial throat radius $R_e$.
Starting from zero in the static case, the scaled quadrupole moment $Q/R_e^3$
increases monotonically to its maximum of $Q/R_e^3=1/8$,
when the throat velocity $v_e$ approaches the speed of light.
Also shown is the scaled quadrupole moment for the Kerr black holes,
$Q/R_e^3 = (v_e/(1+v_e^2))^2/2$. We note that the  scaled quadrupole moment
is larger for the Kerr black holes for any value of the 
horizon resp.~throat velocity $v_e$ (except for $v_e=0$ and $v_e=1$,
where both coincide).

One often considers the dimensionless quadrupole moment  $Q\mu/J^2$,
which is constant for the Kerr solutions, \mbox{$Q\mu/J^2=1$}. 
The dimensionless quadrupole moment $Q\mu_+/J_+^2$ of the 
wormhole solutions increases monotonically from zero in the static limit
to the value of the Kerr solutions in the limit $v_e \to 1$.

The moment of inertia $I=J_+/\omega_0$ of the wormholes is another quantity of interest,
in particular, in comparison with other compact objects.
We exhibit the moment of inertia $I$ for the same set of solutions
in Fig.~\ref{Fig4}.
The scaled moment of inertia $I/R_e^3$ is smaller than the corresponding 
value of the Kerr black holes, $I/R_e^3=1/(2(1+v_e^2))$, 
except in the extremal limit, when they are equal.

The dimensionless moment of inertia 
$J_+/(\omega_0 \mu_+^3)$ is also exhibited in Fig.~\ref{Fig4},
but now versus the dimensionless quadrupole moment $Q\mu_+/J_+^2$.
For comparison, the Kerr solutions are also included, where
$J/(\Omega_{\rm H} \mu^3) = 2(1+\sqrt{1-j^2})$ with $j=J/\mu^2$.
Forming the vertical line at $|Q\mu/J^2|=1$,
they range from $J/(\Omega_{\rm H} \mu^3)=2$ in the extremal
rotating case to $J/(\Omega_{\rm H} \mu^3)=4$, when the
limit $J\to 0$ is taken.
In contrast, for rotating wormholes, 
the dimensionless moment of inertia ${I/\mu_+^3}$ 
diverges in the limit $J_+\to 0$, 
since in the static limit the wormhole mass vanishes, $\mu_+ \to 0$.

\subsection{Geometry and Ergoregion}

\begin{figure}[t!]
\begin{center}
\vspace*{-1.5cm}
\mbox{\hspace{-1.0cm}
\includegraphics[height=.28\textheight, angle =0]{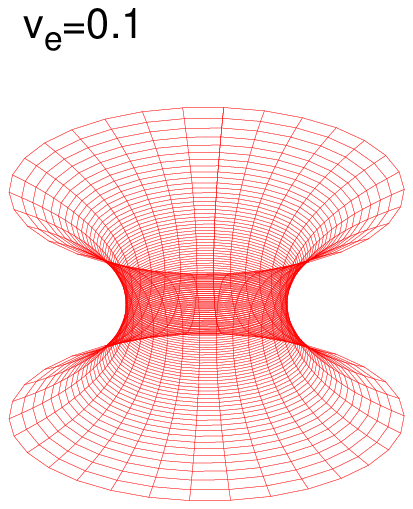}
\hspace{-2cm}
\includegraphics[height=.28\textheight, angle =0]{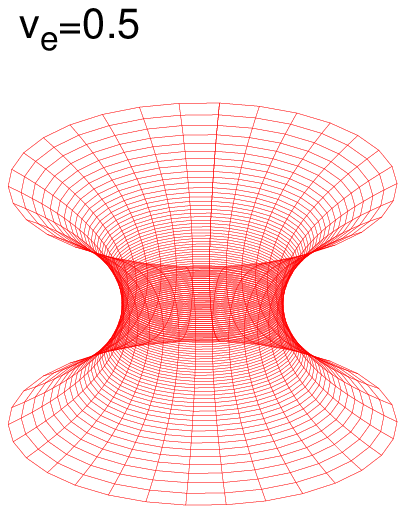}
\vspace*{-1.5cm}}
\mbox{\hspace{-1.0cm}
\includegraphics[height=.28\textheight, angle =0]{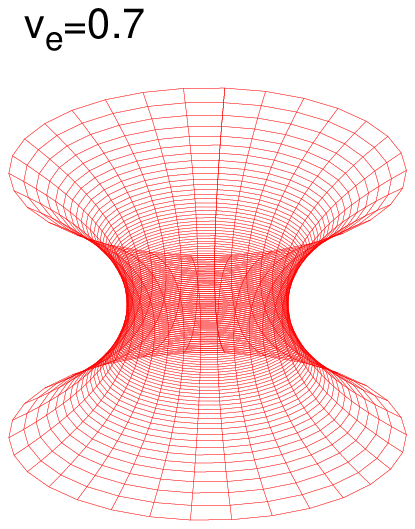}
\hspace{-2cm}
\includegraphics[height=.28\textheight, angle =0]{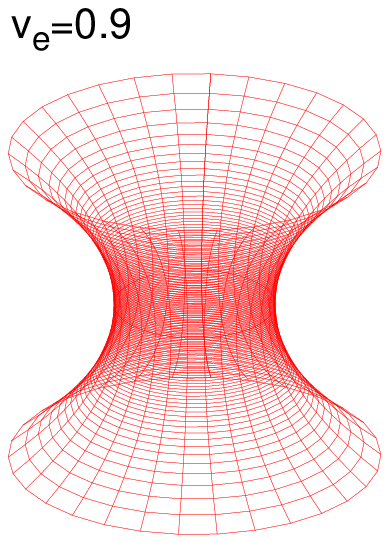}
}
\end{center}

\vspace{-1.5cm}
\caption{Properties of symmetric rotating wormholes at fixed equatorial throat radius $R_e=2$:
isometric embeddings of the equatorial plane for  
rotational velocities $v_e=0.1, \, 0.5, \, 0.7, \, {\rm and} \, 0.9$.
\label{Fig5}
}
\end{figure}

The shape of wormholes can be visualized with the help of
embedding diagrams. 
Let us consider the wormhole metric
at fixed $t$ in the equatorial plane, $\theta=\pi/2$
and embed it isometrically in Euclidean space
\begin{equation}
\label{embed}
d s^2 = e^{-f+\nu} d \eta^2 +e^{-f} h d \phi ^2
= d \rho^2 + \rho^2 d\phi^2 + d z^2  \ ,
\end{equation}
with $\rho = \rho(\eta)$, $z=z(\eta)$.
Comparison yields
\begin{equation}
\left(\frac{d\rho}{d \eta}\right)^2 + \left(\frac{dz}{d\eta}\right)^2 =e^{-f+\nu} \ , 
\quad \rho^2= e^{-f} h\ ,
\end{equation}
which leads to $z(\eta)$.
We exhibit examples of such embedding diagrams in Fig.~\ref{Fig5}
for symmetric rotating wormholes at fixed equatorial throat radius $R_e=2$
with increasing rotational velocities $v_e=0.1, \, 0.5, \, 0.7, \, {\rm and} \, 0.9$.

\begin{figure}[t!]
\begin{center}
\mbox{(a) \hspace*{0.48\textwidth}(b)}\\
\mbox{\hspace{-1.0cm}
\includegraphics[height=.25\textheight, angle =0]{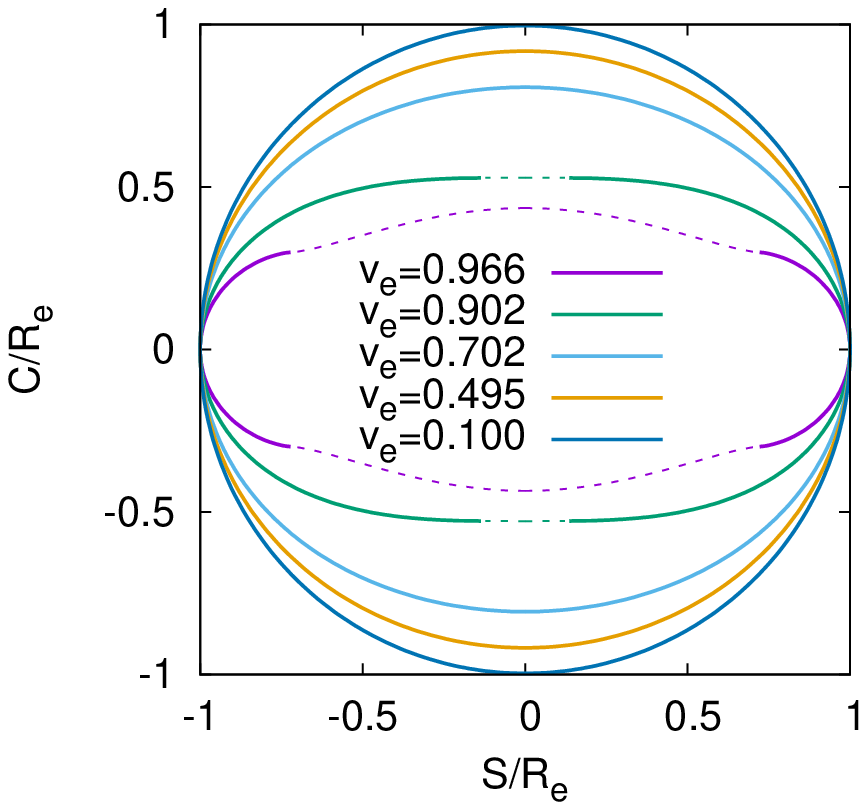} 
\hspace{-0.3cm}
\includegraphics[height=.24\textheight, angle =0]{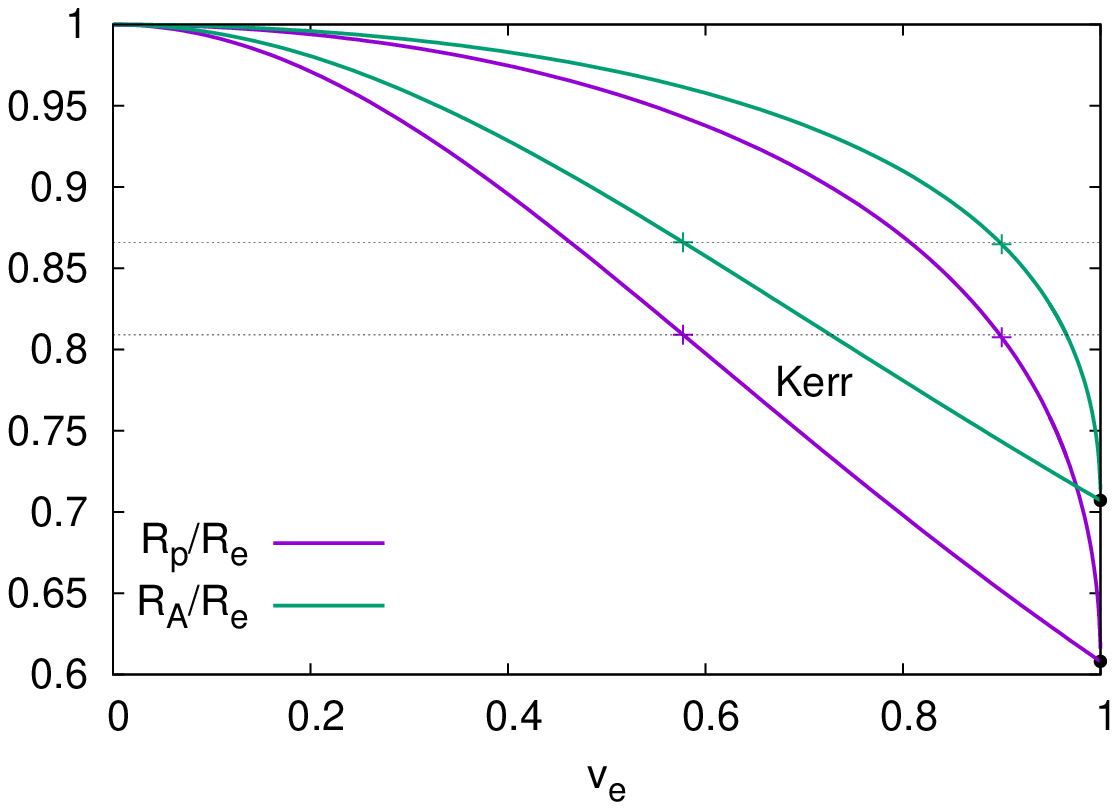}
}
\end{center}
\vspace{-0.5cm}
\caption{
Properties of symmetric rotating wormholes at fixed equatorial throat radius $R_e$:
(a) isometric embeddings of the hypersurface of the throat for  
rotational velocities $v_e=0.966, \, 0.902, \, 0.702,  \, 0.495,\, {\rm and} \, 0.1$;
the dashed curves correspond to pseudo-Euclidean embedding.
(b) The polar radius $R_p/R_e$ and the areal radius $R_A/R_e$ 
versus $v_e$.
The crosses mark $v_e^{\rm cr}$.
For comparison the respective curves for the family of Kerr solutions
are also shown.
\label{Fig6}
}
\end{figure}

We can also try to visualize the shape of the throat,
i.e., the deformation of the hypersurface $\eta=0$.
For that end we try to embed this hypersurface in
a Euclidean space. However, it is known for
Kerr black holes, that  beyond a critical value of the
rotational velocity $v_e^{\rm cr}$ their horizon can no longer be
embedded in a Euclidean space, and a pseudo-Euclidean space
must be used for the embedding \cite{Smarr:1972kt,Smarr:1973zz}.
Since the family of rotating wormhole solutions approaches an extremal
black hole, such a critical value of the rotational velocity $v_e^{\rm cr}$ 
is expected to arise for rotating wormholes as well.

To obtain the respective embedding, we equate the line element
of the hypersurface $\eta=0$, $t=0$ with the \mbox{(pseudo-)}Euclidean line element
\begin{eqnarray}
\label{embed2}
d s^2 &=& e^{-f+\nu} h d \theta^2 +e^{-f} h \sin^2 \theta d \phi ^2
= d \rho^2 + \rho^2 d\phi^2 + d z^2
\end{eqnarray}
with $\rho = S(\theta)$, $z=C(\theta)$.
Now comparison yields
\begin{equation}
\left(\frac{d S}{d \theta}\right)^2 \pm \left(\frac{dC}{d\theta}\right)^2 =e^{-f+\nu} h \ , 
\quad S^2= e^{-f} h\ .
\end{equation}
We show such embedding diagrams in Fig.~\ref{Fig6}
for the same set of wormholes with rotational velocities
$v_e=0.5, \, 0.7, \, 0.9, \, {\rm and} \, 0.95$.
The critical value corresponds to $v_e^{\rm cr}=0.9$
and marks the onset of a negative Gaussian curvature at the poles.
Beyond $v_e^{\rm cr}$ parts of the hypersurface must be embedded
in pseudo-Euclidean space, which is represented by the dashed curves
in the figure.

We complete this demonstration of the throat geometry
by considering the dependence of the throat radii on the 
rotational velocity $v_e$. 
We show in Fig.~\ref{Fig6}
the ratio of the polar radius to the equatorial radius $R_p/R_e$
for fixed equatorial throat radius $R_e=2$.
As one would expect, this ratio decreases monotically with increasing rotation,
and reaches the corresponding ratio of radii of the extremal Kerr horizon 
in the limit $v_e \to 1$.
The critical value $v_e^{\rm cr}$, where the Gaussian curvature
at the poles turns negative, is indicated by a cross in the figure.
For comparison, also the respective curve for the Kerr solutions is shown.
Interestingly, while the critical velocity $v_e^{\rm cr}=0.577$
is much smaller for the Kerr solution,
the ratio $R_p/R_e$ has the same value for the
wormhole and the Kerr critical velocities,
as indicated in the figure by a thin horizontal line.
The figure also exhibits the ratio of the areal radius to the equatorial radius $R_A/R_e$,
where the analogous findings hold.

\begin{figure}[h!]
\begin{center}
\mbox{(a) \hspace*{0.48\textwidth}(b)}\\
\mbox{\hspace{-0.8cm}
\includegraphics[height=.24\textheight, angle =0]{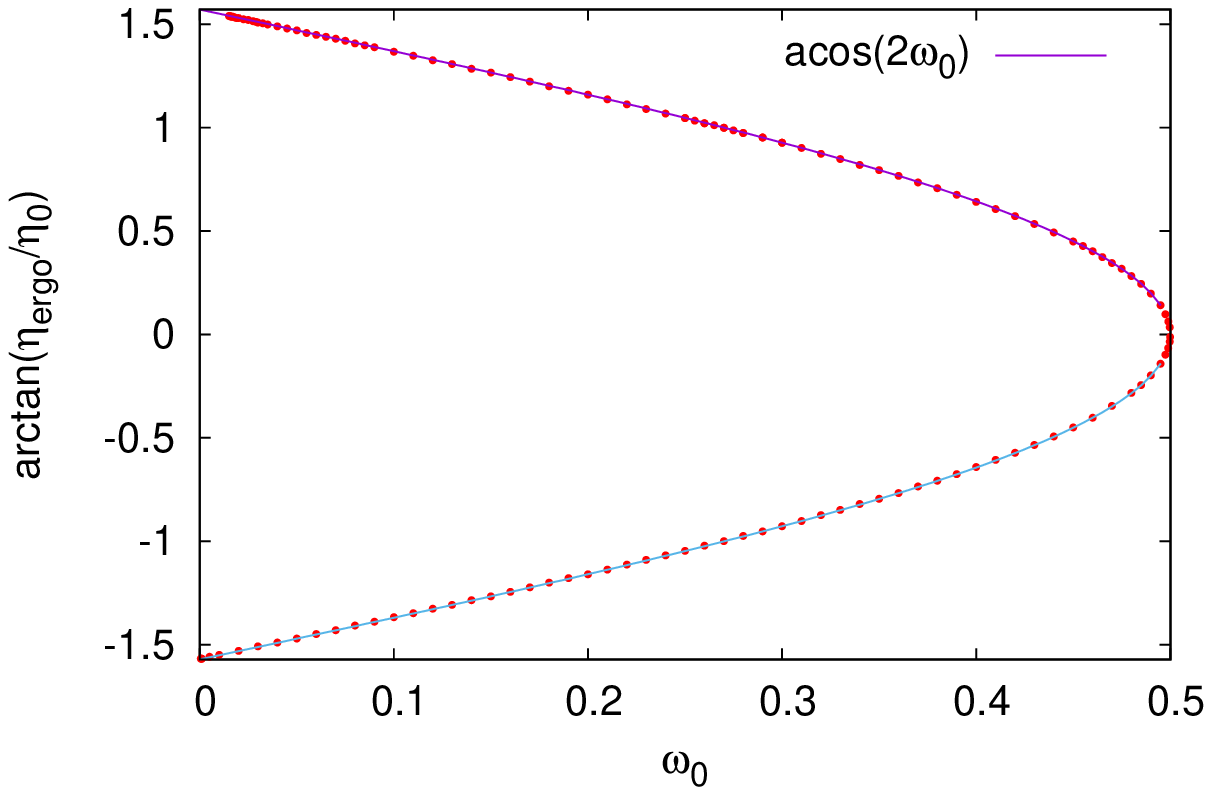}
\includegraphics[height=.24\textheight, angle =0]{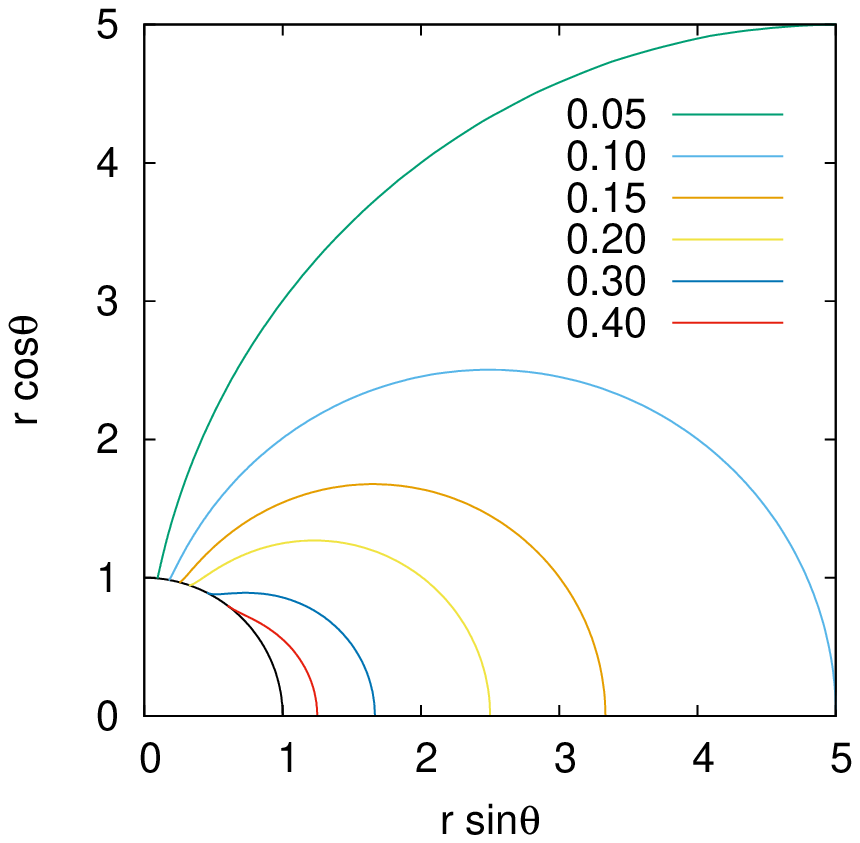}
}
\end{center}
\vspace{-0.5cm}
\caption{
\label{Fig7}
Properties of symmetric rotating wormholes at fixed throat parameter $\eta_0=1$:
(a) the location of the ergosurface $\eta_{\rm ergo}$ in the
equatorial plane
is shown versus the
angular velocity of the throat $\omega_0$
and compared to the function $\arccos(2\omega_0)$;
(b) one quadrant of the ergosurface and the throat 
in coordinates $r=\sqrt{\eta^2+\eta_0^2}$ and $\theta$
for fixed $\phi$ and several values of $\omega_0$.
}
\end{figure}

Let us finally address the ergoregion of rotating wormholes.
The ergoregion is defined by the condition $g_{tt} \ge 0$,
\begin{equation}
g_{tt}= -e^{f}  +e^{-f} h \sin^2\theta \omega^2 \ge 0 \ ,
\label{ergoreg}
\end{equation}
where the equality defines the ergosurface.
Inspection of the rotating wormhole solutions reveals, that
the location of the ergosurface $\eta_{\rm ergo}$ in the
equatorial plane is closely
related to the angular velocity $\omega_0$ of the throat.
Indeed, the relation
$\arctan(\eta_{\rm ergo}/\eta_0)= \arccos(2 \omega_0)$
holds with high accuracy.
This is demonstrated in Fig~\ref{Fig7}.

For wormholes to possess an ergoregion in that part of the spacetime,
where $\eta \ge 0$, which we have chosen to be asymptotically flat,
they must rotate sufficiently fast. 
Note that the condition for the ergosphere at the throat in the equatorial plane
can be written as 
\begin{equation}
g_{tt}= -e^{f(0,\pi/2)}  +e^{-f(0,\pi/2)} \eta_0^2 \omega_0^2 =0 
\ \ \ \Longleftrightarrow \ \ \
-\frac{\eta_0^2}{R_e^2} (1-R_e v_e)(1+R_e v_e) = 0 \ ,
\label{ergoregeqp}
\end{equation}
which yields $R_e=2$ for $v_e=1/2$.
Thus the throat must rotate with half the velocity of light.
As the rotation velocity $v_e$ increases beyond this value,
the ergoregion in this part of the spacetime increases.
This is demonstrated in Fig.~\ref{Fig7}, where the ergosurfaces
for increasing values of $v_e$ and thus decreasing values of $\omega_0$
are exhibited.
However, there is always an ergoregion
in the other part of the spacetime, where $\eta \le 0$, 
and the function $\omega$ 
tends asymptotically to a finite value.

\subsection{Geodesics}

\begin{figure}[t!]
\begin{center}
\mbox{(a) \hspace*{0.48\textwidth}(b)}\\
\mbox{\hspace{-1.0cm}
\includegraphics[height=.24\textheight, angle =0]{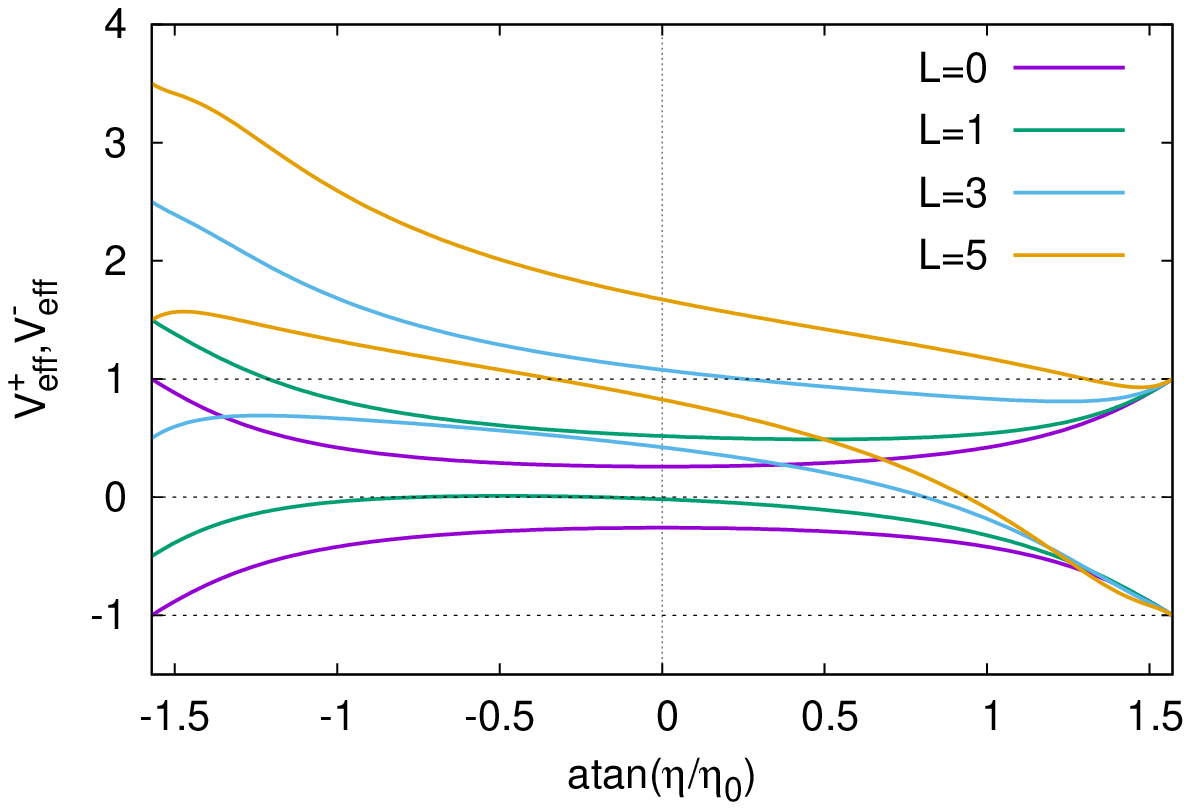}
\hspace{-0.3cm}
\includegraphics[height=.24\textheight, angle =0]{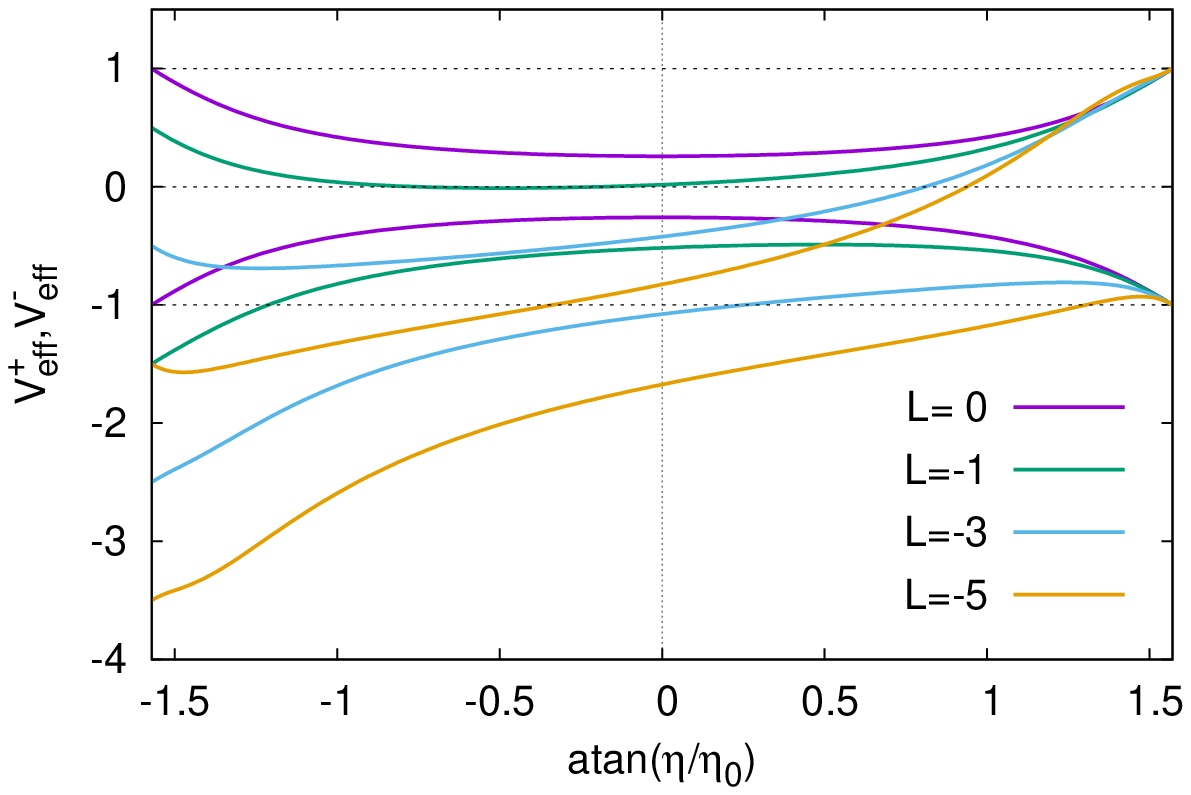}
\vspace{-1.0cm}}\\
\mbox{(c) }\\
\includegraphics[height=.24\textheight, angle =0]{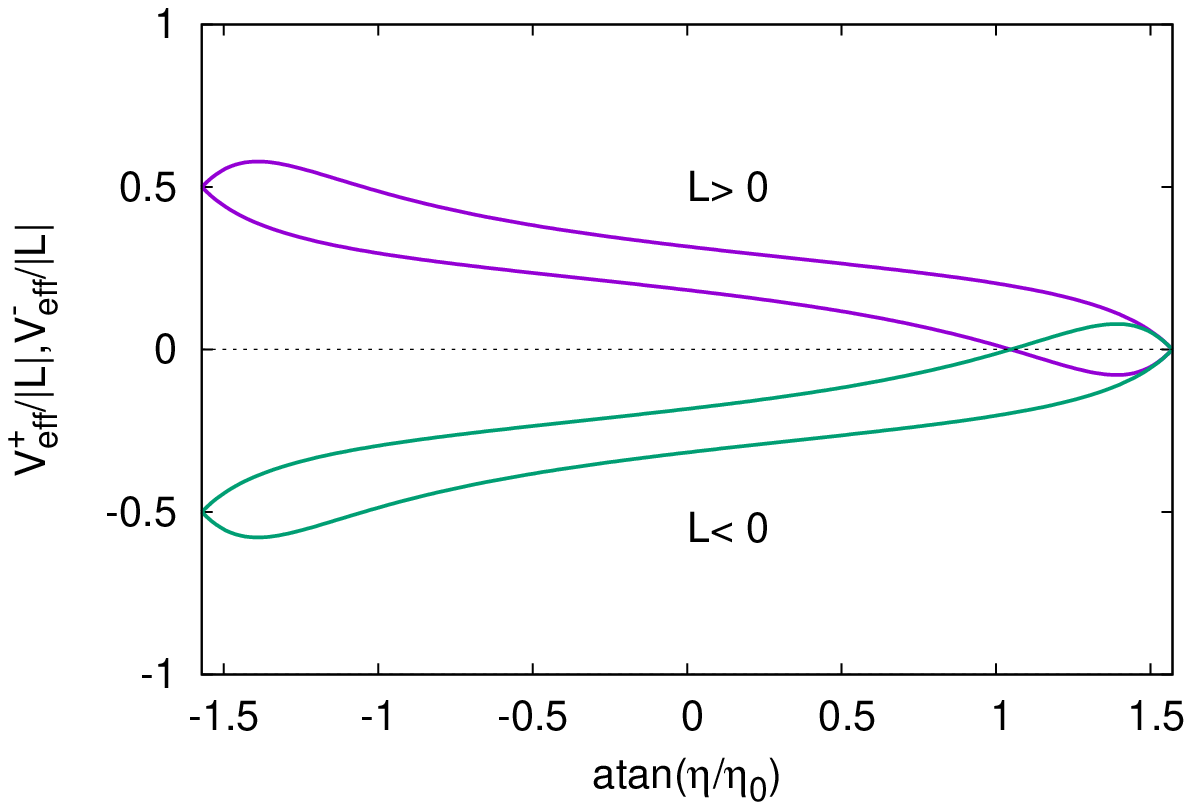}
\end{center}
\vspace{-0.0cm}
\caption{
\label{Fig8}
Geodesics of symmetric rotating wormholes at fixed equatorial throat radius $R_e=2$:
effective potentials $V_{\rm eff}^\pm$ for wormholes 
with rotational velocity {$v_e=0.78$} versus the compactified radial coordinate
$\arctan(\eta/\eta_0)$
for prograde motion (a) and retrograde motion (b)
for several values of the
angular momentum $L$ of a massive particle.
(c) $V_{\rm eff}^\pm/|L|$ for massless particles for prograde and retrograde motion.
}
\end{figure}

We now analyze the motion of particles and light in these rotating
wormhole spacetimes.
Interesting aspects emerging from this analysis are
that massive particles can have stable bound states 
in rotating wormhole spacetimes.
Moreover, there are unstable bound states for massless particles 
which indicates the presence of a photon region.

The motion of particles and light is governed by the Lagrangian ${\cal L}$
\begin{equation}
2 {\cal L} = g_{\mu\nu} \dot{x}^\mu\dot{x}^\nu   = \varepsilon \ ,
\label{lagmot}
\end{equation}
where the dot denotes the derivative with respect to an affine parameter,
and $\varepsilon =-1$ for particles and zero for light.
The symmetries of the spacetime lead to two cyclic coordinates
and thus two constants of motion,
the energy $E$  and the angular momentum $L$.

Let us for simplicity consider only motion in the equatorial plane, $\theta=\pi/2$.
The Lagrangian then simplifies to
\begin{equation}
 2 {\cal L}= -e^{f} \dot t^2 +e^{-f} 
\left( e^{\nu} \dot \eta^2
                    + h \left(\dot \phi -\omega \dot t\right)^2\right) = \varepsilon\ ,
\label{lineelpi2}
\end{equation}
and the conserved charges become
\begin{equation}
E= \dot t \left(e^f - e^{-f} h \omega^2\right)+ e^{-f} h \omega \dot \phi \ , \ \ \
L= - e^{-f} h \omega \dot t + e^{-f} h \dot \phi \ .
\label{EL}
\end{equation}
Solving these equations for $\dot t$ and $\dot \phi$ yields
\begin{equation}
\dot t = e^{-f}\left(E-\omega L\right) \ , \ \ \
\dot \phi = e^{-f}\left(\omega (E-\omega L)+e^{2f}\frac{L}{h} \right) \ .
\label{dottdotphi}
\end{equation}
Insertion of these expressions into the Lagrangian leads to
\begin{equation}
 e^{-f} e^\nu \dot \eta^2 
-e^{-f} \left((E-\omega L)^2-e^f\frac{L^2}{h}\right) = \varepsilon \ .
\end{equation}

When solving this equation for $\dot \eta^2$, it is instructive to introduce 
effective potentials $V_{\rm eff}^\pm$ as follows
\begin{equation}
\dot \eta^2 =  e^{-\nu }
\left(E-V_{\rm eff}^+(L,\eta)\right)\left(E-V_{\rm eff}^-(L,\eta)\right) = \Xi(\eta) 
 \ ,
\label{veffs}
\end{equation}
where we abbreviated the right hand side by $\Xi(\eta)$, and
\begin{equation}
V_{\rm eff}^\pm   =  \omega L \pm \sqrt{ e^{2f} \frac{L^2}{h} -\varepsilon e^f}
\label{veffpm}
\end{equation}
implying $V_{\rm eff}^+(L,\eta) \geq V_{\rm eff}^-(L,\eta)$.
In fact the condition
$\dot{\eta}^2 = \Xi(\eta) \geq 0$
leads to a restriction for the allowed energies and angular
momenta of the particles, 
$E\geq V_{\rm eff}^+(L,\eta)$ or $E\leq V_{\rm eff}^-(L,\eta)$.

\begin{figure}[t!]
\begin{center}
\mbox{\hspace{-1.0cm}
\includegraphics[height=.28\textheight, angle =0]{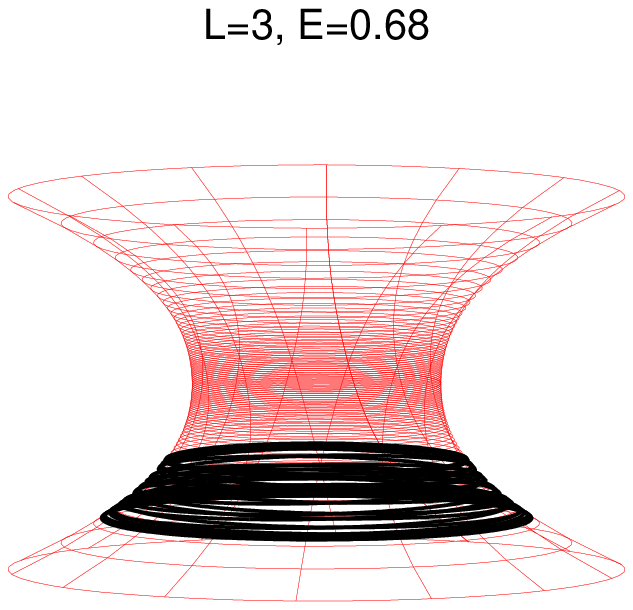}
\hspace{-0.3cm}
\includegraphics[height=.28\textheight, angle =0]{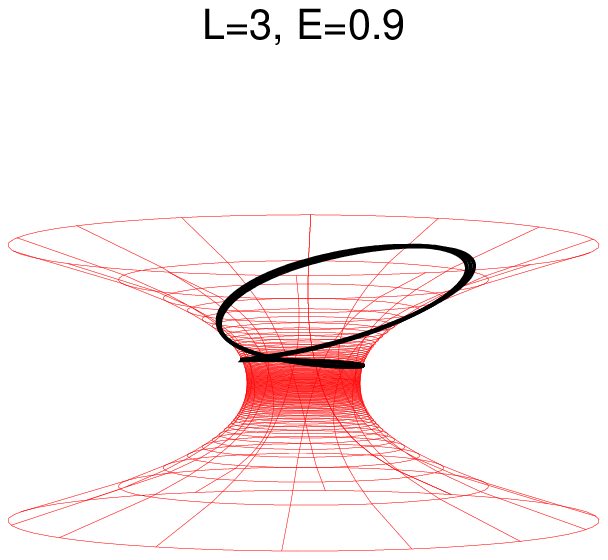}
\vspace{-1.0cm}}
\includegraphics[height=.28\textheight, angle =0]{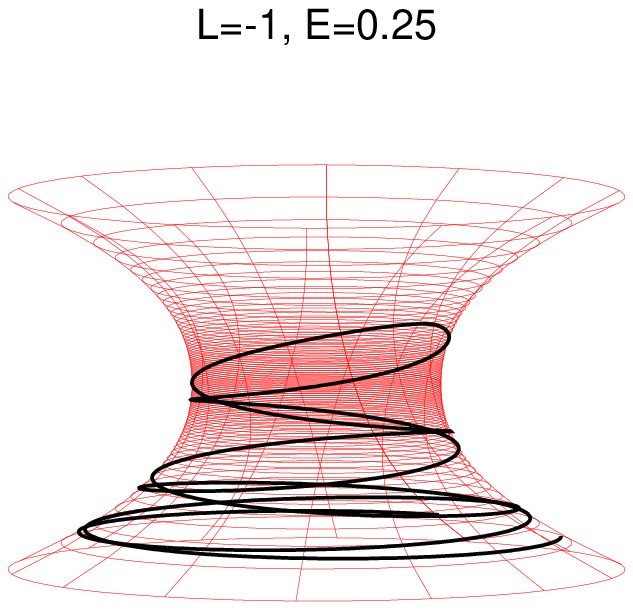}
\end{center}
\vspace{-1.5cm}
\caption{
\label{Fig9}
Geodesics of symmetric rotating wormholes at fixed equatorial throat radius $R_e$:
embedding diagrams of
bound orbits of a massive test particle in the equatorial plane of a symmetric fast rotating
wormhole ($v_e=0.78$) 
for several values of the particle energy $E$ and angular momentum $L$.
}
\end{figure}

Thus the study of these effective potentials allows for a classification of the possible orbits
in the equatorial plane.
We exhibit in Fig.~\ref{Fig8} examples of the effective potentials
for massive and massless particles.
In particular, we observe that there are bound orbits of massive 
particles in such rotating wormhole spacetimes.
This is in contrast to the static Ellis wormhole, which does not
possess bound orbits.
For rotating wormholes there are even three types of bound orbits,
those that always remain within a single universe, those that
oscillate between the two universes, and those that remain at the 
throat ( $L=0$ only).
Two examples of such single universe
bound orbits for a massive particle are shown in Fig.~\ref{Fig9}
together with a two-world bound orbit.

For the case of massless particles the effective potentials
shown in Fig.~\ref{Fig8} are
divided by the modulus of the angular momentum. We note that
$V_{\rm eff}^+$ possesses a local maximum for retrograde motion
for $\eta>0$. This indicates that there is a photon region 
in this asymptotically flat universe.
$V_{\rm eff}^+$ additionally possesses a local maximum for prograde
motion for $\eta<0$, which indicates the presence of a photon region
in the other universe, as well.
However,
whereas the motion there looks like prograde motion in our coordinates,
where the $\eta>0$ universe is asymptotically flat, it corresponds
to retrograde motion, when an appropriate coordinate transformation
is performed, making the $\eta <0$ universe asymptotically flat.

\begin{figure}[h!]
\begin{center}
\mbox{(a) \hspace*{0.48\textwidth}(b)}\\
\mbox{\hspace{-1.0cm}
\includegraphics[height=.24\textheight, angle =0]{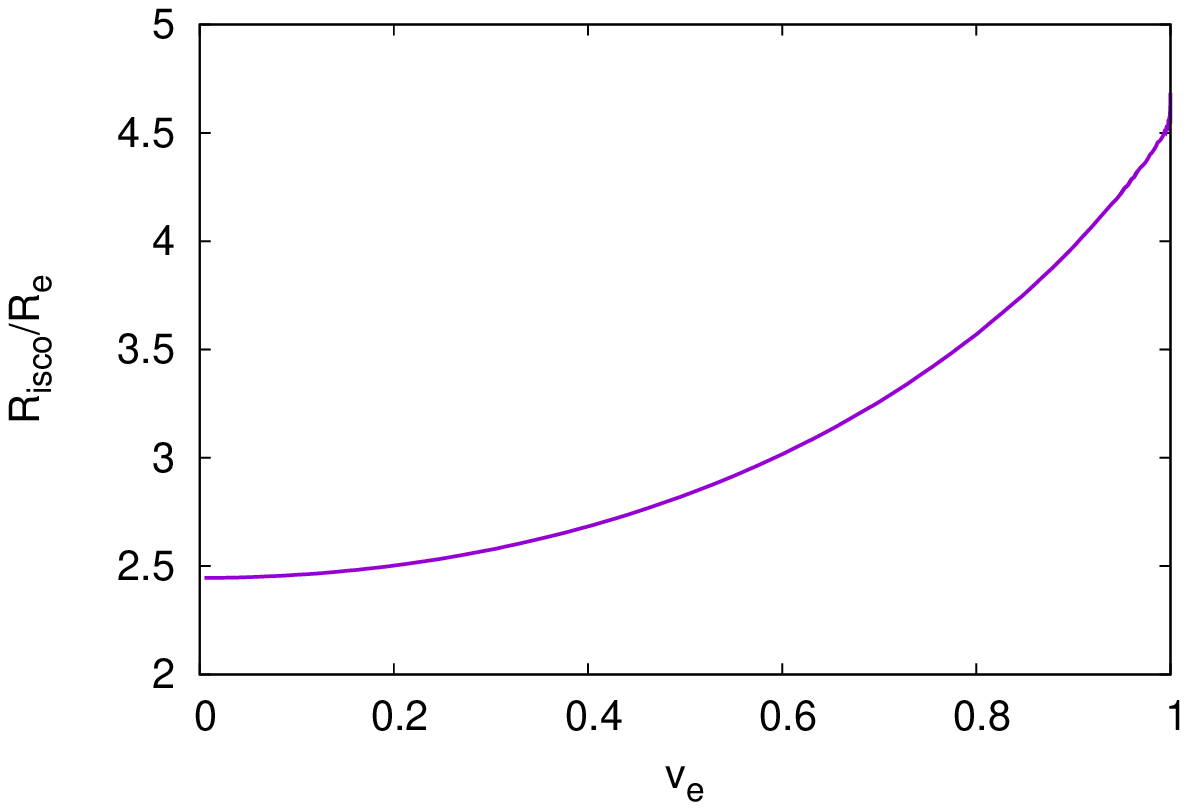}
\hspace{-0.3cm}
\includegraphics[height=.24\textheight, angle =0]{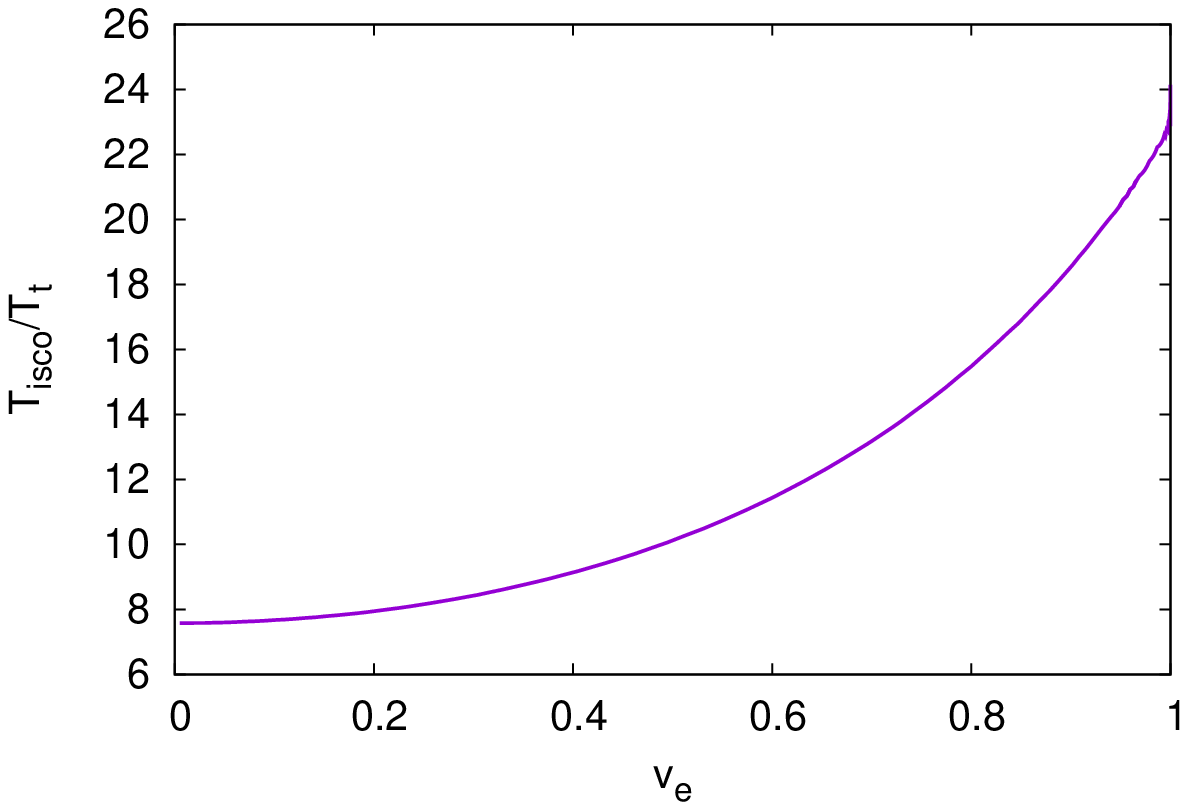}
}
\end{center}
\vspace{-0.5cm}
\caption{
\label{Fig10}
Geodesics of symmetric rotating wormholes at fixed equatorial throat radius $R_e$:
the ratio $R_{\rm ISCO}/R_e$ of the ISCO radius to the throat radius (a)
and the corresponding ratio of the orbital period $T_{\rm ISCO}/T_e$ (b)
for retrograde motion of massive particles in the equatorial plane
versus the rotational velocity $v_e$ of the throat in the equatorial plane.
}
\end{figure}

Of particular interest are the
innermost stable circular orbits (ISCOs).
To find the circular orbits we need to solve for 
$\Xi(\eta_{_{\rm CO}})=\Xi'(\eta_{_{\rm CO}})=0$, while stability requires at the same time
$\Xi''(\eta_{_{\rm CO}}) < 0$. ISCOs are then characterized by the change of stability,
i.~e., $\Xi''(\eta_{_{\rm ISCO}}) =0$.
For these rotating wormholes there are both corotating and counterrotating ISCOs,
just like for the Kerr black holes.
Whereas the ISCO of the corotating orbits resides at the throat,
the location of the ISCO of the counterrotating orbits depends on the rotation velocity.
Here the ratio $R_{\rm ISCO}/R_e$ increases almost by a factor of two
from the static limit to the limiting extremal Kerr black hole,
as seen in Fig.~\ref{Fig10}. The figure also displays
the ratio of the orbital period $T_{\rm ISCO}/T_e$,
where $T_e$ denotes the orbital period of the throat (or the horizon of the
extremal black hole), which increases roughly
by a factor of three from the static limit to the extremal Kerr black hole.

\section{Non-Symmetric Wormholes}

We now turn to the discussion of the properties of non-symmetric rotating 
Ellis wormholes.
In the non-symmetric case for fixed $\eta_0$
the wormhole solutions depend on two parameters, $\gamma$ and $\omega_{-\infty}$.
By varying these, the domain of existence can be mapped out.
In the following we will first discuss the global charges of these
non-symmetric rotating wormholes, 
and subsequently we will address their geometry.

\subsection{Global Properties}

Let us start our discussion of the
properties of non-symmetric rotating wormholes with their global properties
mass $\mu_+$, angular momentum $J_+$ and scalar charge $D$, obtained
from the asymptotic expansion at $\eta \to \infty$.
To demonstrate these properties, we fix the throat parameter $\eta_0=1$
and consider families of solutions, where the remaining two parameters
$\gamma$ and $\omega_{-\infty}$ are varied.

\begin{figure}[t!]
\begin{center}
\mbox{(a) \hspace*{0.48\textwidth}(b)}\\
\mbox{
\includegraphics[height=.23\textheight, angle =0]{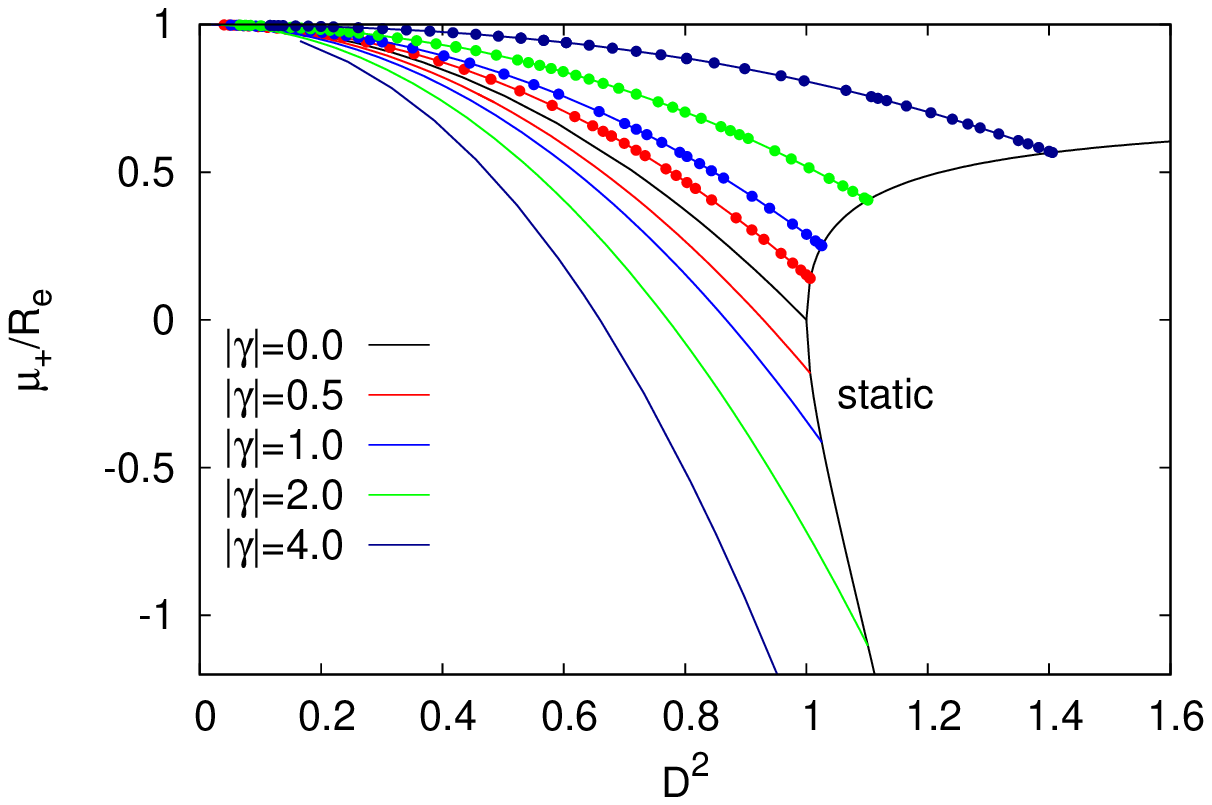}
\includegraphics[height=.23\textheight, angle =0]{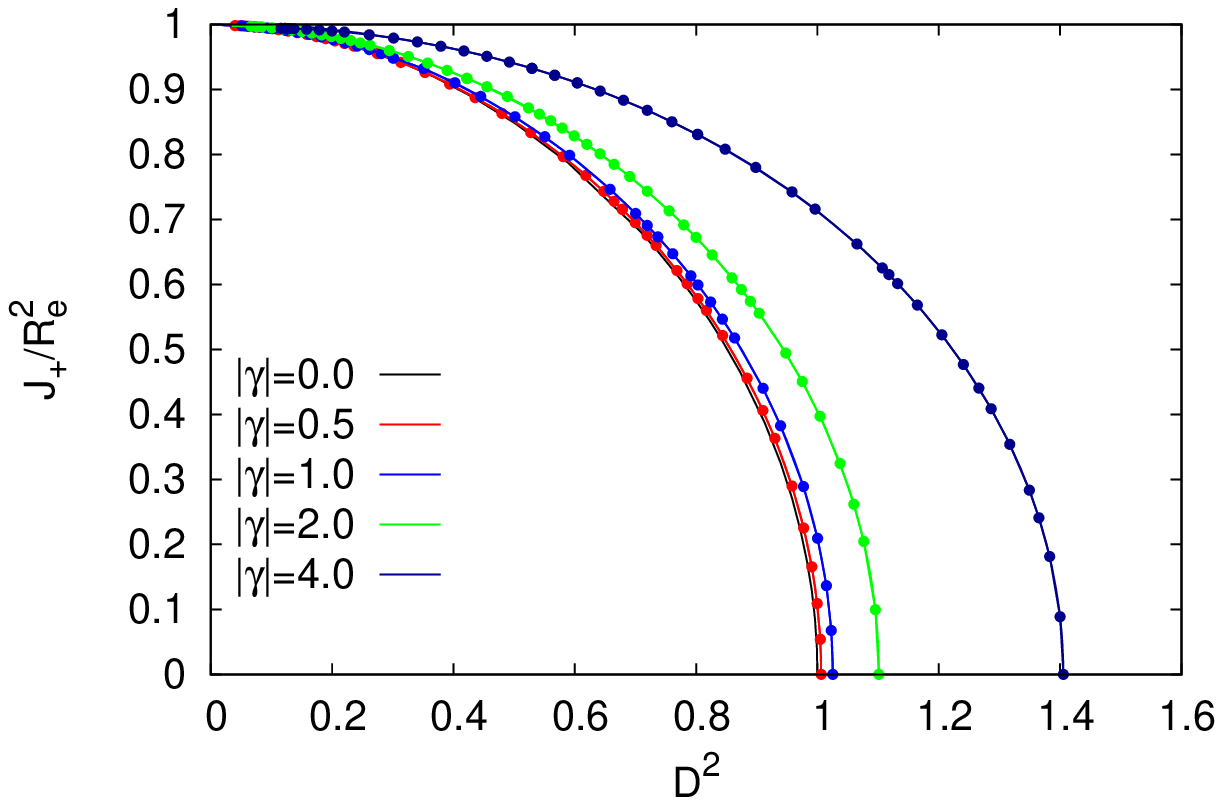}
}
\mbox{(c) \hspace*{0.48\textwidth}(d)}\\
\mbox{
\includegraphics[height=.23\textheight, angle =0]{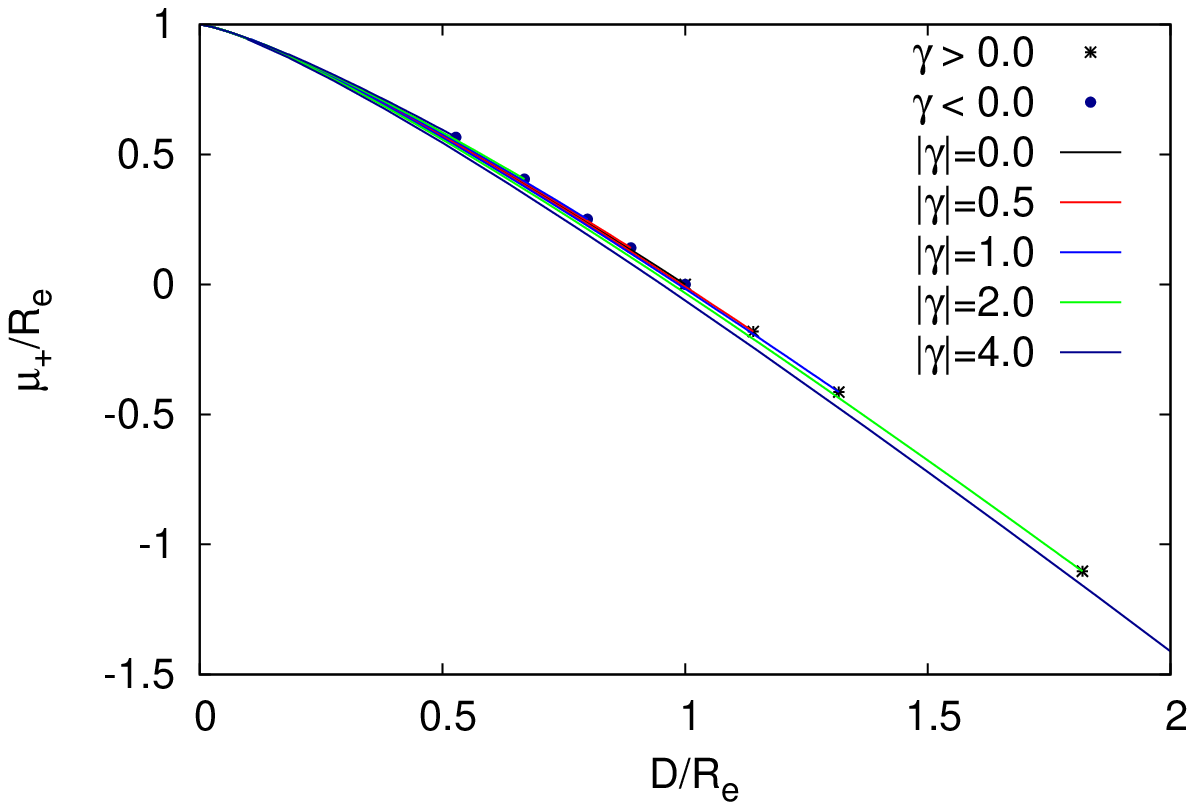}
\includegraphics[height=.23\textheight, angle =0]{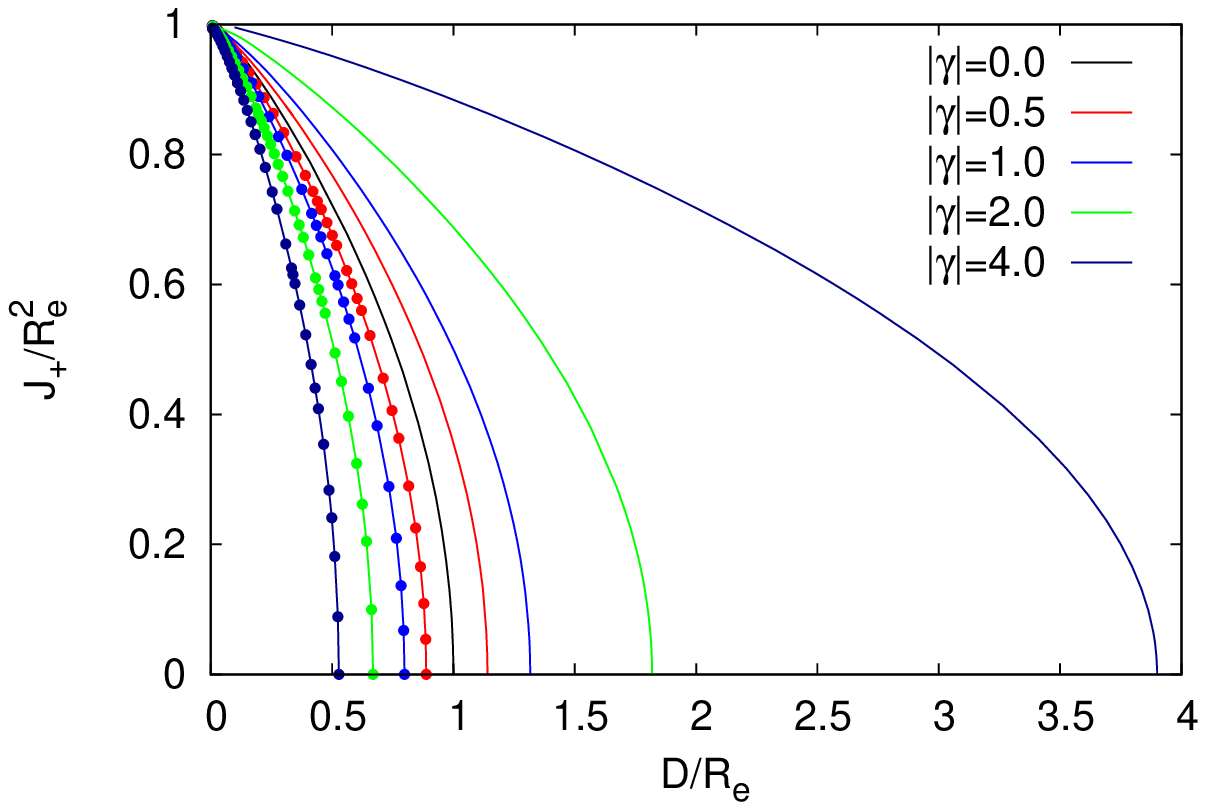}
}\\
\mbox{(e) }\\
\mbox{
\includegraphics[height=.23\textheight, angle =0]{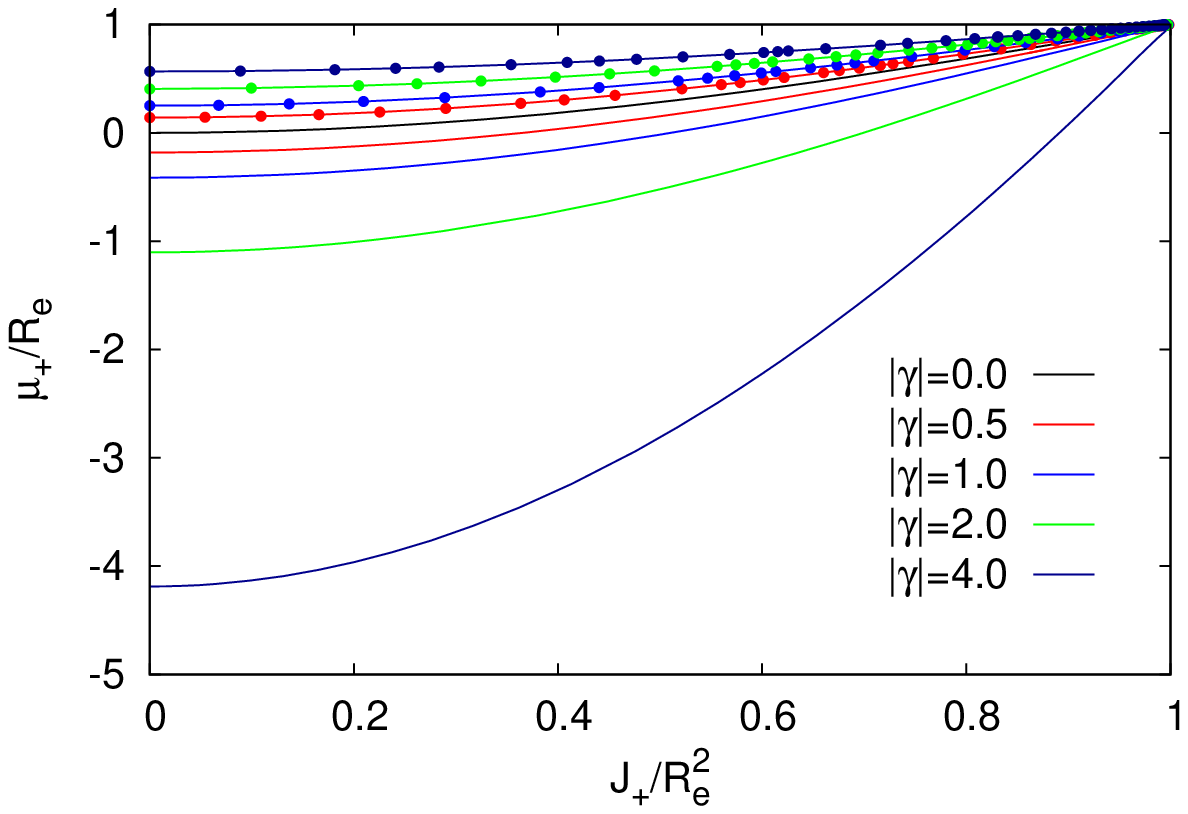}
}
\end{center}
\vspace{-0.5cm}
\caption{
Properties of non-symmetric rotating wormholes at fixed throat parameter $\eta_0=1$:
(a) the mass $\mu_+/R_e$ scaled by the equatorial throat radius
versus the squared scalar charge $D^2$, 
(b) same as (a) for the angular momentum $J_+/R_e^2$ scaled by the squared 
equatorial throat radius,
(c) the scaled mass $\mu_+/R_e$ versus the scaled scalar charge $D/R_e$,
(d) the scaled angular momentum $J_+/R_e^2 $ versus the scaled scalar charge $D/R_e$,
(e) the scaled mass $\mu_+/R_e$ versus the scaled angular momentum $J/R_e^2 $.
The parameter $\gamma$ is varied in the range $-4 \le \gamma \le 4$
with negative values indicated by dots.
}
\label{Fig11}
\end{figure}

In Fig.~\ref{Fig11}a we exhibit
the mass $\mu_+/R_e$ scaled by the equatorial throat radius
versus the squared scalar charge $D^2$ 
for several fixed values of the parameter $\gamma$ 
in the range $-4 \le \gamma \le 4$,
where the negative values of $\gamma$ are indicated in the figure by dots.
The curves are obtained by varying the value of $\omega_{-\infty}$
for a fixed value of $\gamma$, starting from the static case 
$\omega_{-\infty}=0$.

The symmetric static solution has vanishing mass and scalar charge $D=1$,
while the mass of the rotating solutions increases monotonically towards
the mass of the limiting extremal Kerr black hole as the
scalar charge decreases monotonically to zero.
As seen in the figure, for fixed $\gamma$
and fixed equatorial throat radius $R_e$, the mass always increases
monotonically from the corresponding static value to the
extremal Kerr value, while the scalar charge decreases monotonically to zero.
In particular, for negative $\gamma$ the mass is always positive.
This is in contrast to positive $\gamma$, where the mass becomes negative
in a part of the domain of existence. However, independent of $\gamma$,
all families of non-symmetric wormhole solutions 
approach the extremal Kerr solution.

When we consider the 
angular momentum $J_+/R_e^2$ scaled by the squared
equatorial throat radius, exhibited in Fig.~\ref{Fig11}b
versus the squared scalar charge $D^2$,
we see the same monotonic behavior from the static solutions
towards the limiting extremal Kerr black hole.
The surprise encountered here, is that for a given scalar charge
the value of the scaled angular momentum does not depend on the
sign of $\gamma$.

To map the domain of existence it is favorable to consider
the scaled scalar charge $D/R_e$ (instead of $D^2$).
The scaled mass $\mu_+/R_e$ and the scaled angular momentum $J_+/R_e^2$
are shown versus $D/R_e$ in Figs.~\ref{Fig11} %
(c) and (d), respectively.
The scaled mass is then located in a thin band, reaching 
to large values of $D/R_e$ for large positive values of $\gamma$,
while shrinking towards the extremal Kerr limit for large negative 
values of $\gamma$.
The scaled angular momentum $J_+/R_e^2$ on the other hand 
now distinguishes between positive and negative values of $\gamma$,
when considered versus $D/R_e$. Fig.~\ref{Fig11}e finally
shows the scaled mass $\mu_+/R_e$ versus the scaled angular momentum $J_+/R_e^2$.
Here the limiting behavior of the scaled mass for large negative $\gamma$ 
can be anticipated, which corresponds to $\mu_+/R_e \to 1$.

\begin{figure}[t!]
\begin{center}
\mbox{(a) \hspace*{0.48\textwidth}(b)}\\
\mbox{
\includegraphics[height=.23\textheight, angle =0]{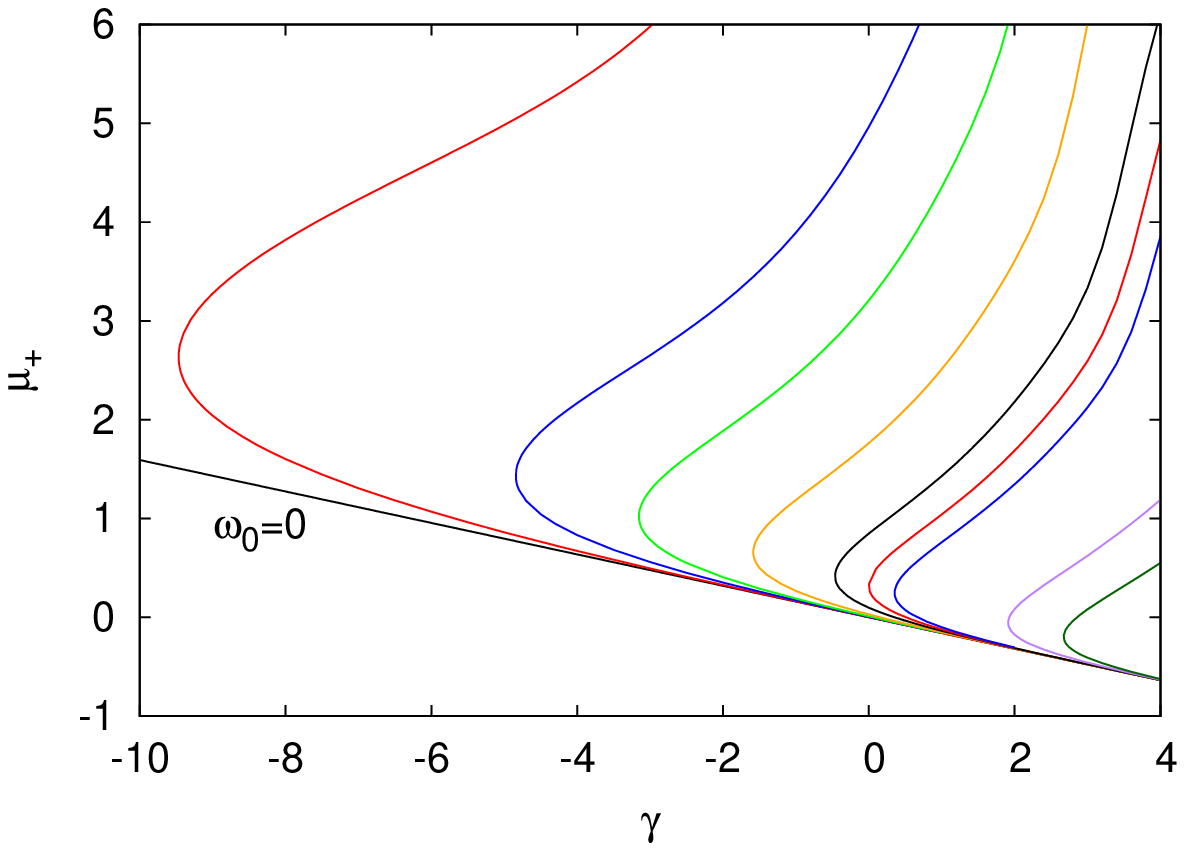}
\includegraphics[height=.23\textheight, angle =0]{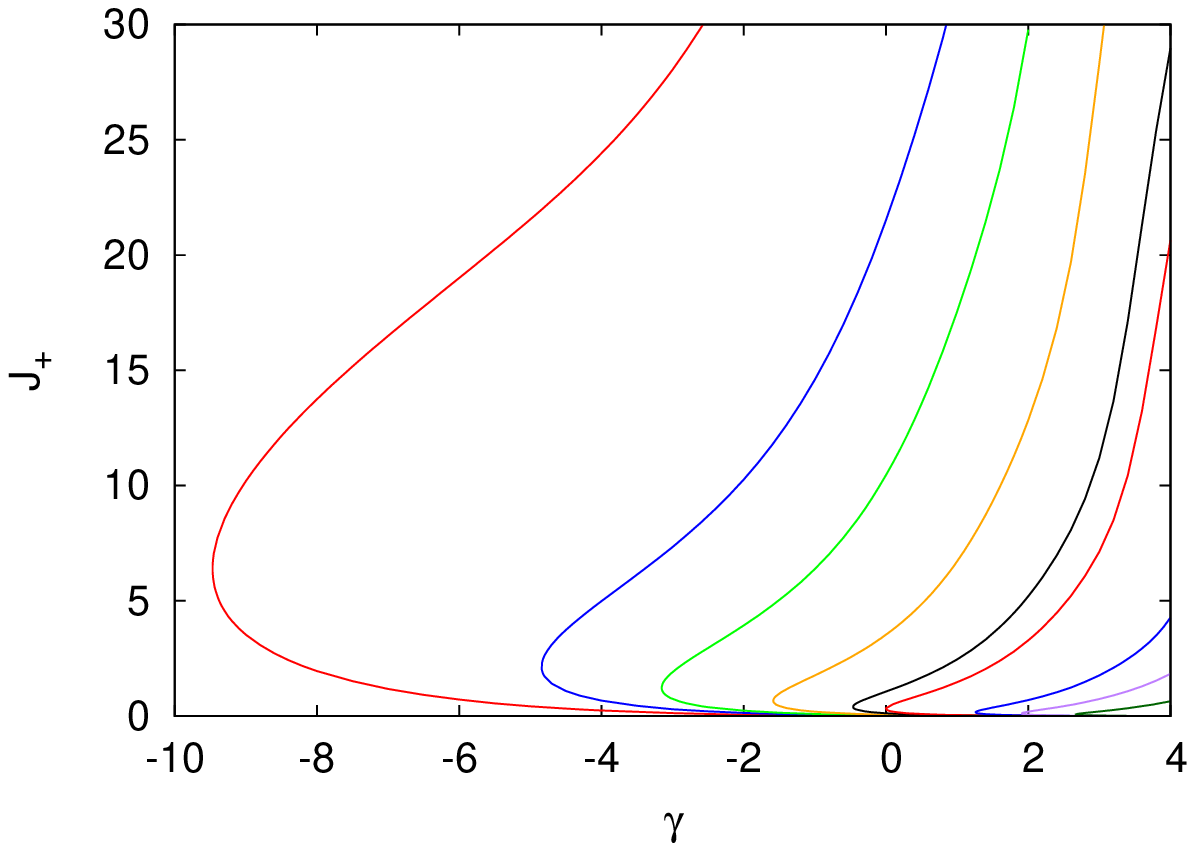}
}
\mbox{(c) \hspace*{0.48\textwidth}(d)}
\mbox{
\includegraphics[height=.23\textheight, angle =0]{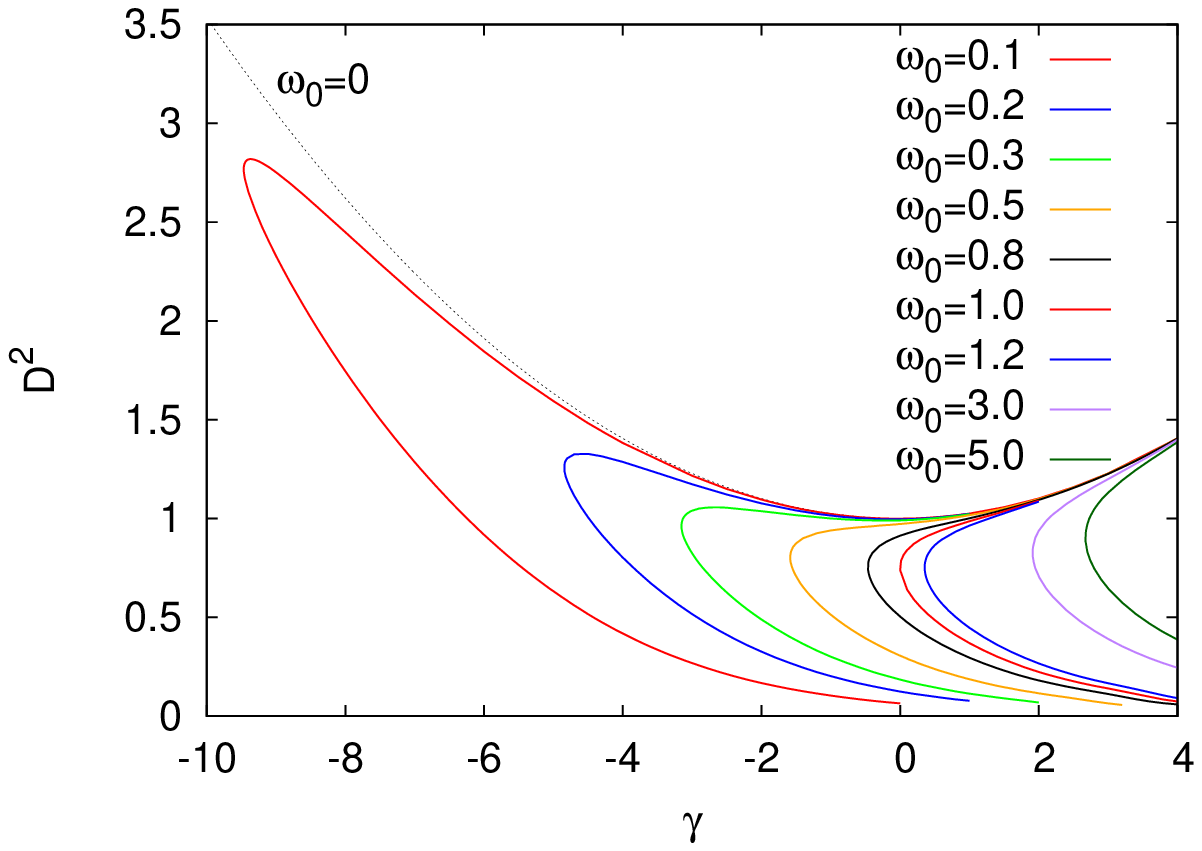}
\includegraphics[height=.23\textheight, angle =0]{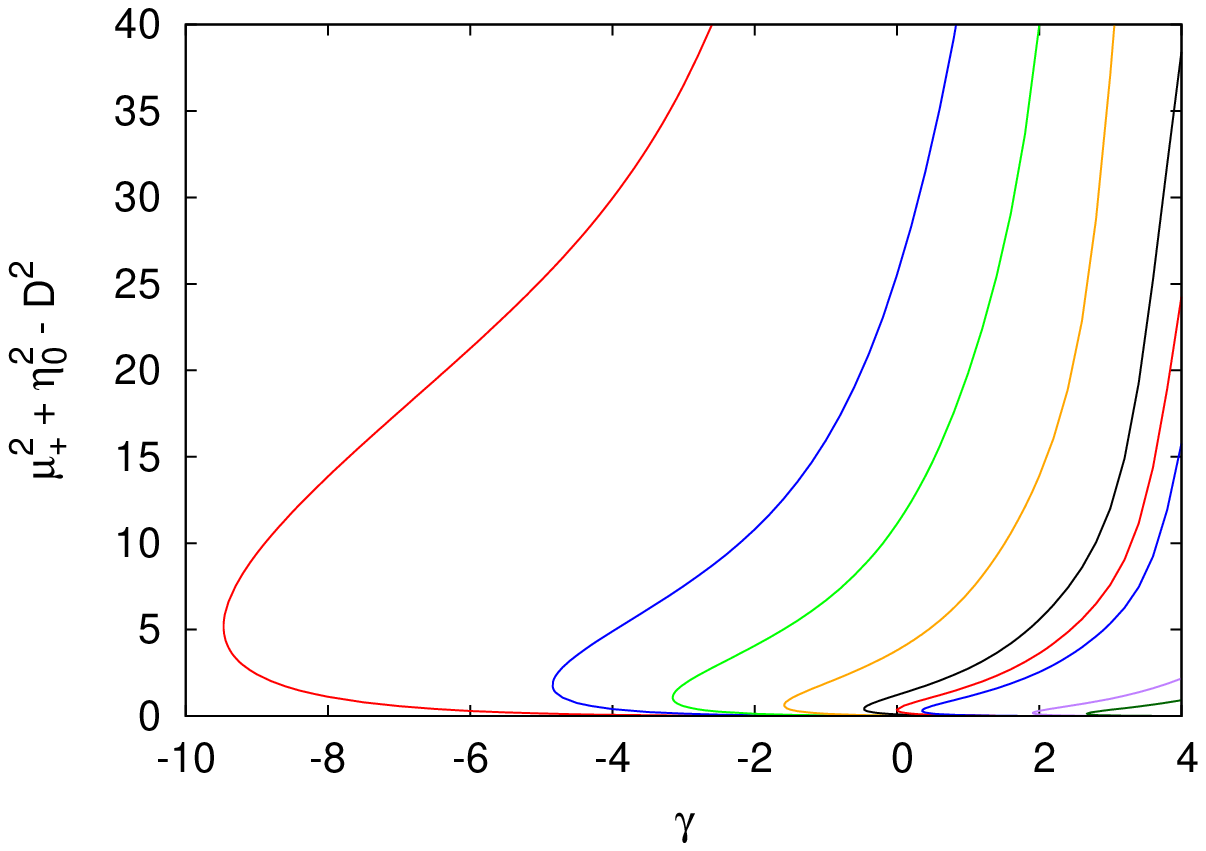}
}
\end{center}
\vspace{-0.5cm}
\caption{
Properties of non-symmetric rotating wormholes at fixed throat parameter $\eta_0=1$:
(a) the mass $\mu_+$,
(b) the angular momentum $J_+$
(c) the squared scalar charge $D^2$, and
(d) the coefficient $c_2= \mu_+^2 + \eta_0^2 - D^2$ versus $\gamma$.
The parameter $\omega_{-\infty}$ is varied in the range $0 \le \omega_{-\infty} \le 5$.
}
\label{Fig12}
\end{figure}

To demonstrate the dependence on the parameter $\omega_{-\infty}$,
we exhibit in Fig.~\ref{Fig12} the mass $\mu_+$ (a), 
the angular momentum $J_+$ (b),
the squared scalar charge $D^2$ (c), and the 
coefficient Eq.~(\ref{j-m-rel}) $c_2 = \mu_+^2 + \eta_0^2 - D^2$  
(d) versus the parameter $\gamma$
for fixed values of the parameter $\omega_{-\infty}$
in the range $0 \le \omega_{-\infty} \le 5$.
Recall, that $\gamma=0$ represents the symmetric case, while
$\omega_{-\infty}=0$ represents the static case.
Again the asymmetry between positive and negative values of $\gamma$
is evident.

\begin{figure}[h!]
\begin{center}
\mbox{(a) \hspace*{0.48\textwidth}(b)}\\
\mbox{
\includegraphics[height=.23\textheight, angle =0]{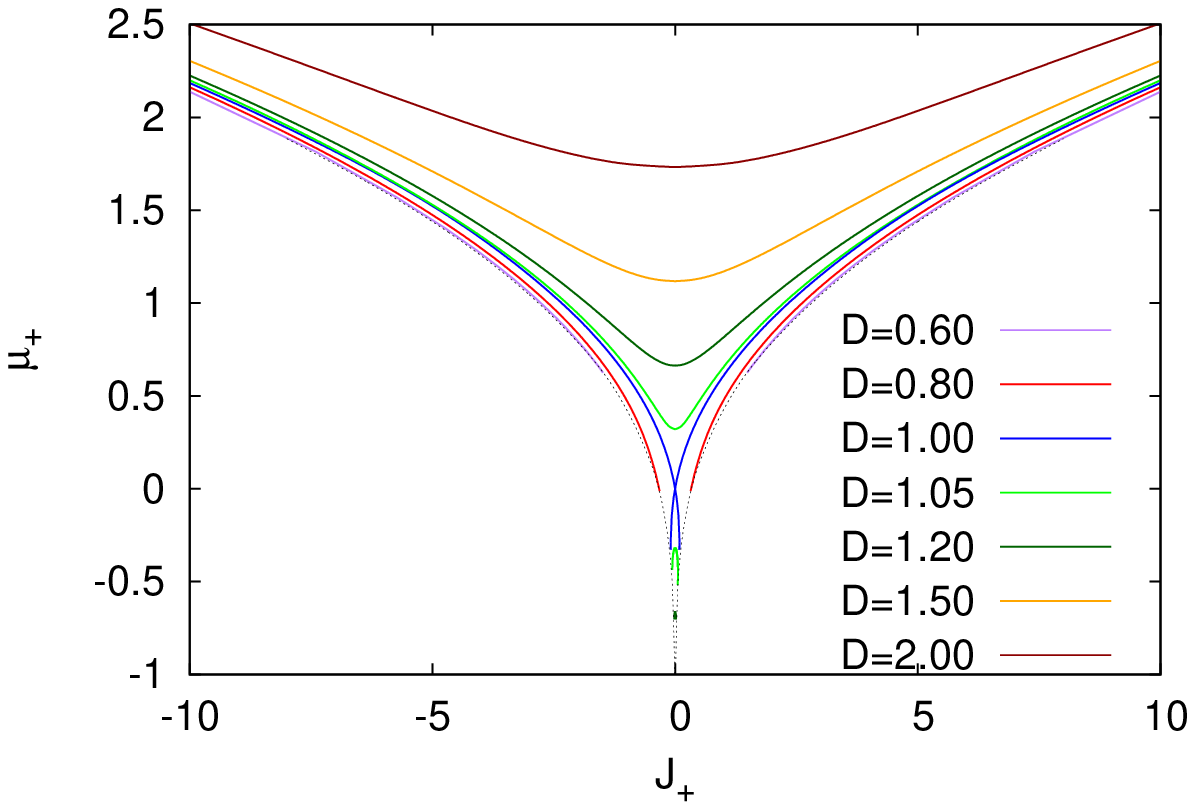}
\includegraphics[height=.23\textheight, angle =0]{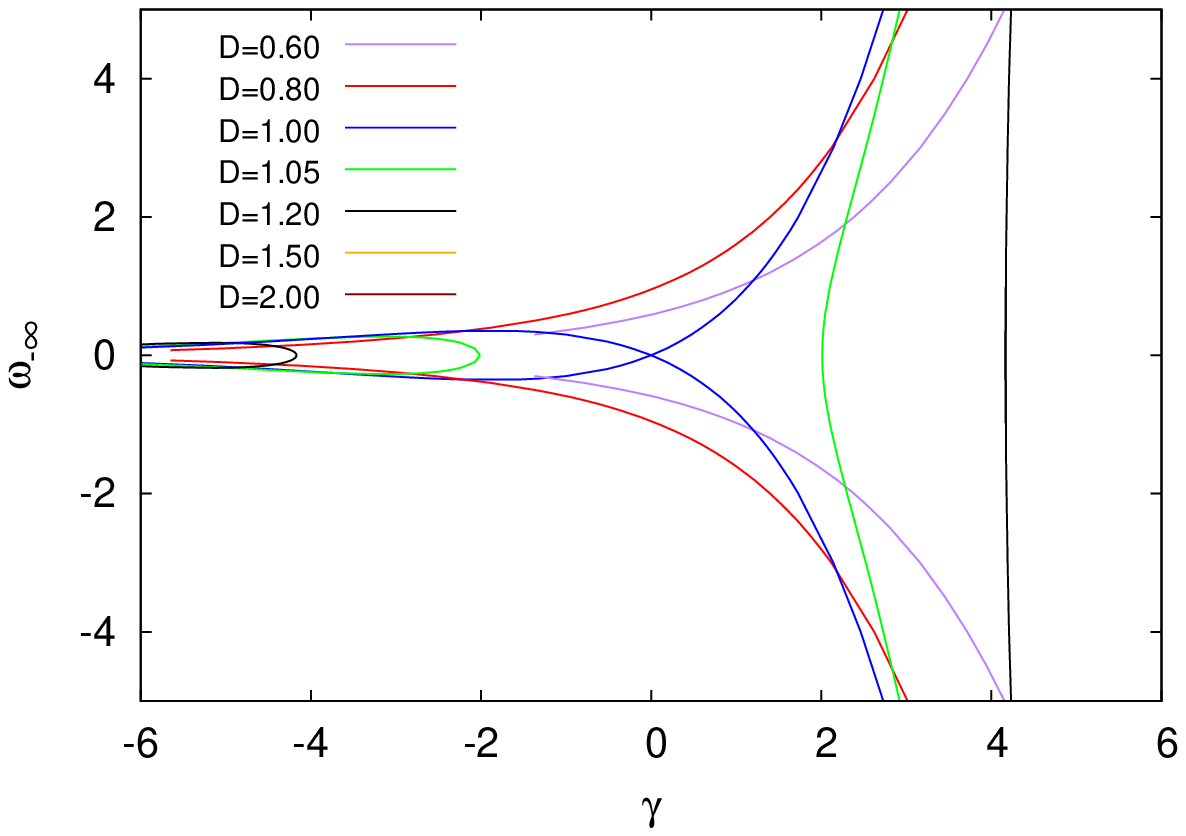}
}
\end{center}
\vspace{-0.5cm}
\caption{
Properties of non-symmetric rotating wormholes at fixed throat parameter $\eta_0=1$:
(a) the mass $\mu_+$ versus the angular momentum $J_+$,
(b) the parameter $\omega_{-\infty}$ versus the parameter $\gamma$.
The scalar charge $D$ is varied in the range $0.6 \le D \le 2$.
The thin dotted curves in (a) delimit the domain of existence.
}
\label{Fig13}
\end{figure}

As a final point of interest concerning the global charges, 
let us consider the mass $\mu_+$ versus the
angular momentum $J_+$ for fixed scalar charge $D$,
exhibited in Fig.~\ref{Fig13}a,
where $D$ is varied in the range $0.6 \le D \le 2$.
The corresponding values of the parameters $\omega_{-\infty}$
and $\gamma$ are shown in Fig.~\ref{Fig13}b.
Here we observe an interesting bifurcation phenomenon,
where the critical value corresponds to $D_{\rm cr}=1$.
For a given value of $D< D_{\rm cr}$, there are two $M$
curves, located symmetrically w.r.t.~$J_+$, which end
at respective limiting values.
As $D$ increases towards $D_{\rm cr}$, the two curves
approach each other closer and closer, until at
$D_{\rm cr}$ the two curves cross.
This happens precisely at $\mu_+=J_+=0$.
For $D> D_{\rm cr}$, the pattern changes, and 
there emerge a higher mass curve and a low mass curve.
The figure also exhibits two thin dotted curves,
which delimit the domain of existence.

\subsection{Velocity of the throat}

\begin{figure}[t!]
\begin{center}
\mbox{(a) \hspace*{0.48\textwidth}(b)}\\
\mbox{
\includegraphics[height=.23\textheight, angle =0]{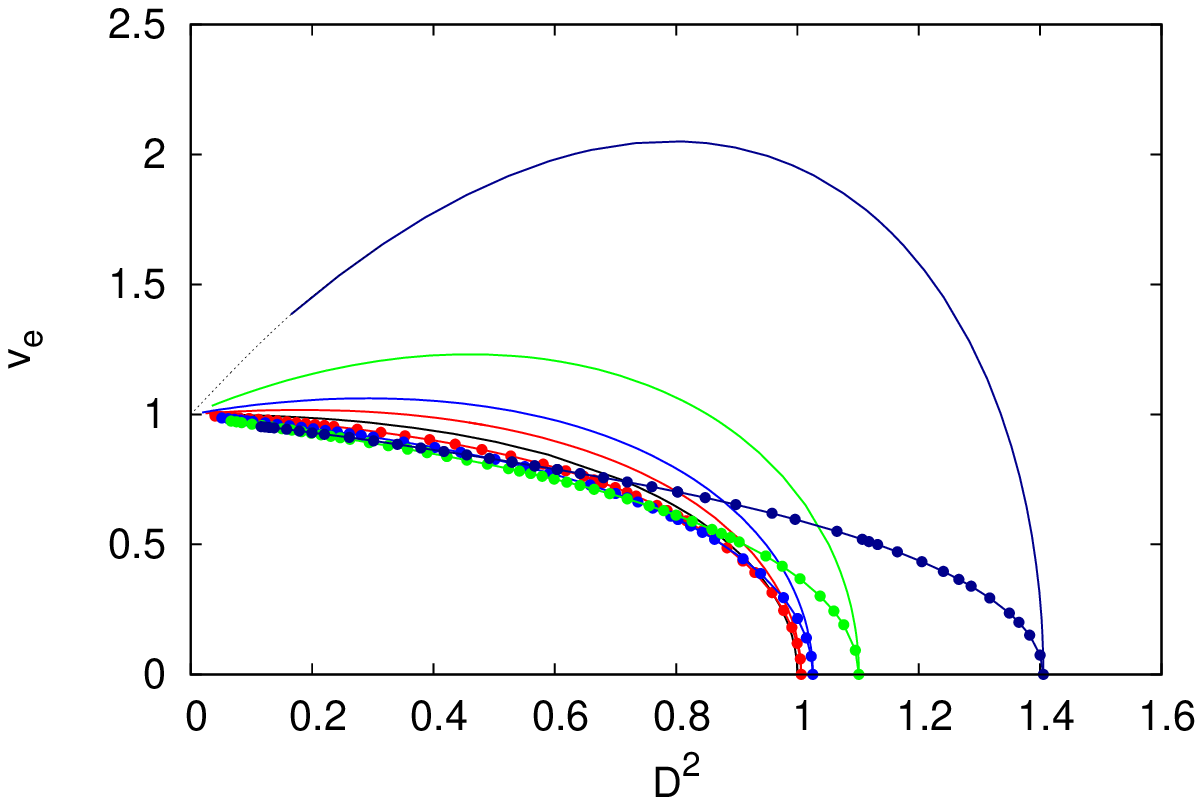}
\includegraphics[height=.23\textheight, angle =0]{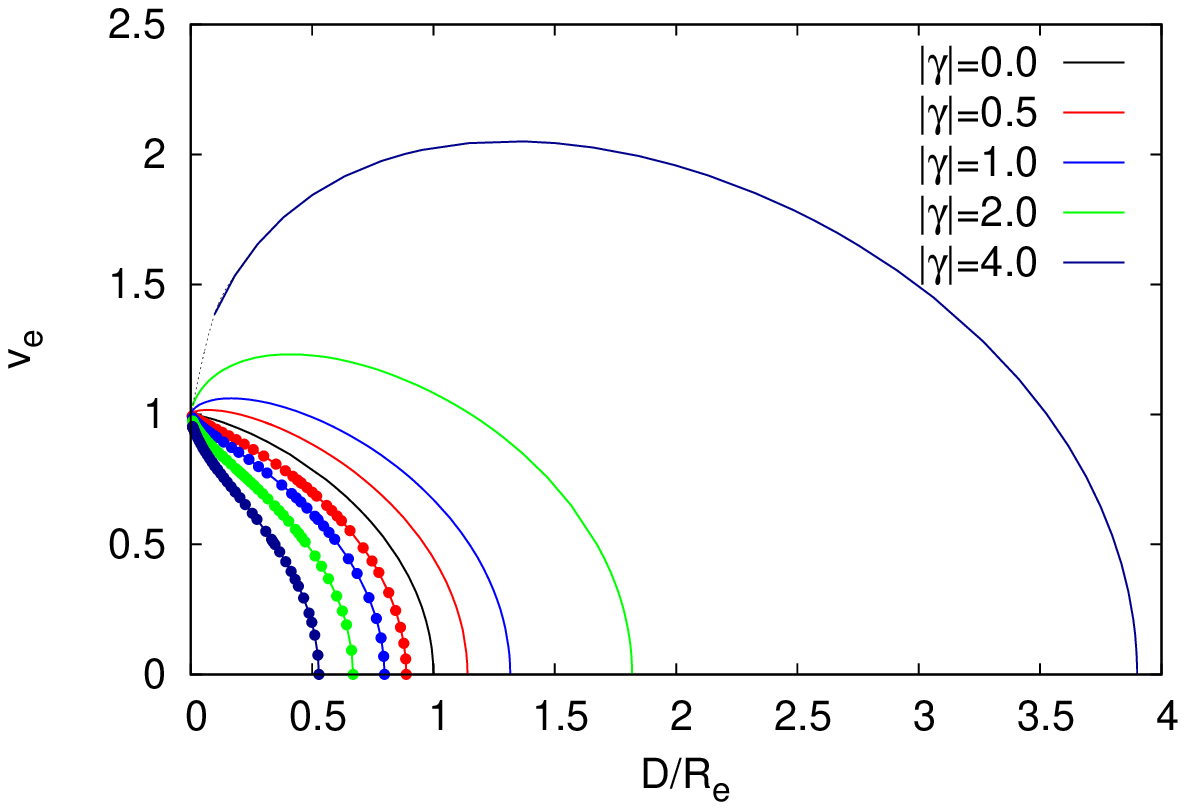}
}
\end{center}
\vspace{-0.5cm}
\caption{
Properties of non-symmetric rotating wormholes at fixed throat parameter $\eta_0=1$:
the rotational velocity in the equatorial plane $v_e= {R_e\Omega}$
versus the squared scalar charge $D^2$ (a) and versus the
scaled scalar charge $D/R_e$ (b)
The parameter $\gamma$ is varied in the range $-4 \le \gamma \le 4$
with negative values indicated by dots.
The thin dotted lines correspond to extrapolations towards
the limiting extremal Kerr black hole.
}
\label{Fig14}
\end{figure}

When the throat has angular velocity $\Omega=\omega_0$, 
its rotational velocity
in the equatorial plane is given by
$v_e= {R_e\Omega}$.
As seen in the previous section,
symmetric wormholes always satisfy $v_e \le 1$.
This continues to hold for non-symmetric rotating wormholes with 
negative values of $\gamma$.
However, non-symmetric rotating wormholes with 
positive values of $\gamma$ can exceed this bound.
This is seen in Fig.~\ref{Fig14}a, where we exhibit the
rotational velocity $v_e$ versus the square of the scalar charge, $D^2$,
for several families of non-symmetric rotating wormholes, 
characterized by fixed values of the parameter $\gamma$.

Since the rotational velocity $v_e$ of the extremal Kerr black hole
precisely saturates the bound, $v_e=1$, 
and since all families of non-symmetric rotating wormholes possess the
extremal Kerr black hole as their limiting configuration,
the quantity $v_e$ does no longer possess a monotonic behavior
for the families of rotating wormholes with fixed $\gamma>0$.
Consequently, $v_e$ is no longer the quantity of choice
to demonstrate the physical properties of non-symmetric rotating wormholes.
In this case the scalar charge, squared or scaled, represents a
preferable physical parameter, since it always changes monotonically.
The rotational velocity $v_e$ is shown versus the scaled scalar charge, $D/R_e$,
in Fig.~\ref{Fig14}b.

Let us now address the origin of the fact, that 
the rotational velocity $v_e$ exceeds the speed of light,
i.e., $v_e>1$,
for $\gamma>0$ wormholes in a part of their parameter space
adjoining the extremal Kerr black hole.
For those wormholes the metric function $\omega$,
whose value at the throat enters the expression for the
rotational velocity $v_e$, increases strongly
towards the throat, assuming very large values at the throat
(and even larger values beyond the throat).
At the same time, the metric coefficient $g_{00}$
is monotonically decreasing from its
asymptotic value of $g_{00}(\infty) = -1$
to large negative values at the throat.
(Note, that we exhibit examples of the metric functions for such wormholes
in the Appendix.)
Thus the wormholes exhibit strong antigravitational features,
which are mainly associated with their negative mass.
Consequently time is running much faster
for an observer at the throat than for an observer 
far from the wormhole.
For an observer close to the throat the rotational velocity $v_e$
never exceeds the speed of light.

\subsection{Location of the throat}

\begin{figure}[t!]
\begin{center}
\mbox{
\includegraphics[height=.23\textheight, angle =0]{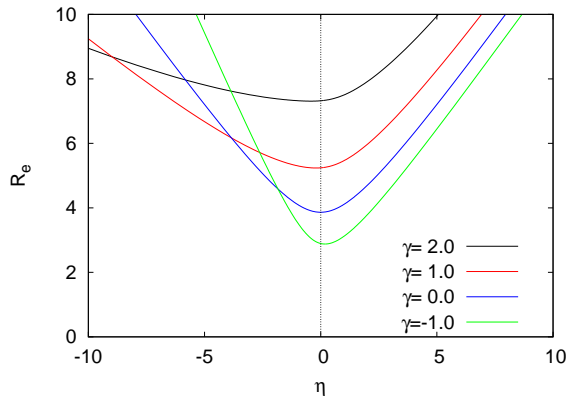}
}
\end{center}
\vspace{-0.5cm}
\caption{
Properties of non-symmetric rotating wormholes at fixed throat parameter $\eta_0=1$:
the equatorial radius $R_e$ versus the radial coordinate $\eta$
for spacetimes with $\omega(-\infty)=0.5$ (on the upper branch) and
$\gamma=f(-\infty) = 2.0$, $1.0$, $0.0$ and $-1.0$.
The minimal value of $R_e$ represents the respective location of the throat.
}
\label{Fig15}
\end{figure}

Let us next address the throat of these
non-symmetric rotating wormholes.
In particular, we would like to know, where we find the throat 
in the equatorial plane, i.e., at which coordinate $\eta_{e}$
it is located there, and whether the throat coordinate $\eta_{t}$
depends on the polar angle $\theta$.
To approach the first question, we
show in Fig.~\ref{Fig15} examples of the 
equatorial radius $R_e$ versus the radial coordinate $\eta$ for 
several wormhole solutions. 
These spacetimes are characterized by the parameter value $\omega(-\infty)=0.5$ 
(located on the upper branch) and the values of the asymmetry parameter
$\gamma=f(-\infty) = 2.0$, $1.0$, $0.0$ and $-1.0$.
Their scalar charge is given by
$D=0.33984$, $0.43155$, $0.55277$, and $0.71636$, respectively.

The location $\eta_{e}$ of the throat in the equatorial plane 
is given by the minimum of the equatorial radius $R_e$.
We note that the location of the throat $\eta_{e}$
depends only weakly on the parameter $\gamma$ in the considered range,
$-1 \leq \gamma \leq 2$.
For negative $\gamma$ the throat is located at small positive $\eta_{e}$,
while for positive $\gamma$ it resides at small negative $\eta_{e}$.
We observe that,
in contrast to the throat location $\eta_{e}$,
the throat radius $R_e$ depends strongly on $\gamma$,
and increases with increasing $\gamma$.

In Boyer-Lindquist coordinates, the event horizon of a rotating black hole
is characterized by a single value of the radial coordinate.
Likewise, the throat of a symmetric rotating wormhole is 
characterized by a single value of the radial coordinate.
Let us now see, whether an analogous result holds for the throat of
a non-symmetric rotating wormhole.
To that end we analyze the coordinates of the minimal surface
defining the throat. In particular, we calculate the value of the radial
throat coordinate $\eta_{t}$ as a function of the polar angle $\theta$,
%
and find only a very slight dependence of $\eta_{t}$ on $\theta$.
For instance, the ratio $\eta_{t}/\eta_{e}$ is always very close to one,
with deviations less than 0.1\% 
and thus within the numerical accuracy.
While we cannot answer, whether $\eta_{t}$ 
is indeed a constant,
we can state
that a constant $\eta_{t}$ represents at least 
a good approximation.
From a geometrical point of view, 
however, the throat is highly deformed
for a fast rotating wormhole.

\subsection{Geodesics}

\begin{figure}[t!]
\begin{center}
\mbox{\hspace*{2.cm}(a) \hspace*{0.48\textwidth}(b)}\\
\mbox{
\includegraphics[height=.28\textheight, angle =0]{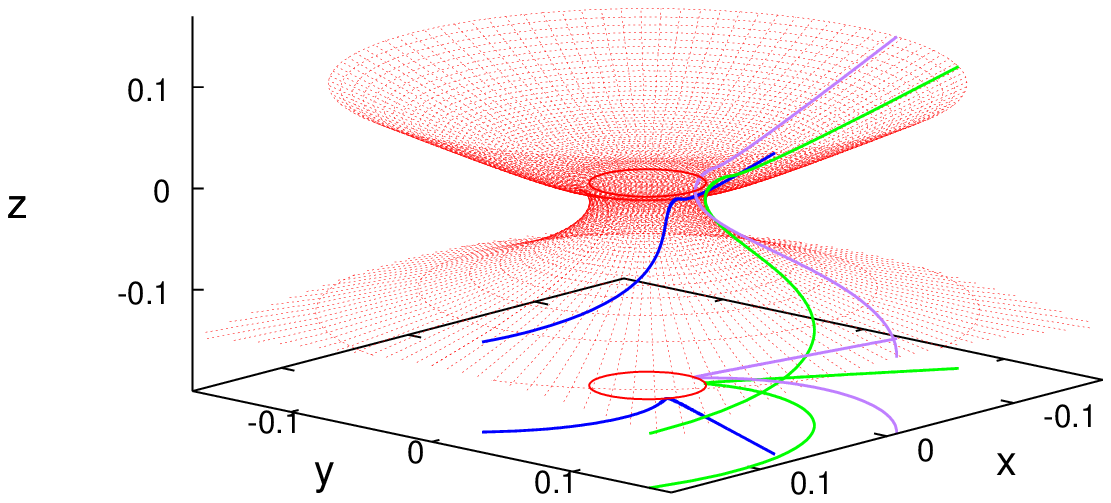}
\hspace*{-1.0cm}
\includegraphics[height=.23\textheight, angle =0]{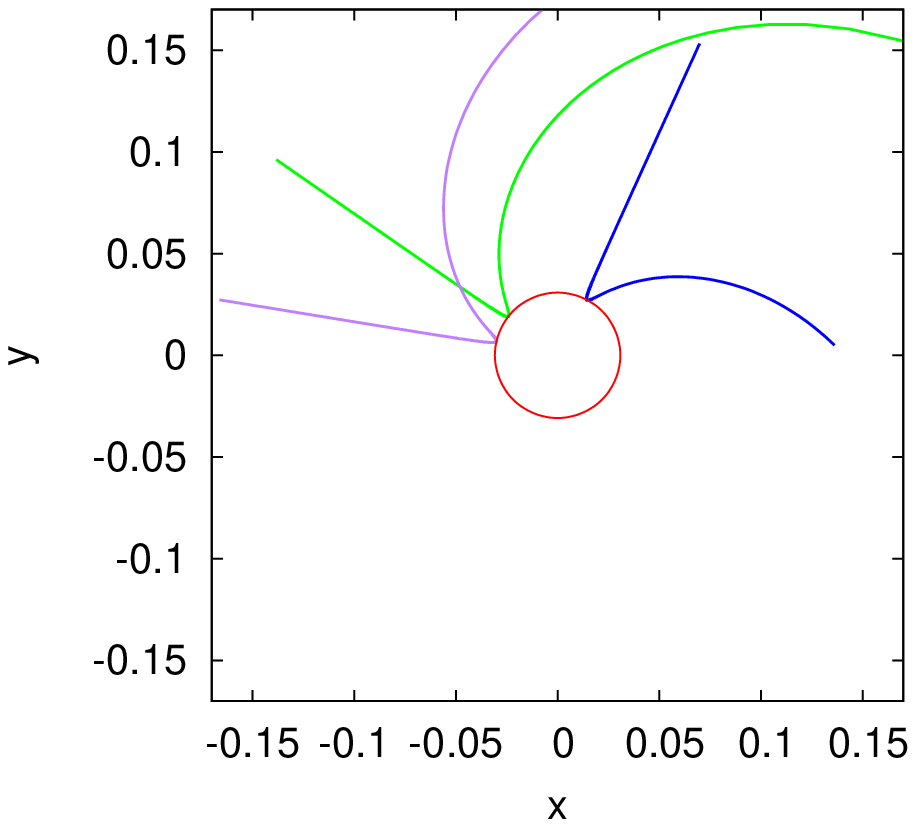}
}
\end{center}
\vspace{-0.5cm}
\caption{
Geodesics of non-symmetric rotating wormholes:
massless particles with zero angular momentum starting at
coordinates $x_0=0.136$, $0.335$, $0.399$ and $y_0=0$ on the second manifold,
passing through the throat
for a wormhole solution with $\omega_{-\infty} =1000$, $\gamma=10.008$.
Embedding diagram (a) and projection in the equatorial plane (b).
}
\label{Fig16}
\end{figure}

As our last quest let us address the trajectories of particles and light
in these non-symmetric rotating wormhole spacetimes.
As in the symmetric case we find stable bound orbits of massive 
particles and unstable bound orbits of massless particles.
However, we now focus on the orbits of particles in wormhole spacetimes,
whose equatorial throat velocity $v_e$ exceeds the speed of light,
since this is a new phenomenon as compared to the symmetric case.
As an example, we exhibit in Fig.~\ref{Fig16} the orbits
of massless particles with zero angular momentum starting at the
coordinates $x_0=20$, 50, 60 on the second manifold, 
which pass through the throat of a
wormhole solution with parameters $\omega_{-\infty} =1000$, $\gamma=10.008$,
and $v_e \approx 8$. Fig.~\ref{Fig16}a shows the orbits in an embedding
of the wormhole spacetime, while Fig.~\ref{Fig16}b gives a projection of 
the orbits in the equatorial plane.

In such a spacetime
the particle orbits appear rather strange at first sight.
Moving with zero angular momentum in the
negative $\eta$ region towards the throat, 
the particle orbits are strongly dragged 
along by the rapidly rotating space time
until the particles pass the throat.  Beyond the throat
the particles then follow more or less straight lines,
as they move across the positive $\eta$ region towards infinity.
In this region the function $\omega$ decreases rapidly 
outside the throat and vanishes asymptotically.
So the dragging of the radial geodesics diminishes fast, when away from the 
throat.

\section{Conclusions}

The construction of Lorentzian rotating wormholes, which are
generalizations of the Ellis wormholes,
has been a challenge for a very long time.
With this study we have provided the first detailed investigation
of rotating Ellis wormholes in General Relativity.
These wormhole solutions are supported by a phantom field,
and can be either symmetric or non-symmetric.
Whereas the static solutions are known in closed form,
the non-perturbative rotating wormholes have been obtained
by numerical integration of the coupled system of
Einstein-phantom equations, subject to appropriate boundary conditions.
The latter guarantee that the solutions are globally regular and
asymptotically flat on one side, while for the second side 
a coordinate transformation is necessary to achieve asymptotic flatness.

The global charges of the wormhole solutions can be obtained
from their asymptotic expansions. For symmetric wormholes
the mass is the same on both sides of the wormhole.
The same holds true for the angular momentum. However,
this is no longer the case, when the asymmetry parameter $\gamma$
differs from zero.
The throat of symmetric wormholes resides at the center
of our coordinate system, where the radial coordinate $\eta$ is zero.
The corresponding surface represents a minimal surface,
which changes from a round sphere in the static limit
to a strongly deformed axially symmetric surface with negative Gaussian
curvature at the poles, when the rotational velocity 
becomes sufficiently high.
For non-symmetric wormholes the minimal surface defining the throat
may still be described by 
a surface of constant radial coordinate $\eta$
(at least within our numerical accuracy).
But, depending on the sign of the asymmetry parameter $\gamma$,
the location of the throat is shifting
towards positive or negative values of $\eta$. 

Many of the properties revealed by our calculations are rather
surprising, for instance,
\vspace{-0.21cm}
\begin{itemize}
\itemsep=-4pt
\item
  rotating wormholes satisfy a Smarr type relation,
\item
  rotating wormholes possess an extremal Kerr black hole
  as limiting configuration,
\item
   for rotating wormholes
   the crossing from positive to negative Gaussian curvature 
   at the poles of the throat occurs
   precisely at the same ratio of radii as for Kerr black holes,
\item
   the dimensionless quadrupole moment of rotating wormholes can strongly
   exceed the Kerr value,
\item
   the ergoregion of the rotating wormholes is very different from the
   expectations raised by the Teo wormhole,
\item
   non-symmetric wormholes can possess a rotational velocity 
   of the throat which exceeds the speed of light,
\item
   unlike the static Ellis wormholes, 
   the rotating wormholes do possess bound orbits.
\end{itemize}  
\vspace{-0.15cm}

Furthermore, we conjecture that the observations 
that (i) the violation of the NEC decreases with increasing 
global charge, while (ii)
the wormholes approach an extremal black hole solution
as a limiting configuration,
represent generic features of wormholes.
For the present study, the global charge is angular momentum,
and the extremal black hole corresponds to a Kerr black hole.
As shown before,
electrically charged static Ellis wormholes in four dimensions
approach with increasing charge in an analogous fashion 
an extremal Reissner-Nordstr\"om black hole
\cite{Hauser:2013jea},
while rotating wormholes (with equal angular momenta)
in five dimensions tend with increasing angular momenta
towards the corresponding extremal Myers-Perry black hole
\cite{Dzhunushaliev:2013jja}.

So far we have not addressed the important issue of the
stability of these rotating wormhole solutions.
It has been shown before, that the static Ellis wormholes are unstable,
both in four
\cite{Shinkai:2002gv,Gonzalez:2008wd,Gonzalez:2008xk}
(see also \cite{Konoplya:2016hmd})
and higher dimensions
\cite{Torii:2013xba}.
On the other hand, arguments have been given which indicate
that wormholes might be stabilized by rotation \cite{Matos:2005uh}.
Indeed, in the case of 
five-dimensional rotating wormholes it has been shown, 
that the unstable mode
of the static solutions disappears, when the rotation
is sufficiently fast 
\cite{Dzhunushaliev:2013jja}.
We are planning to do the corresponding stability analysis
also for four-dimensional rotating wormholes, to see
whether rotation will eliminate the unstable mode
as well in four dimensions.
On the other hand, stability of wormholes 
could also be achieved, when the Einstein-Hilbert action
of General Relativity is
replaced by a more general action with
higher curvature terms
\cite{Kanti:2011jz,Kanti:2011yv}.

\begin{acknowledgments}
We gratefully acknowledge support by the DFG within the Research
Training Group 1620 ``Models of Gravity''
and by FP7, Marie Curie Actions, People, 
International Research Staff Exchange Scheme (IRSES-606096).
We gratefully acknowledge discussions with E.~Radu.
\end{acknowledgments}

\section{Appendix}

\begin{figure}[h!]
\begin{center}
\mbox{(a) \hspace*{0.48\textwidth}(b)}\\
\mbox{
\includegraphics[height=.23\textheight, angle =0]{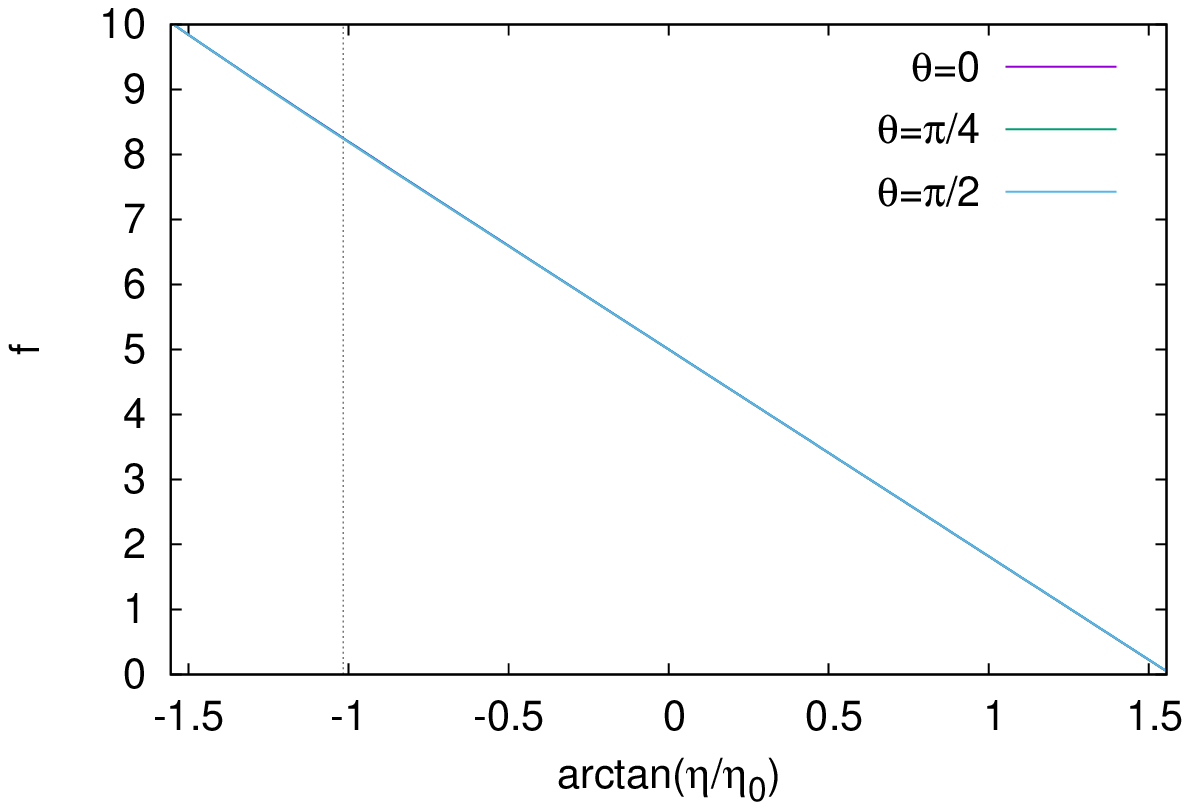}
\includegraphics[height=.23\textheight, angle =0]{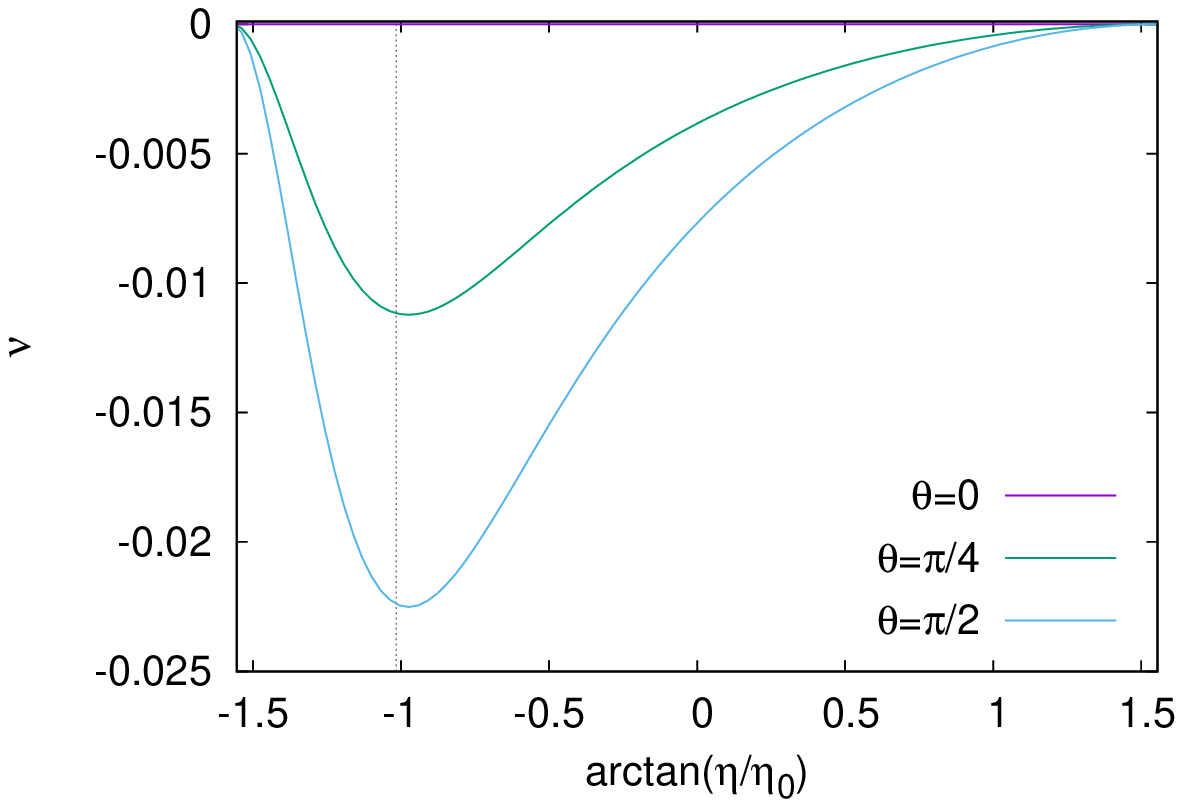}
}
\mbox{(c) \hspace*{0.48\textwidth}(d)}
\mbox{
\includegraphics[height=.23\textheight, angle =0]{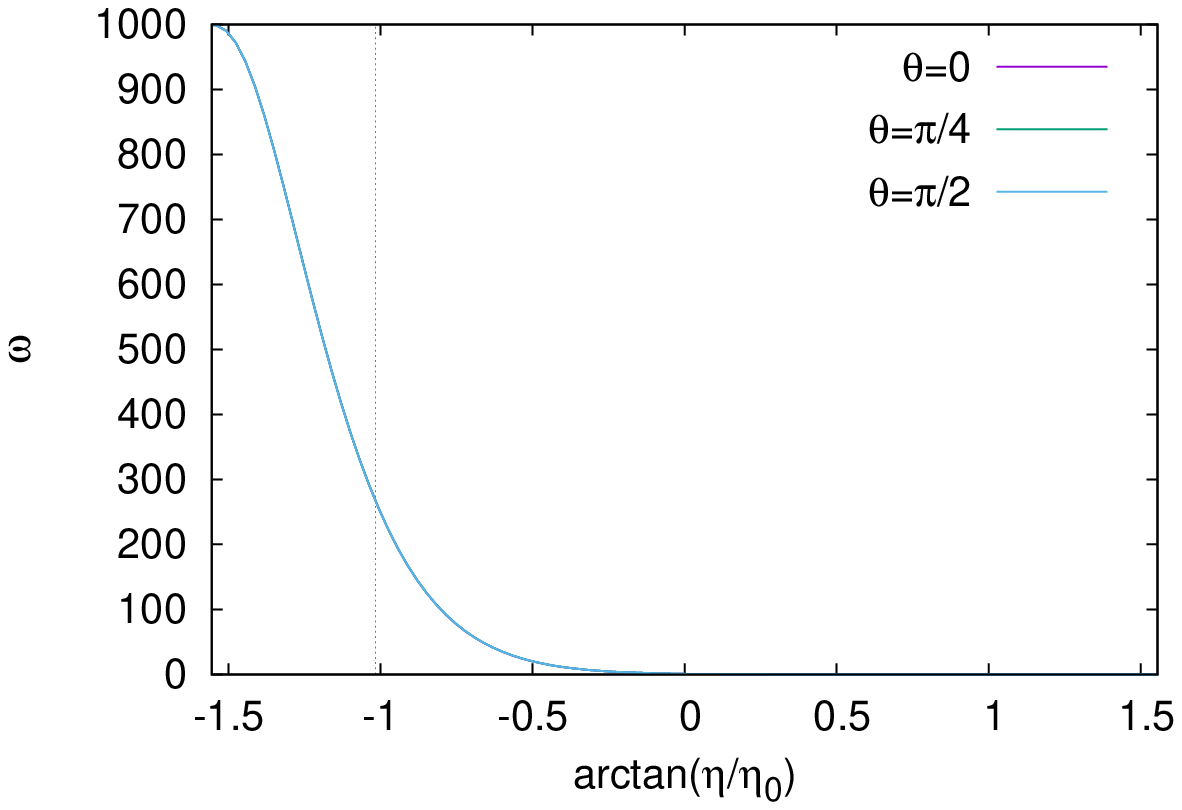}
\includegraphics[height=.23\textheight, angle =0]{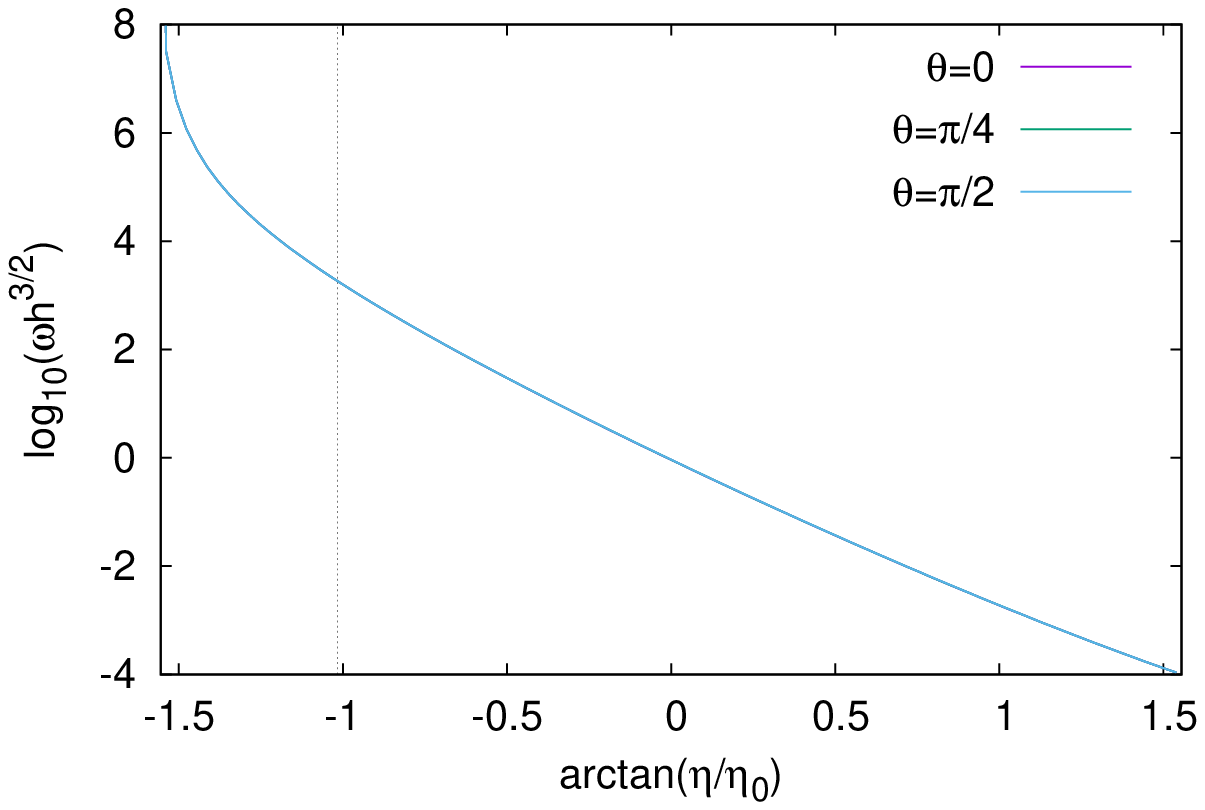}
}
\mbox{(e) \hspace*{0.48\textwidth}(f)}
\mbox{
\includegraphics[height=.23\textheight, angle =0]{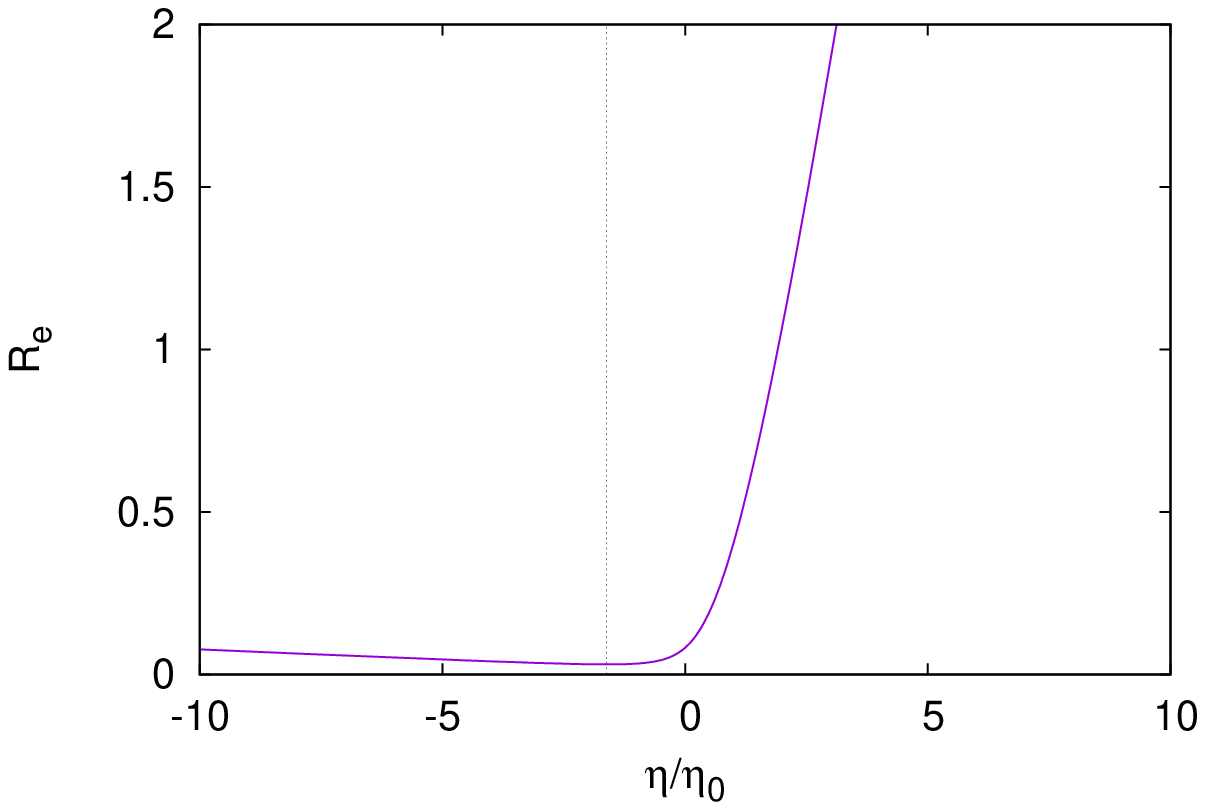}
\includegraphics[height=.23\textheight, angle =0]{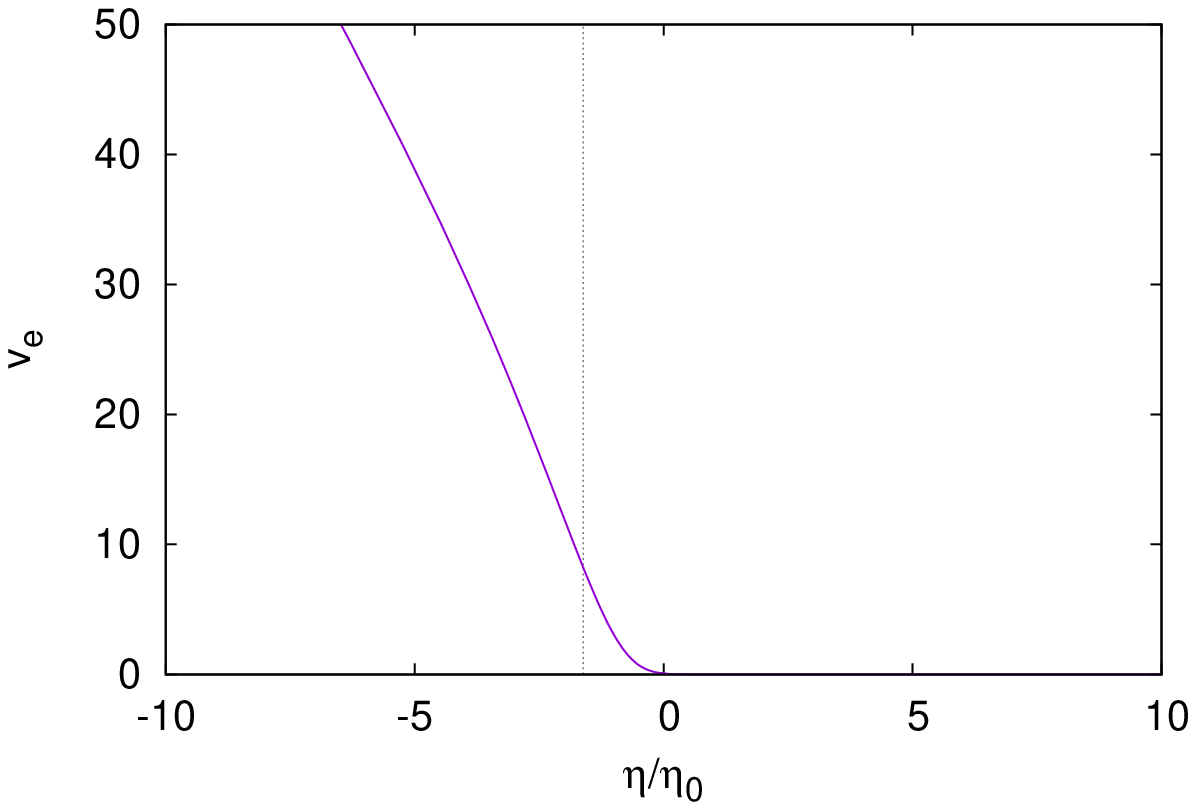}
}
\end{center}
\vspace{-0.5cm}
\caption{
Non-symmetric rotating wormhole with $v_e>1$: 
the metric functions $f$ (a), $\nu$ (b), $\omega$ (c),
the combination $\omega h^{3/2}$ (d), 
versus the compactified radial coordinate
$\arctan(\eta/\eta_0)$, and 
the equatorial radius $R_e$ (e) and the rotational velocity
$v_e = R_e \omega $ (f) versus the scaled radial coordinate $\eta/\eta_0$.
The thin vertical line marks the location of the throat.
}
\label{Fig17}
\end{figure}

We here demonstrate the behaviour of the metric functions for
asymmetric wormholes with rotational velocity $v_e>1$.
Fig.~\ref{Fig17} exhibits the 
functions $f$ (a), $\nu$ (b), $\omega$ (c)
versus the compactified radial coordinate
$\arctan(\eta/\eta_0)$
for a solution with parameters $\gamma=10.008$, $\omega_{-\infty}=1000$.
The function $f$ hardly deviates from the corresponding function
of the static wormhole, and the function $\nu$ remains close to one.
Still the function $\omega$ assumes very large values towards the
throat and beyond the throat.
The combination $\omega h^{3/2}$ is exhibited in (d) to 
demonstrate the fall-off of $\omega$ for $\eta \to + \infty$.

For this wormhole spacetime also the equatorial radius $R_e(\eta)$ (e) and the
rotational velocity in the equatorial plane $v_e(\eta)$ (f)
are shown versus the scaled radial coordinate
$\eta/\eta_0$,
demonstrating that $v_e(\eta)$ exceeds the velocity of light
beyond a certain value of the radial coordinate.

\end{document}